\documentclass[11pt,letterpaper]{article}
\usepackage{putex}
\usepackage{graphicx}
\usepackage{latexsym,amsmath,amsfonts,amssymb,amsthm}
\usepackage{bbm}
\usepackage{cite}
\usepackage{indentfirst}
\usepackage{booktabs,array}
\usepackage{placeins}
\usepackage{mathtools}
\usepackage{cancel}
\usepackage[dvipsnames]{xcolor}
\usepackage{tikz}
\usepackage{adjustbox}
\usepackage{genyoungtabtikz}
\usepackage{enumitem}
\usepackage{setspace}
\usetikzlibrary{decorations.markings}
\usepackage{graphicx}
\usepackage[margin=10pt,font=small,labelfont=bf]{caption}
\usepackage{subcaption}
\usepackage{xcolor}
\usepackage{float}
\theoremstyle{definition}

\usepackage{microtype}
\usepackage{hyperref}

\hypersetup{pdfborderstyle={/S/U/W 0.5}}

\setlist{itemsep=2pt plus 1pt minus 1pt, topsep=2pt plus 1pt minus 1pt}

\renewenvironment{thebibliography}[1]{%
\begin{oldthebibliography}{#1}%
\setlength\itemsep{-2mm}%
}%
{%
\end{oldthebibliography}%
}

\newcommand{\ie}{\textsl{i.e.\@}}

\numberwithin{equation}{section}

\newcommand{\nn}{\nonumber}

\DeclareMathOperator{\Tr}{Tr}

\DeclareMathOperator*{\Res}{Res}

\DeclareMathOperator*{\SumInt}{%
\mathchoice%
  {\ooalign{$\displaystyle\sum$\cr\hidewidth$\displaystyle\int$\hidewidth\cr}}
  {\ooalign{\raisebox{.14\height}{\scalebox{.7}{$\textstyle\sum$}}\cr\hidewidth$\textstyle\int$\hidewidth\cr}}
  {\ooalign{\raisebox{.2\height}{\scalebox{.6}{$\scriptstyle\sum$}}\cr$\scriptstyle\int$\cr}}
  {\ooalign{\raisebox{.2\height}{\scalebox{.6}{$\scriptstyle\sum$}}\cr$\scriptstyle\int$\cr}}
}

\newcommand{\nl}{n^\text{L}}
\newcommand{\nla}{n^\text{L}_A}
\newcommand{\nr}{n^\text{R}}
\newcommand{\nra}{n^\text{R}_A}

\newcommand{\vnl}{\vec n^\text{L}}
\newcommand{\vnr}{\vec n^\text{R}}

\newcommand{\typeI}{{\color{black!30!green}type I }}
\newcommand{\typeII}{{\color{black!30!red}type II }}

\begin{document}


\title{\begin{LARGE}
{Intersecting Surface Defects and Instanton Partition Functions}
\end{LARGE}}

\authors{Yiwen Pan$^1$ and Wolfger Peelaers$^2$ 
\medskip\medskip\medskip\medskip\medskip\medskip\medskip\medskip
 }

\institution{UU}{${}^1$
Department of Physics and Astronomy, Uppsala University, \cr
$\;\,$ Box 516, SE-75120 Uppsala, Sweden}
\institution{NHETC}{${}^2$
New High Energy Theory Center, Rutgers University,  \cr
$\;\,$ Piscataway, NJ 08854, USA}

\abstract{We analyze intersecting surface defects inserted in interacting four-dimensional $\mathcal N=2$ supersymmetric quantum field theories. We employ the realization of a class of such systems as the infrared fixed points of renormalization group flows from larger theories, triggered by perturbed Seiberg-Witten monopole-like configurations, to compute their partition functions. These results are cast into the form of a partition function of 4d/2d/0d coupled systems. Our computations provide concrete expressions for the instanton partition function in the presence of intersecting defects and we study the corresponding ADHM model.
}

\preprint{UUITP-31/16}

\maketitle


{
\setcounter{tocdepth}{2}
\setlength\parskip{-0.7mm}
\tableofcontents
}

\section{Introduction}
Half-BPS codimension two defect operators form a rich class of observables in supersymmetric quantum field theories. Their vacuum expectation values, as those of all defect operators, are diagnostic tools to identify the phase of the quantum field theory \cite{Wilson:1974sk,tHooft:1977nqb,Gukov:2013zka}. Various quantum field theoretic constructions of codimension two defects have been proposed and explored in the literature, see for example the review \cite{Gukov:2014gja}. First, one can engineer a defect by defining a prescribed singularity for the gauge fields (and additional vector multiplet scalars) along the codimension two surface, as in \cite{Gukov:2006jk}. Second, a defect operator can be constructed by coupling a quantum field theory supported on its worldvolume to the bulk quantum field theory. The coupling can be achieved by gauging lower-dimensional flavor symmetries with higher-dimensional gauge fields and/or by turning on superpotential couplings. Third, a codimension two defect in a theory $\mathcal T$ can be designed in terms of a renormalization group flow from a larger theory $\widetilde{\mathcal T}$ triggered by a position-dependent, vortex-like Higgs branch vacuum expectation value \cite{Gaiotto:2012xa,Gaiotto:2014ina}.\footnote{The gauged perspective of \cite{Gaiotto:2012xa} is equivalent to considering sectors with fixed winding in a `Higgs branch localization' computation. See \cite{Benini:2012ui,Doroud:2012xw,Fujitsuka:2013fga,Benini:2013yva,Peelaers:2014ima,Pan:2014bwa,Closset:2015rna,Chen:2015fta,Pan:2015hza} for such computations in various dimensions.} Naturally, some defects can be constructed in multiple ways. Nevertheless, it is of importance to study all constructions separately, as their computational difficulties and conceptual merits vary. Such study is helped tremendously by the fact that when placing the theory on a compact Euclidean manifold, all three descriptions are, in principle, amenable to an exact analysis using localization techniques. See \cite{Pestun:2016zxk} for a recent comprehensive review on localization techniques.

The M-theory construction of four-dimensional $\mathcal N=2$ supersymmetric theories of class $\mathcal S$ (of type $A_{N-1}$)\cite{Gaiotto:2009we} allows one to identify the class of concrete defects of interest to this paper: adding additional stacks of M2-branes ending on the main stack of $N$ M5-branes can introduce surface defects in the four-dimensional theory. The thus obtained M2-brane defects are known to be labeled by a representation $\mathcal R$ of $SU(N)$. In \cite{Gomis:2014eya}, the two-dimensional quiver gauge theory residing on the support of the defect and its coupling to the bulk four-dimensional theory were identified in detail. In fact, for the case of defects labeled by symmetric representations two different coupled systems were proposed. For the purposes of this paper, it is important to remark that one of these descriptions can alternatively be obtained from the third construction described in the previous paragraph.\footnote{The fact the application of this Higgsins prescription introduces M2-brane defects labeled by symmetric representations was understood in the original paper \cite{Gaiotto:2012xa}, see for example also \cite{Alday:2013kda}.}

Allowing for simultaneous insertions of multiple half-BPS defects, intersecting each other along codimension four loci, while preserving one quarter of the supersymmetry, enlarges the collection of defects considerably and is very well-motivated. Indeed, in \cite{Gomis:2016ljm} it was conjectured and overwhelming evidence was found in favor of the statement that the squashed four-sphere partition function of theories of class $\mathcal S$ in the presence of intersecting M2-brane defects, wrapping two intersecting two-spheres, is the translation of the insertion of a generic degenerate vertex operator in the corresponding Liouville/Toda conformal field theory correlator through the AGT dictionary \cite{Alday:2009aq,Wyllard:2009hg}, extending and completing \cite{Gomis:2014eya,Alday:2009fs}. Note that such defects are labeled by a pair of representations $(\mathcal R',\mathcal R)$, which is precisely the defining information of a generic degenerate vertex operator in Liouville/Toda theory.\footnote{A generic degenerate momentum reads $\alpha = -b \Omega_{\mathcal R} - b^{-1} \Omega_{\mathcal R'}$, in terms of the highest weight vectors $\Omega_\mathcal{R}, \Omega_{\mathcal R'}$ of irreducible representations $\mathcal R$ and $\mathcal R'$ respectively, and $b$ parametrizes the Virasoro central charge.} 

In \cite{Gomis:2016ljm}, the insertion of intersecting defects was engineered by considering a coupled 4d/2d/0d system. In this description, the defect is engineered by coupling quantum field theories supported on the respective codimension two worldvolumes as well as additional degrees of freedom residing at their intersection to each other and to the bulk quantum field theory. The precise 4d/2d/0d coupled systems describing intersecting M2-brane defects were conjectured. As was also the case for a single defect, intersecting defects labeled by symmetric representations can be described by two different coupled systems.

A localization computation, performed explicitly in \cite{Gomis:2016ljm}, allows one to calculate the squashed four-sphere partition function of such system.\footnote{See also \cite{Lamy-Poirier:2014sea} for a localization computation in the presence of a single defect.} Let $\mathcal T$ denote the four-dimensional theory and let $\tau^{\text{L/R}}$ denote two-dimensional theories residing on the defects wrapping the two-spheres $S^2_{\text{L}}$ and $S^2_{\text{R}}$, which intersect each other at the north pole and south pole. The full partition function then takes the schematic form
\begin{equation}
  Z^{(\mathcal{T},S^2_\text{L} \cup S^2_\text{R} \subset S^4_b)} = \SumInt \ Z_{\text{pert}}^{(\mathcal T,S^4_b)}\ Z_{\text{pert}}^{(\tau^{\text{L}},S^2_\text{L})} \ Z_{\text{pert}}^{(\tau^{\text{R}},S^2_\text{R})} \   Z^{+}_{\text{intersection}}\ Z^{-}_{\text{intersection}}\ \left|Z_{\text{inst}}^{(\mathcal T, \mathbb R^2_{\text{L}}\cup \mathbb R^2_{\text{R}} \subset \mathbb R^4)}\right|^2 \;,\label{generalform4d2d0d}
\end{equation}
where the factors $Z_{\text{pert}}^{(T,\mathbb M)}$ denote the product of the classical action and one-loop determinant of the theory $T$ placed on the manifold $\mathbb M$ (in their Coulomb branch localized form). Furthermore, $Z^{\pm}_{\text{intersection}}$ are the one-loop determinants of the degrees of freedom at the two intersection points respectively, and $|Z_{\text{inst}}^{(\mathcal T, \mathbb R^2_{\text{L}}\cup \mathbb R^2_{\text{R}} \subset \mathbb R^4)}|^2$ are two copies of the instanton partition function, one for the north pole and one for the south pole, describing instantons in the presence of the intersecting surface defects spanning the local coordinate planes $\mathbb R^2_{\text{L}}\cup \mathbb R^2_{\text{R}}$ in $\mathbb R^4$. In \cite{Gomis:2016ljm} the focus was on the already very rich dynamics of 4d/2d/0d systems without four-dimensional gauge fields, thus avoiding the intricacies of the instanton partition functions. In this paper we aim at considering intersecting defects in interacting four-dimensional field theories and addressing the problem of instanton counting in the presence of such defects.\footnote{By taking one of the intersecting defects to be trivial, one can always simplify our results to the case of a single defect. In \cite{Gomis:2014eya} an extensive study was performed of the squashed four-sphere partition function of theories of free hypermultiplets in the presence of a single defect.}

Our approach will be, alternative to that in \cite{Gomis:2016ljm}, to construct theories $\mathcal T$ in the presence of intersecting M2-brane defects labeled by symmetric representations using the aforementioned third strategy, \ie{}, by considering a renormalization group flow from a larger theory $\widetilde {\mathcal T}$ triggered by a position-dependent vacuum expectation value with an intersecting vortex-like profile.\footnote{To be more precise, the configuration that triggers the renormalization group flow is a solution to the (perturbed) Seiberg-Witten monopole equations \cite{Witten:1994cg}, see \cite{Pan:2015hza}.} When the theory $\widetilde{\mathcal{T}}$ is a Lagrangian theory on $S^4_b$, this Higgsing prescription offers a straightforward computational tool to calculate the partition function $Z^{(\mathcal{T},S^2_\text{L} \cup S^2_\text{R} \subset S^4_b)}$ of $\mathcal T$ in the presence of said intersecting defects. In more detail, it instructs one to consider the residue of a certain pole of the partition function $Z^{(\mathcal{\tilde{T}}, S^4_b)}$, which can be calculated by considering pinching poles of the integrand of the matrix integral computing $Z^{(\widetilde{\mathcal{T}}, S^4_b)}$. The result involves intricate sums over a restricted set of Young diagrams, which we subsequently cast in the form of a coupled 4d/2d/0d system as in \eqref{generalform4d2d0d}, by reorganizing the sums over the restricted diagrams into the integrals over gauge equivariant parameters and sums over magnetic fluxes of the partition functions of the two-dimensional theories $\tau^\text{L/R}$. This step heavily relies on factorization properties of the summand of instanton partition functions, which we derive in appendix \ref{appendix:IPF-factorization}, when evaluated at special values of their gauge equivariant parameter. More importantly, we obtain concrete expressions for the instanton partition function, computing the equivariant volume of the instanton moduli space in the presence of intersecting codimension two singularities, and their corresponding ADHM matrix model.

\begin{figure}[t]
  \centering
  \includegraphics[width=0.6\textwidth]{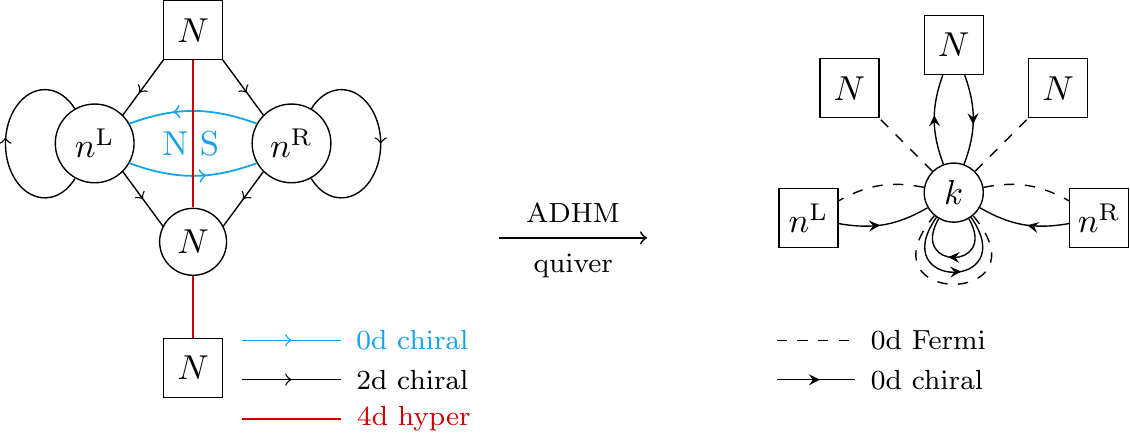}
  \caption{\label{fig:ADHM} On the left, the coupled 4d/2d/0d quiver gauge theory realizing the insertion, in four-dimensional $\mathcal N=2$ SQCD, of intersecting M2-brane surface defects labeled by symmetric representations of rank $n^{\text{R}}$ and $n^{\text{L}}$ respectively is depicted. The zero-dimensional multiplets are denoted using two-dimensional $\mathcal N=(0,2)$ quiver notation reduced to zero dimensions. Various superpotential couplings are turned on, in direct analogy to the ones given in detail in \cite{Gomis:2016ljm}. On the right, the ADHM model for $k$-instantons of the left theory is shown. The model preserves the dimensional reduction to zero dimensions of two-dimensional $\mathcal N=(0,2)$ supersymmetry. We used the corresponding quiver conventions. A J-type superpotential equal to the sum of the $U(k)$ adjoint bilinears formed out of the two pairs of chiral multiplets is turned on for the adjoint Fermi multiplet. The flavor charges carried by the various multiplets are also compatible with a quadratic J- or E-type superpotential for the Fermi multiplets charged under $U(n^{\text{L/R}})$.\protect\footnotemark{}}
\end{figure}

The main result of the paper, thus obtained, is the $S^4_b$-partition function of a four-dimensional $\mathcal N=2$ $SU(N)$ gauge theory with $N$ fundamental and $N$ antifundamental hypermultiplets,\footnote{While there is no distinction between a fundamental and antifundamental hypermultiplet, it is a useful terminology to keep track of the respective quiver gauge theory nodes. We choose to call the right/upper node of each link the fundamental one.\label{footnote:afund-fund}} \ie{}, SQCD, in the presence of intersecting M2-brane surface defects, labeled by $n^{\text{R}}$ and $n^{\text{L}}$-fold symmetric representations respectively. It takes the form \eqref{generalform4d2d0d} and can be found explicitly in \eqref{def:SQCDA-defect-matrix-model}. To be more precise, the coupled system we obtain involves chiral multiplets as zero-dimensional degrees of freedom, \ie{}, it coincides with the one described in conjecture 4 of \cite{Gomis:2016ljm} with four-dimensional $\mathcal{N} = 2$ SQCD. The left subfigure in figure \ref{fig:ADHM} depicts the 4d/2d/0d coupled system under consideration. 
We derive the instanton partition function $Z_{\text{inst}}^{(\mathcal T, \mathbb R^2_{\text{L}}\cup \mathbb R^2_{\text{R}} \subset \mathbb R^4)}$ in the presence of intersecting planar surface defects and find it to take the form
\begin{equation}
Z_{\text{inst}}^{(\mathcal T, \mathbb R^2_{\text{L}}\cup \mathbb R^2_{\text{R}} \subset \mathbb R^4)} = \sum_{\vec Y} {{q}^{|\vec Y|}} \ z_{{\text{vect}}}^{{\mathbb{R}^4}}(\vec Y)\  z_{{\text{afund}}}^{{\mathbb{R}^4}}(\vec Y)\ z_{{\text{fund}}}^{{\mathbb{R}^4}}(\vec Y) \  z_{{\text{defect}}}^{\mathbb{R}_{\text{L}}^2}(\vec Y)\ z_{{\text{defect}}}^{\mathbb{R}_{\text{R}}^2}(\vec Y)\;,
\end{equation}
where we omitted all gauge and flavor equivariant parameters. It is expressed as the usual sum over $N$-tuples $\vec Y$ of Young diagrams. The summand contains the new fctors $z_{{\text{defect}}}^{\mathbb{R}_{\text{L/R}}^2}$, which can be found explicitly in \eqref{zdefect}, capturing the contributions to the instanton counting of the additional zero-modes in the presence of intersecting surface defects, in addition to the standard factors $z_{{\text{vect}}}^{{\mathbb{R}^4}}$, $z_{{\text{fund}}}^{{\mathbb{R}^4}}$ and $z_{{\text{afund}}}^{{\mathbb{R}^4}}$ describing the contributions from the vector multiplet and $N+N$ hypermultiplets. The coefficient of $q^k$ of the above result can be derived from the ADHM model for $k$-instantons depicted in the right subfigure of figure \ref{fig:ADHM}. We have confirmed this ADHM model by analyzing the brane construction of said instantons, see section \ref{sec:IPFwDef} for all the details. In section \ref{sec:conclusions} we present conjectural generalizations of the instanton counting in the case of generic intersecting M2-brane defects.

\footnotetext{The partition function is insensitive to the presence of superpotential couplings.}The paper is organized as follows. We start in section \ref{section:Higgsing_Prescription} by briefly recalling the Higgsing prescription to compute squashed sphere partition functions in the presence of (intersecting) M2-brane defects labeled by symmetric representations. We also present its brane realization. In section \ref{section:free-hyper} we implement the prescription for the case where $\mathcal T$ is a four- or five-dimensional theory of $N^2$ free hypermultiplets placed on a squashed sphere. The vacuum expectation value in $\mathcal T$ of intersecting M2-brane defects on the sphere has been computed in \cite{Gomis:2016ljm} from the point of view of the 4d/2d/0d or 5d/3d/1d coupled system and takes the form \eqref{generalform4d2d0d} (without the instanton contributions). For the case of symmetric representations, we reproduce this expression directly, and provide a derivation of a few details that were not addressed in \cite{Gomis:2016ljm}. We notice that the superpotential constraints of the coupled system on the parameters appearing in the partition function are reproduced effortlessly in the Higgsing computation thanks to the fact that they have a common origin in the theory $\widetilde{\mathcal T}$, which in this case is SQCD. These relatively simple examples allow us to show in some detail the interplay of the various ingredients of the Higgsed partition function of theory $\widetilde{\mathcal T}$, and how to cast it in the form \eqref{generalform4d2d0d}. In section \ref{section: interacting theories} we turn our attention to inserting defects in four-dimensional $\mathcal N=2$ SQCD. We apply the Higgsing prescription to an $SU(N)\times SU(N)$ gauge theory with bifundamental hypermultiplets and for each gauge group an additional $N$ fundamental hypermultiplets, and cast the resulting partition function in the form \eqref{generalform4d2d0d}. As a result we obtain a sharp prediction for the instanton partition function in the presence of intersecting surface defects. This expression provides concrete support for the ADHM matrix model that we obtain in section \ref{sec:IPFwDef} from a brane construction. We present our conclusions and some future directions in section \ref{sec:conclusions}. Five appendices contain various technical details and computations.

\section{Higgsing and codimension two defects}\label{section:Higgsing_Prescription}

In this section we start by briefly recalling the Higgsing prescription to compute the partition function of a theory $\mathcal T$ in the presence of (intersecting) defects placed on the squashed four/five-sphere  \cite{Gaiotto:2012xa,Gaiotto:2014ina}. We also consider the brane realization of this prescription, which provides a natural bridge to the description of intersecting surface defects in terms of a 4d/2d/0d (or 5d/3d/1d) coupled system as in \cite{Gomis:2016ljm}.

\subsection{The Higgsing prescription}\label{subsec:Higgsing_Prescription}
We will be interested in four/five-dimensional quantum field theories with eight supercharges.\footnote{The localization computations we will employ throughout this paper rely on a Lagrangian description, but the Higgsing prescription is applicable outside the realm of Lagrangian theories. We will restrict attention to (Lagrangian) four-dimensional $\mathcal N=2$ supersymmetric quantum field theories of class $\mathcal S$ and their five-dimensional uplift.} Let us for concreteness start by considering four-dimensional $\mathcal N=2$ supersymmetric theories. Consider a theory $\mathcal T$ whose flavor symmetry contains an $SU(N)$ factor, and consider the theory of $N^2$ free hypermultiplets, which has flavor symmetry $USp(2N^2)\supset SU(N)\times SU(N)\times U(1)$. By gauging the diagonal subgroup of the $SU(N)$ flavor symmetry factor of the former theory with one of the $SU(N)$ factors of the latter theory, we obtain a new theory $\widetilde{\mathcal T}$. As compared to $\mathcal T$, the theory $\widetilde{\mathcal T}$ has an extra $U(1)$ factor in its flavor symmetry group. We denote the corresponding mass parameter as $\check M.$

The theory $\widetilde{\mathcal T}$ can be placed on the squashed four-sphere $S^4_b$,\footnote{\label{definitionsquashedfoursphere}We consider $S^4_b$ defined through the embedding equation in five-dimensional Euclidean space $\mathbb R^5 = \mathbb R\times \mathbb C^2$ with coordinates $x,z_1,z_2$ 
\begin{equation*}
\frac{x^2}{r^2} +  \frac{|z_1|^2}{ \ell^2} + \frac{|z_2|^2}{\tilde \ell^2} = 1\;,
\end{equation*} 
in terms of parameters $r,\ell,\tilde \ell$ with dimension of length. The squashing parameter $b$ is defined as $b^2=\frac{\ell}{\tilde \ell}$. The isometries of $S^4_b$ are given by $U(1)^{\text{R}}\times U(1)^{\text{L}}$, which act by rotating the $z_1$ and $z_2$ plane respectively. The fixed locus of $U(1)^{\text{R}}$ is a squashed two-spheres: $S^2_{\text{R}} = S^4_b \big|_{z_1 = 0}$ and, similarly, the fixed locus of $U(1)^{\text{L}}$ is $S^2_{\text{L}} = S^4_b \big|_{z_2 = 0}$. The two-spheres $S^2_{\text{R}}$ and $S^2_{\text{L}}$ intersect at their north pole and south pole, \ie{}, the points with coordinates $z_1=z_2=0$ and $x_0=\pm r$.} and its partition function can be computed using localization techniques \cite{Pestun:2007rz,Hama:2012bg}. Let us denote the supercharge used to localize the theory as $\mathcal Q$. Its square is given by
\begin{equation}
\mathcal Q^2 = b^{-1} \mathcal M_{\text{R}} + b \mathcal M_{\text{L}} - (b+b^{-1}) \mathcal R + i\sum_J M_J F_J + \text{gauge transformation}\;,
\end{equation}
where $\mathcal M_{\text{R/L}}$ are generators of the $U(1)^{\text{R/L}}$ isometries of $S^4_b$ (see footnote \ref{definitionsquashedfoursphere}), $\mathcal R$ is the $SU(2)_{\mathcal R}$ Cartan generator and $F_J$ are the Cartan generators of the flavor symmetry algebra. The coefficients $M_J$ are mass parameters rescaled by $\sqrt{\ell\tilde\ell}$, where $\ell$ and $\tilde \ell$ are two radii of the squashed sphere (see footnote \ref{definitionsquashedfoursphere}), to make them dimensionless. Localization techniques simplify the computation of the $S^4_b$ partition function to the calculation of one-loop determinants of quadratic fluctuations around the localization locus given by arbitrary constant values for $\Sigma^{\widetilde{\mathcal T}}$, the imaginary part of the vector multiplet scalar of the total gauge group.\footnote{More precisely, this is the ``Coulomb branch localization'' locus. Alternatively, one can perform a ``Higgs branch localization'' computation, see \cite{Chen:2015fta,Pan:2015hza}.} The final result for the $S^4_b$ partition function of the theory $\widetilde{\mathcal T}$ is then
\begin{equation}\label{S4bpartitionfunction}
Z^{(\widetilde{\mathcal T},S^4_b)}(M) = \int \text{d}\Sigma^{\widetilde{\mathcal T}}\  Z_{\text{cl}}^{(\widetilde{\mathcal T},S^4_b)}(\Sigma^{\widetilde{\mathcal T}})\  Z_{\text{1-loop}}^{(\widetilde{\mathcal T},S^4_b)}(\Sigma^{\widetilde{\mathcal T}},M) \ |Z_{\text{inst}}^{(\widetilde{\mathcal T},\mathbb R^4)}(q,\Sigma ,M^\epsilon )|^2 \;,
\end{equation}
where $Z_{\text{cl}}^{(\widetilde{\mathcal T},S^4_b)}$ denotes the classical action evaluated on the localization locus, $Z_{\text{1-loop}}^{(\widetilde{\mathcal T},S^4_b)}$ is the one-loop determinant and $|Z_{\text{inst}}^{(\widetilde{\mathcal T},\mathbb R^4)}(q,\Sigma ,M^\epsilon )|^2$ are two copies of the Nekrasov instanton partition function \cite{Nekrasov:2002qd,Nekrasov:2003rj}, capturing the contribution to the localized path integral of instantons residing at the north and south pole of $S^4_b$.

In \cite{Gaiotto:2012xa,Gaiotto:2014ina}, it was argued, by considering the physics at the infrared fixed point of the renormalization group flow triggered by a position dependent Higgs branch vacuum expectation value for the baryon constructed out of the hypermultiplet scalars, which carries charges $\mathcal M_{\text{L}} = -n^{\text{L}}, \mathcal M_{\text{R}} = -n^{\text{R}}, \mathcal R = N/2$ and $\check F = N$, that the partition function $Z^{(\widetilde{\mathcal T},S^4_b)}(M)$ necessarily has a pole when
\begin{equation}\label{poleposition}
i \check M = \frac{b+b^{-1}}{2} + b^{-1} \frac{n^{\text{R}}}{N} + b \frac{n^{\text{L}}}{N}\;.
\end{equation}
Moreover, the residue of the pole precisely captures the partition function of the theory $\mathcal T$ in the presence of M2-brane surface defects labeled by $n^{\text{R}}$-fold and $n^{\text{L}}$-fold symmetric representations respectively up to the left-over contribution of the hypermultiplet that captures the fluctuations around the Higgs branch vacuum. These defects wrap two intersecting two-spheres $S^2_{\text{R/L}},$ the fixed loci of $U(1)^{\text{R/L}}$.

The pole at \eqref{poleposition} of $Z^{(\widetilde{\mathcal T},S^4_b)}(M)$ finds its origin in the matrix integral \eqref{S4bpartitionfunction} because of poles of the integrand pinching the integration contour. To see this, let us separate out the $SU(N)$ gauge group that gauges the free hypermultiplet to $\mathcal T$, and split $\Sigma^{\widetilde{\mathcal T}}$ accordingly: $\Sigma^{\widetilde{\mathcal T}} = (\Sigma^{\mathcal T}, \Sigma),$ where $\Sigma^{\mathcal T}$ is the vector multiplet scalar of the full gauge group of theory $\mathcal T,$ and $\Sigma$ the $SU(N)$ vector multiplet scalar. We can then rewrite \eqref{S4bpartitionfunction} as
\begin{multline}\label{S4bpartitionfunctionsplit}
Z^{(\widetilde{\mathcal T},S^4_b)}(M) = \int \text{d}\Sigma^{\mathcal T}\ \int \text{d}\Sigma\ Z_{\text{cl}}^{(\widetilde{\mathcal T},S^4_b)}(\Sigma^{\mathcal T},\Sigma)\  Z_{\text{1-loop}}^{({\mathcal T},S^4_b)}(\Sigma^{\mathcal T},\Sigma,M) \ |Z_{\text{inst}}^{(\widetilde{\mathcal T},\mathbb R^4)}(q,\Sigma^{\mathcal T},\Sigma,M^\epsilon)|^2 \\
\times \prod_{\substack{A,B=1\\A\neq B}}^{N} \Upsilon_b(i(\Sigma_A - \Sigma_B))\  \prod_{A=1}^{N}\prod_{I=1}^{N}\Upsilon_b\left(i(\Sigma_A-M_I-\check M)+\frac{Q}{2}\right)^{-1}\;.
\end{multline}
The first factor in the second line is the one-loop determinant of the $SU(N)$ vector multiplet, while the second factor is the contribution of the $N^2$ extra hypermultiplets, organized into $N$ $SU(N)$ fundamental hypermultipets.\footnote{See appendix \ref{app:Special functions} for the definition and some useful properties of the various special functions that are used throughout the paper.} Here $M_I, I=1,\ldots, N$ denote the mass parameters associated to the $SU(N)$ flavor symmetry (with $\sum_I M_I =0$). The integrand of the $\Sigma$-integral has poles (among many others) located at
\begin{equation}\label{polepositionspinching}
i\Sigma_A = i M_{\sigma(A)} + i \check M - n^{\text{R}}_A b^{-1} - n^{\text{L}}_A b - \frac{b+b^{-1}}{2} \qquad \text{with} \qquad n^{\text{R/L}}_A\geq 0\;, \quad A=1,\ldots,N\;,
\end{equation}
where $\sigma$ denotes a permutation of $N$ variables. These poles arise from the one-loop determinant of the extra hypermultiplets. When the $U(1)$ mass parameter $\check M$ takes the value of \eqref{poleposition}, they pinch the integration contour if
\begin{equation}\label{partitions}
n^{\text{R}} = \sum_{A=1}^{N} n^{\text{R}}_A\;, \qquad  n^{\text{L}} = \sum_{A=1}^{N} n^{\text{L}}_A\;, 
\end{equation}
since we only have $N-1$ independent $SU(N)$ integration variables. Note that the residue of the pole of $Z^{(\widetilde{\mathcal T},S^4_b)}$ at \eqref{poleposition} is equal to the sum over all partitions of $n^{\text{R}},n^{\text{L}}$ in \eqref{partitions} of the residue of the $\Sigma$-integrand of $Z^{(\widetilde{\mathcal T},S^4_b)}$ at the pole position \eqref{polepositionspinching} when treating the $\Sigma_A$ as $N$ independent variables.\footnote{Upon gauging the additional $U(1)$ flavor symmetry and turning on a Fayet-Iliopoulos parameter, which coincides with the gauged setup of \cite{Gaiotto:2012xa,Gaiotto:2014ina}, the residues of precisely these poles were given meaning in the ``Higgs branch localization'' computation of \cite{Pan:2015hza} in terms of Seiberg-Witten monopoles.}

A similar analysis can be performed for five-dimensional $\mathcal N=1$ theories. The theory $\widetilde {\mathcal T}$ can be put on the squashed five-sphere $S^5_{\vec\omega}$,\footnote{\label{definitionsquashedfivesphere}The squashed five-sphere $S^5_{\vec\omega = (\omega_1,\omega_2,\omega_3)}$ is given by the locus in $\mathbb C^3$ satisfying
\begin{equation}
\omega_1^2 |z_1|^2 + \omega_2^2 |z_2|^2 + \omega_3^2 |z_3|^2 = 1\;.
\end{equation}
Its isometries are $U(1)^{(1)}\times U(1)^{(2)} \times U(1)^{(3)},$ which act by rotations on the three complex planes respectively. The fixed locus of $U(1)^{(\alpha)}$ is the squashed three-sphere $S^3_{(\alpha)} = S^5_{\vec\omega} \big |_{z_\alpha=0},$ while the fixed locus of $U(1)^{(\alpha)}\times U(1)^{(\beta\neq \alpha)}$ is the circle $S^1_{(\alpha\cap \beta)} = S^5_{\vec\omega} \big |_{z_\alpha=z_\beta=0}.$ The notation indicates that it appears as the intersection of the three-spheres $S^3_{(\alpha)}$ and $S^3_{(\beta)}.$ A convenient visualization of the five-sphere and its fixed loci under one or two of the $U(1)$ isometries is as a $T^3$-fibration over a solid triangle, where on the edges one of the cycles shrinks and at the corners two cycles shrink simultanously.} and its partition function can again be computed using localization techniques \cite{Hosomichi:2012ek,Kallen:2012va,Kim:2012ava,Imamura:2012bm,Lockhart:2012vp,Kim:2012qf}. The localizing supercharge $\mathcal Q$ squares to
\begin{equation}
\mathcal Q^2 = \sum_{\alpha=1}^3 \omega_\alpha \mathcal M_{(\alpha)} -(\omega_1+\omega_2+\omega_3) \mathcal R + i \sum_J M_J F_J + \text{gauge transformation}\;,
\end{equation}
where $\mathcal M_{(\alpha)}$ are the generators of the $U(1)^{(1)}\times U(1)^{(2)} \times U(1)^{(3)}$ isometry of the squashed five-sphere $S^5_{\vec\omega}$ (see footnote \ref{definitionsquashedfivesphere}). The localization locus consists of arbitrary constant values for the vector multiplet scalar $\Sigma^{\widetilde{\mathcal T}}$, hence the partition function reads
\begin{equation}\label{S5omegapartitionfunction}
  Z^{(\widetilde{\mathcal T},S^5_{\vec\omega})}(M) = \int \text{d}\Sigma^{\widetilde{\mathcal T}}\  Z_{\text{cl}}^{(\widetilde{\mathcal T},S^5_{\vec\omega})}(\Sigma^{\widetilde{\mathcal T}})\  Z_{\text{1-loop}}^{(\widetilde{\mathcal T},S^5_{\vec\omega})}(\Sigma^{\widetilde{\mathcal T}},M) \ |Z_{\text{inst}}^{(\widetilde{\mathcal T},\mathbb R^4\times S^1)}(q,\Sigma^{\widetilde{\mathcal{T}}} ,M^\omega )|^3 \;.
\end{equation}
One can argue that $Z^{(\widetilde{\mathcal T},S^5_{\vec\omega})}(M)$ has a pole at
\begin{equation}\label{polepositionS5}
i \check M = \frac{\omega_1+\omega_2+\omega_3}{2} + \sum_{i=1}^3 \omega_\alpha \frac{n^{(\alpha)}}{N}\;,
\end{equation}
whose residue computes the $S^5_{\vec\omega}$ partition function of $\mathcal T$ in the presence of codimension two defects labeled by $n^{(\alpha)}$-fold symmetric representations and wrapping the three-spheres $S^3_{(\alpha)}$ obtained as the fixed loci of the $U(1)^{(\alpha)}$ isometries (see footnote \ref{definitionsquashedfivesphere}), respectively. These three-spheres intersect each other in pairs along a circle. Again, this pole arises from pinching the integration contour by poles of the one-loop determinant of the $N^2$ hypermultiplets located at
\begin{equation}\label{polepositionsS5pinching}
i\Sigma_A = i M_{\sigma(A)} + i \check M -  \sum_{\alpha=1}^3 n_A^{(\alpha)} \omega_{\alpha} - \frac{\omega_1+\omega_2+\omega_3}{2} \qquad \text{with} \qquad n_A^{(\alpha)}\geq 0\;, \quad A=1,\ldots,N\;,
\end{equation}
if $n^{(\alpha)} = \sum_{A=1}^{N} n_A^{(\alpha)}$. The residue of $Z^{(\widetilde{\mathcal T},S^5_{\vec\omega})}(M)$ at the pole given in \eqref{polepositionS5} equals the sum over partitions of the integers $n^{(\alpha)}$ of the residue of the integrand at the pole position \eqref{polepositionsS5pinching} with the $\Sigma_A$ treated as independent variables.\footnote{In \cite{Pan:2014bwa}, these residues were interpreted as the contribution to the partition function of K-theoretic Seiberg-Witten monopoles.}

\subsection{Brane realization}\label{subsec:brane realization}
\begin{figure}[t]
  \centering
  \includegraphics[width=\textwidth]{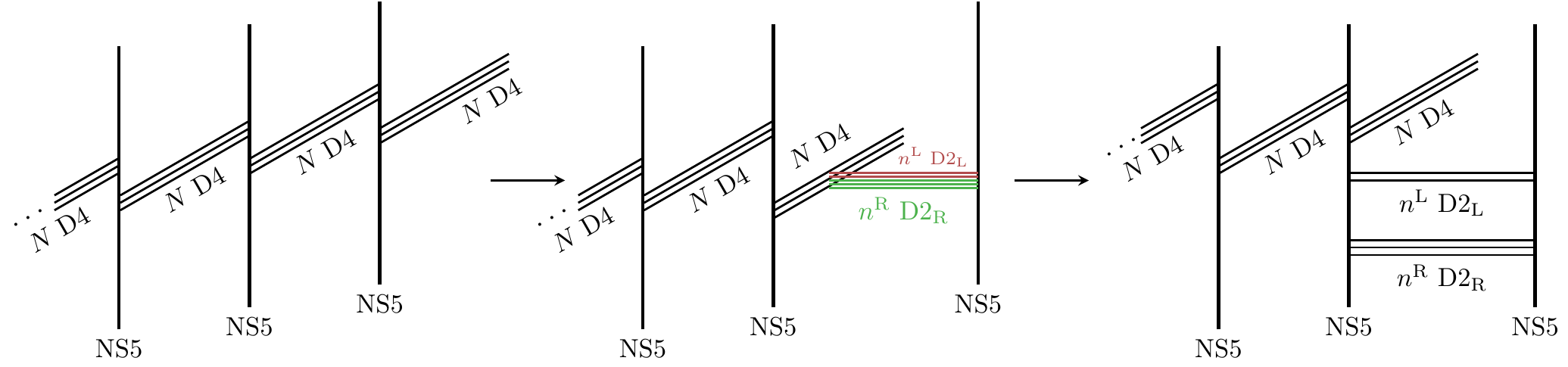}
  \caption{\label{fig:manynodequiver_higgsed} Gauging the diagonal subgroup of the $SU(N)$ flavor symmetry carried by the right-hand stack of D4-branes and an $SU(N)$ subgroup of the flavor symmetry of an additional $N^2$ free hypermultiplets amounts to adding an additional NS5-brane on the right end of the brane array. This leads to the figure on the left. Higgsing the system as in subsection \ref{subsec:Higgsing_Prescription} corresponds to pulling away this NS5-brane from the main stack, while stretching $n^{\text{R}}$ D2$_\text{R}$ and $n^{\text{L}}$ D2$_\text{L}$-branes in between it and the D4-branes, producing the middle figure. The final figure represents the system in the Coulomb phase.}
\end{figure}

To sharpen one's intuition of the Higgsing prescription outlined in the previous subsection, one may look at its brane realization \cite{Gaiotto:2014ina}. Consider a four-dimensional $\mathcal N=2$ gauge theory $\mathcal T$ described by the linear quiver and corresponding type IIA brane configuration\footnote{\label{branedirections}The branes in this figure as well as those in figure \ref{fig:manynodequiver_higgsed} and the following discussion span the following dimensions:
\begin{center}
  \begin{tabular}{l | cccccccccc}
              & $1$ & $2$ & $3$ & $4$ & $5$ & $6$ & $7$ & $8$ & $9$ & $10$ \\
    \hline
    NS5                   & --- & --- & --- & --- & --- & --- &     &     &     &     \\
    \ \ D4                & --- & --- & --- & --- &     &     & --- &     &     &     \\
    \ \ D2$_\text{L}$     & --- & --- &     &     &     &     &     &     &     & --- \\
    \ \ D2$_\text{R}$     &     &     & --- & --- &     &     &     &     &     & --- \\
    \ \ D0                &     &     &     &     &     &     & --- &     &     &     \\
  \end{tabular}
\end{center}
}
\begin{center}
\includegraphics[width=0.8\textwidth]{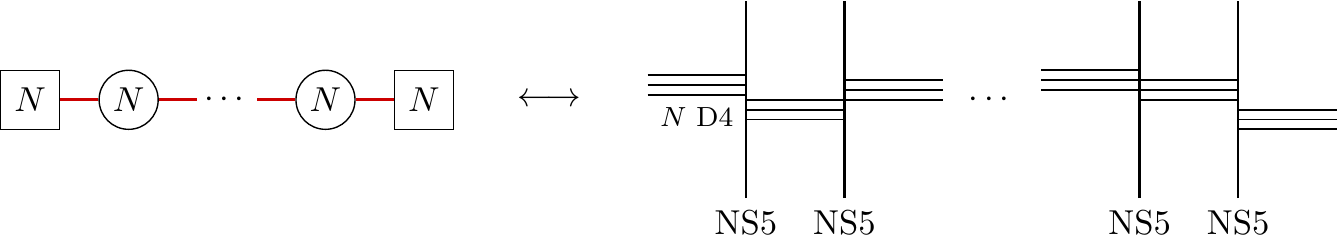}
\end{center}
Gauging in a theory of $N^2$ hypermultiplets amounts to adding an additional NS5-brane on the right end of the brane array. The Higgsing prescription of the previous subsection is then trivially implemented by pulling away this additional NS5-brane (in the 10-direction of footnote \ref{branedirections}), while suspending $n^{\text{R}}$ D2$_\text{R}$ and $n^{\text{L}}$ D2$_\text{L}$-branes between the displaced NS5-brane and the right stack of D4-branes, see figure \ref{fig:manynodequiver_higgsed}. 

\begin{figure}[t!]
  \centering
  \includegraphics[width=0.4\textwidth]{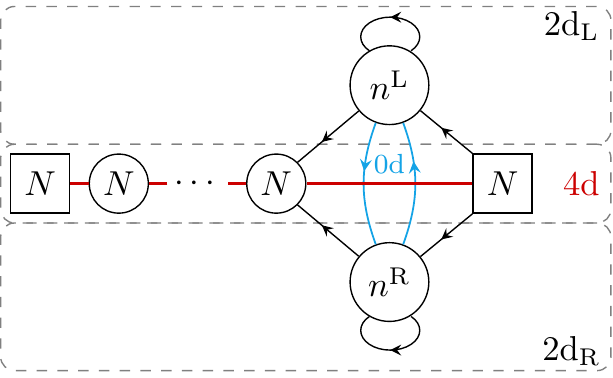}
  \caption{\label{fig:InsertSymmetrics} Coupled 4d/2d/0d quiver gauge theory realizing intersecting M2-brane surface defects labeled by symmetric representations, of rank $n^{\text{R}}$ and $n^{\text{L}}$ respectively, in a four-dimensional $\mathcal N=2$ linear quiver gauge theory. The two-dimensional degrees of freedom, depicted in $\mathcal N=(2,2)$ quiver notation, are coupled to the four-dimensional ones through cubic and quartic superpotential couplings. The explicit superpotentials can be found in \cite{Gomis:2016ljm}. The zero-dimensional degrees of freedom, denoted using two-dimensional $\mathcal N=(0,2)$ quiver notations dimensionally reduced to zero dimensions, with solid lines representing chiral multiplets, participate in E- and J-type superpotentials. }
\end{figure}

Various observations should be made. First of all, the brane picture in figure \ref{fig:manynodequiver_higgsed} was also considered in \cite{Gomis:2016ljm} to describe intersecting M2-brane surface defects labeled by $n^{\text{R}}$ and $n^{\text{L}}$-fold symmetric representations respectively. Its field theory realization is described by a coupled 4d/2d/0d system, described by the quiver in figure \ref{fig:InsertSymmetrics} (see \cite{Gomis:2016ljm}). 
Note that the two-dimensional theories, residing on the D2$_\text{R}$ and D2$_\text{L}$-branes, are in their Higgs phase, with equal Fayet-Iliopoulos parameter $\xi_{\text{FI}}$ proportional to the distance (in the 7-direction) between the displaced NS5-brane and the next right-most NS5-brane. Before Higgsing, this distance was proportional to the inverse square of the gauge coupling of the extra $SU(N)$ gauge node:
\begin{equation}
\xi_{\text{FI}} = \frac{4\pi}{g_{\text{YM}}^2}\;.
\end{equation}
In particular, the Higgsing prescription will produce gauge theory results in the regime where $\xi_{\text{FI}}$ is positive, and where the defect is inserted at the right-most end of the quiver. In this paper we will restrict attention to this regime. Note however that sliding the displaced NS5-brane along the brane array in figure \ref{fig:manynodequiver_higgsed} implements hopping dualities \cite{Gadde:2013ftv,Gomis:2014eya} (see also \cite{Assel:2015oxa, Closset:2016arn}), which in the quiver gauge theory description of figure \ref{fig:InsertSymmetrics} translate to coupling the defect world volume theory to a different pair of neighboring nodes of the four-dimensional quiver, while not changing the resulting partition function.

In \cite{Gomis:2016ljm}, a first-principles localization computation was performed to calculate the partition function of the coupled 4d/2d/0d system when placed on a squashed four-sphere, with the defects wrapping two intersecting two-spheres $S^2_{\text{R/L}},$ the fixed loci of $U(1)^{\text{R/L}}$, in the case of non-interacting four-dimensional theories. Our aim in the next section will be to reproduce these results from the Higgsing point of view. When the four-dimensional theory contains gauge fields, the localization computation needs as input the Nekrasov instanton partition function in the presence of intersecting planar surface defects, which modify non-trivially the ADHM data. The Higgsing prescription does not require such input, and in section \ref{section: interacting theories} we will apply it to $\mathcal N=2$ SQCD. This computation will allow us to extract the modified ADHM integral.

The brane realization of figure \ref{fig:manynodequiver_higgsed} already provides compelling hints about how the ADHM data should be modified. In this setup, instantons are described by D0-branes stretching between the NS5-branes. Their worldvolume theory is enriched by massless modes (in the Coulomb phase, \ie{}, when $\xi_{\text{FI}}=0$), if any, arising from open strings stretching between the D0-branes and the D2$_\text{R}$ and D2$_\text{L}$-branes. These give rise to the dimensional reduction of a two-dimensional $\mathcal N=(2,2)$ chiral multiplet to zero dimensions, or equivalently, the dimensional reduction of a two-dimensional $\mathcal N=(0,2)$ chiral multiplet and Fermi multiplet. We will provide more details about the instanton counting in the presence of defects in section \ref{sec:IPFwDef}. Our Higgsing computation of section \ref{section: interacting theories} will provide an independent verification of these arguments.

\section{Intersecting defects in theory of \texorpdfstring{$N^2$}{N2} free hypermultiplets }\label{section:free-hyper}
In this section we work out in some detail the Higgsing computation for the case where $\mathcal T$ is a theory of free hypermultiplets. We will find perfect agreement with the description of intersecting M2-brane defects labeled by symmetric representations in terms of a 4d/2d/0d (or 5d/3d/1d) system\cite{Gomis:2016ljm}. Our computation also provides a derivation of the Jeffrey-Kirwan-like residue prescription used to evaluate the partition function of the coupled 4d/2d/0d (or 5d/3d/1d) system, and of the flavor charges of the degrees of freedom living on the intersection. In the next section we will consider the case of interacting theories $\mathcal T$.

\subsection{Intersecting codimension two defects on \texorpdfstring{$S^5_{\vec{\omega}}$}{S5}}
\begin{figure}[t]
  \centering
  \includegraphics[width=0.5\textwidth]{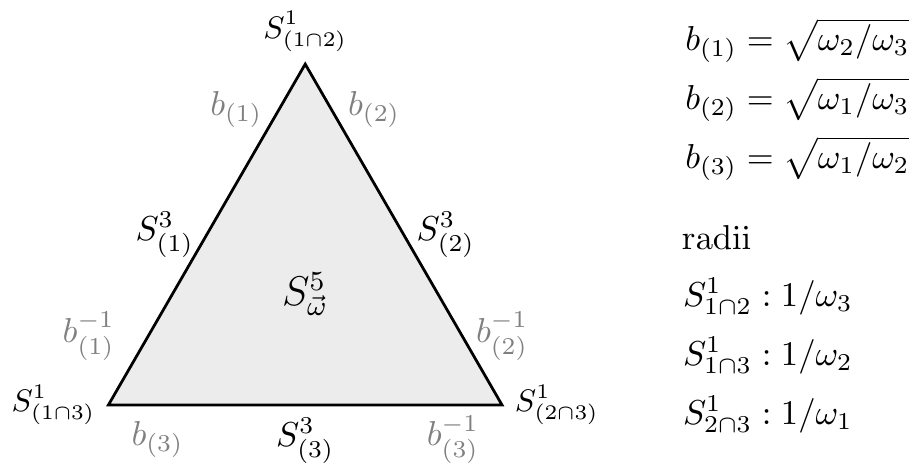}
  \caption{\label{figure:T3fibration} A convenient representation of $S^5_{\vec \omega}$ in terms of a $T^3$-fibration over a triangle. Each edge represents a three-sphere invariant point-wise under one of the $U(1)$ isometries, and each vertex represents an $S^1$, where two $S^3$'s intersect, invariant point-wise under two $U(1)$ isometries. Each $S^1$ has two tubular neighborhoods of the form $S^1 \times \mathbb{R}^2$ in the two intersecting $S^3$'s, with omega-deformation parameters given in terms of $b_{(\alpha)}^{\pm 1}$, as shown in the figure.}
\end{figure}

As a first application of the Higgsing prescription of the previous section, we consider the partition function of a theory of $N^2$ free hypermultiplets on $S^5_{\vec{\omega}}$ in the presence of intersecting codimension two defects wrapping two of the three-spheres $S^3_{(\alpha)}$ fixed by the $U(1)^{(\alpha)}$ isometry (see footnote \ref{definitionsquashedfivesphere}, and also figure \ref{figure:T3fibration}),
say $S^3_{(1)}$ and $S^3_{(2)}$. Our aim will be to cast the result in the manifest form of the partition function of a 5d/3d/1d coupled system, as in \cite{Gomis:2016ljm}. We consider this case first since the fact that the intersection $S^3_{(1)}\cap S^3_{(2)}=S^1_{(1\cap 2)}$ has a single connected component is a simplifying feature that will be absent in the example of $S^4_b$ in the next subsection. 

\subsubsection{\texorpdfstring{$S^5_{\vec\omega}$}{S5} partition function of \texorpdfstring{$\widetilde{\mathcal T}$}{\widetilde T}}
Our starting point, the theory $\widetilde{\mathcal T}$, is described by the quiver
\begin{center}
\includegraphics[width=.2\textwidth]{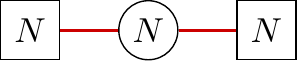}\;.
\end{center}
That is, it is an $SU(N)$ gauge theory with $N$ fundamental and $N$ anti-fundamental hypermultiplets, 
 \ie{}, $\mathcal N=2$ SQCD.\footnote{Recall our terminology of footnote \ref{footnote:afund-fund}.} The $S^5_{\vec\omega}$-partition function of $\widetilde{\mathcal T}$ is computed by the matrix integral \eqref{S5omegapartitionfunction} \cite{Qiu:2013aga,Qiu:2014oqa,Hosomichi:2012ek,Kallen:2012va,Kim:2012ava,Imamura:2012bm,Lockhart:2012vp,Kim:2012qf}
\begin{equation}
Z^{(\widetilde{\mathcal T},S^5_{\vec\omega})}(M,\tilde M) =  \int \text{d}\Sigma \ Z_{\text{cl}}^{(\widetilde{\mathcal T},S^5_{\vec\omega})}(\Sigma)\ Z_{\text{1-loop}}^{(\widetilde{\mathcal T},S^5_{\vec\omega})}(\Sigma, M,\tilde M)\ |Z_{\text{inst}}^{(\widetilde{\mathcal T},\mathbb R^4\times S^1)}(q,\Sigma ,M^\epsilon ,\tilde M^\epsilon )|^3 \;.
\end{equation}
The explicit expression for the classical action is given by
\begin{equation}\label{S5classicalaction_SQCD}
Z_{\text{cl}}^{(\widetilde{\mathcal T},S^5_{\vec\omega})}(\Sigma) = \exp\left[-\frac{8\pi^3}{\omega_1\omega_2\omega_3 g^2_{\text{YM}}}\Tr \Sigma^{2}\right]\;,
\end{equation}
while the one-loop determinant $Z_{\text{1-loop}}^{(\widetilde{\mathcal T},S^5_{\vec\omega})}$ is the product of the one-loop determinants of the $SU(N)$ vector multiplet, the $N$ fundamental hypermultiplets and the $N$ antifundamental hypermultiplets:
\begin{align}\label{S5oneloop_SQCD}
&Z_{\text{1-loop}}^{(\widetilde{\mathcal T},S^5_{\vec\omega})}(\Sigma, M,\tilde M) = Z_{{\text{vect}}}^{S^5_{\vec\omega}}(\Sigma )\ Z_{\text{fund}}^{S^5_{\vec\omega}}(\Sigma ,M)\ Z_{\text{afund}}^{S^5_{\vec\omega}}(\Sigma ,\tilde M) \\
&=\frac{\prod_{\substack{A , B = 1\\A\neq B}}^N {{S_3}(i({\Sigma_A} - {\Sigma_B})\ |\ \vec\omega )}}{\prod_{A = 1}^N \prod_{I = 1}^N  {{S_3}(i({\Sigma_A} - {M_I}) + | \vec\omega  |/2\ |\ \vec\omega )} \ \prod_{A = 1}^N \prod_{J = 1}^N  {{S_3}(i( - {\Sigma_A} + {{\tilde M}_J}) + |\vec\omega|/2\ |\ \vec\omega )} }\;,
\end{align}
written in terms of the triple sine function. Here we used the notation $|\vec\omega|=\omega_1+\omega_2+\omega_3$. Note that we did not explicitly separate the masses for the $SU(N)\times U(1)$ flavor symmetry, but instead considered $U(N)$ masses. Finally, there are three copies of the K-theoretic instanton partition function, capturing contributions of instantons residing at the circles kept fixed by two out of three $U(1)$ isometries. Concretely, one has
\begin{multline}\label{IPFS5_SQCD}
|Z_{\text{inst}}^{(\widetilde{\mathcal T},\mathbb R^4\times S^1)}(q,\Sigma ,{M^\omega },{{\tilde M}^\omega }){|^3}   \equiv \; Z_{{\text{inst}}}^{(\widetilde{\mathcal T},\mathbb R^4\times S^1_{(2\cap 3)})}\Big(q_1,\frac{\Sigma }{{{\omega _1}}},\frac{{{M^\omega }}}{{{\omega _1}}},\frac{{{{\tilde M}^\omega }}}{{{\omega _1}}},\frac{{2\pi }}{{{\omega _1}}},\frac{{{\omega _3}}}{{{\omega _1}}},\frac{{{\omega _2}}}{{{\omega _1}}}\Big) \\
\times Z_{{\text{inst}}}^{(\widetilde{\mathcal T},\mathbb R^4\times S^1_{(1\cap 3)})}\Big(q_2,\frac{\Sigma }{{{\omega _2}}},\frac{{{M^\omega }}}{{{\omega _2}}},\frac{{{{\tilde M}^\omega }}}{{{\omega _2}}},\frac{{2\pi }}{{{\omega _2}}},\frac{{{\omega _3}}}{{{\omega _2}}},\frac{{{\omega _1}}}{{{\omega _2}}}\Big) \  Z_{{\text{inst}}}^{(\widetilde{\mathcal T},\mathbb R^4\times S^1_{(1\cap 2)})}\Big(q_3,\frac{\Sigma }{{{\omega _3}}},\frac{{{M^\omega }}}{{{\omega _3}}},\frac{{{{\tilde M}^\omega }}}{{{\omega _3}}},\frac{{2\pi }}{{{\omega _3}}},\frac{{{\omega _1}}}{{{\omega _3}}},\frac{{{\omega _2}}}{{{\omega _3}}}\Big) \;,
\end{multline}
where $q_\alpha = \exp\left(-\frac{8\pi^2}{g_{YM}^2}\frac{2\pi}{\omega_\alpha}\right)$. Each factor can be written as a sum over an $N$-tuple of Young diagrams \cite{Nekrasov:2002qd, Nekrasov:2003rj}
\begin{equation}
\vec Y=(Y_1, Y_2,\ldots, Y_N)\;, \quad \text{with} \quad Y_A=(Y_{A1}\geq Y_{A2}\geq \ldots \geq Y_{AW_{Y_A}}\geq Y_{A(W_{Y_A}+1)} = \ldots = 0 )
\end{equation}
of a product over the contributions of vector and matter multiplets: 
\begin{multline}\label{IPFR4S1_SQCD}
Z_{{\text{inst}}}^{(\widetilde{\mathcal T},{\mathbb{R}^4} \times {S^1})}\Big( {q,\frac{\beta }{{2\pi }}\Sigma,\frac{\beta }{{2\pi }}{M^\omega },\frac{\beta }{{2\pi }}{{\tilde M}^\omega },\beta ,{\epsilon _1},{\epsilon _2}} \Big)\\
=\sum_{\vec Y} {q^{|\vec Y|}}z_{\text{vect}}^{{\mathbb{R}^4} \times {S^1}}\left(\vec Y, {\frac{\beta }{{2\pi }}\Sigma} \right) z_{\text{fund}}^{{\mathbb{R}^4} \times {S^1}}\left(\vec Y, {\frac{\beta }{{2\pi }}\Sigma,\frac{\beta }{{2\pi }}{M^\omega }} \right)z_{\text{afund}}^{{\mathbb{R}^4} \times {S^1}}\left(\vec Y, {\frac{\beta }{{2\pi }}\Sigma,\frac{\beta }{{2\pi }}{{\tilde M}^\omega }} \right)   \;.
\end{multline}
Here we have omitted the explicit dependence on $\epsilon_1, \epsilon_2$ in all factors $z^{{\mathbb{R}^4} \times {S^1}}$. The instanton counting parameter $q$ is given by $q=\exp\left(-\frac{8\pi^2\beta}{g_{YM}^2}\right),$ and $|\vec Y|$ denotes the total number of boxes in the $N$-tuple of Young diagrams. The expression for $z_{{\text{fund}}}$ reads
\begin{equation}\label{IPFR4S1_fund}
z_{\text{fund}}^{{\mathbb{R}^4} \times {S^1}}\left(\vec Y, {\frac{\beta }{{2\pi }}\Sigma,\frac{\beta }{{2\pi }}{M^\omega }} \right)=\prod_{A = 1}^N \prod_{I = 1}^N \prod_{r = 1}^\infty  \prod_{s=1}^{Y_{Ar}} 2i\sinh \pi i \left(\frac{\beta }{{2\pi }}(i\Sigma _A - iM^\omega_I) + r\epsilon_1  + s\epsilon_2 \right) \;,
\end{equation}
while those of $z_{{\text{vect}}}^{{\mathbb{R}^4} \times {S^1}}$ and $z_{{\text{afund}}}^{{\mathbb{R}^4} \times {S^1}}$ are given in \eqref{zvectIPF}-\eqref{z(a)fundIPF} in appendix \ref{appendix:IPF-factorization}.\footnote{In appendix \ref{appendix:IPF-factorization} we have simultaneously performed manipulations of four-dimensional and five-dimensional instanton partition functions, which is possible after introducing the generalized factorial with respect to a function $f(x)$, defined in appendix \ref{subapp: generalized factorials}, with $f(x)$ in four and five dimensions given in \eqref{fin4dand5d}. } Note that the masses that enter in \eqref{IPFR4S1_SQCD} are slightly shifted (see \cite{Okuda:2010ke}):
\begin{equation}
  {M^\omega } \equiv M - \frac{i}{2}(\omega_1 + \omega_2 + \omega_3)\;, \qquad   {\tilde M^\omega } \equiv \tilde M - \frac{i}{2}(\omega_1 + \omega_2 + \omega_3)\;.
\end{equation}

\subsubsection{Implementing the Higgsing prescription}
As outlined in the previous section, to introduce intersecting codimension two defects wrapping the three-spheres $S^3_{(1)}$ and $S^3_{(2)}$ and labeled by the $n^{(1)}$-fold and $n^{(2)}$-fold symmetric representation respectively, we should consider the residue at the pole position \eqref{polepositionsS5pinching} with $n^{(3)}=0$ (and hence $n_A^{(3)}=0$ for all $A=1,\ldots, N$)\footnote{Recall that we have regrouped the mass for the $U(1)$ flavor symmetry and those for the $SU(N)$ flavor symmetry into $U(N)$ masses.}
\begin{equation}\label{poleS3S3}
i\Sigma_A = i M_{\sigma(A)} -  n_A^{(1)} \omega_{1} -  n_A^{(2)} \omega_{2}  - \frac{\omega_1+\omega_2+\omega_3}{2} \qquad \text{for} \qquad  A=1,\ldots,N\;,
\end{equation}
while treating $\Sigma_A$ as $N$ independent variables, and sum over all partitions $\vec n^{(1)}$ of $n^{(1)}$ and $\vec n^{(2)}$ of $n^{(2)}$. As before, $\sigma(A)$ is a permutation of $A = 1, ..., N$ which we take to be, without loss of generality, $\sigma(A) = A$. At this point let us introduce  the notation that ``$\rightarrow$'' means evaluating the residue at the pole \eqref{poleS3S3} and removing some spurious factors. As we aim to cast the result in the form of a matrix integral describing the coupled 5d/3d/1d system, we try to factorize all contributions accordingly in pieces depending only on information of either three-sphere $S^3_{(1)}$ or $S^3_{(2)}$. As we will see, the non-factorizable pieces nicely cancel against each other, except for a factor that will ultimately describe the one-dimensional degrees of freedom residing on the intersection.

It is straightforward to work out the residue at the pole position \eqref{poleS3S3}. The classical action \eqref{S5classicalaction_SQCD} and the one-loop determinant \eqref{S5oneloop_SQCD} become, using recursion relations for the triple sine functions (see \eqref{recursion-triple-sine}),\footnote{Here we omitted on the right-hand side the left-over hypermultiplet contributions mentioned in the previous section as well as the classical action evaluated on the Higgs branch vacuum at infinity, \ie, on the position-independent Higgs branch vacuum.}
\begin{equation}\label{S5cl1loop_atpole_SQCD}
Z_{\text{cl}}^{(\widetilde{\mathcal T},S^5_{\vec\omega})}\ Z_{\text{1-loop}}^{(\widetilde{\mathcal T},S^5_{\vec\omega})} \rightarrow  Z_{\text{1-loop}}^{(\mathcal T,S^5_{\vec\omega})}\ \Big(Z^{S^3_{(1)}}_{\text{cl}|\vec n^{(1)}}\ Z^{S^3_{(1)}}_{\text{1-loop}|\vec n^{(1)}}\Big)\ \Big(Z^{S^3_{(2)}}_{\text{cl}|\vec n^{(2)}}\ Z^{S^3_{(2)}}_{\text{1-loop}|\vec n^{(2)}}\Big) \Big(Z^{\widetilde{\mathcal T};\vec n^{(1)},\vec n^{(2)}}_{\text{cl,extra}}\ Z^{\widetilde{\mathcal T};\vec n^{(1)},\vec n^{(2)}}_{\text{1-loop,extra}}\Big)\;.
\end{equation}
Let us unpack this expression a bit. First, $Z_{\text{1-loop}}^{(\mathcal T,S^5_{\vec\omega})}$ is the one-loop determinant of $N^2$ free hypermultiplets, which constitute the infrared theory $\mathcal T.$ It reads
\begin{equation}
  Z_{\text{1-loop}}^{(\mathcal T,S^5_{\vec\omega})} = \prod_{A=1}^N\prod_{J=1}^N \frac{1}{S_3(-iM_A + i \tilde M_J + |\vec \omega| \ |\ \vec \omega)} = \prod_{A=1}^N\prod_{J=1}^N \frac{1}{S_3(iM_A - i \tilde M_J  \ |\ \vec \omega)}\;.
\end{equation}
Note that the masses of the $N^2$ free hypermultiplets, represented by a two-flavor-node quiver, are $M_{AJ} = M_A - \tilde M_J + i \frac{|\vec \omega|}{2}$. Recall that $\frac{1}{N}\sum_{J=1}^N i \tilde M_J = i \check{ \tilde M},$ while $ \frac{1}{N}\sum_{A=1}^N iM_A = \frac{|\vec \omega|}{2} + \frac{n^{(1)}}{N} \omega_1 + \frac{n^{(2)}}{N}\omega_2$. Second, we find the classical action and one-loop determinant of squashed three-sphere partition functions of a three-dimensional $\mathcal N=2$ supersymmetric $U(n^{(\alpha)})$ gauge theory with $N$ fundamental and $N$ antifundamental chiral multiplets and one adjoint chiral multiplet, \ie{}, the quiver gauge theory
\begin{center}
\includegraphics[width=.2\textwidth]{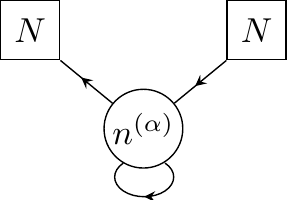}
\end{center}
We will henceforth call this theory `SQCDA.'\footnote{Note that the rank of the gauge group is the rank of one of the symmetric representations labeling the defects supported on the codimension two surfaces, or in other words, it can be inferred from the precise coefficients of the pole of the $\widetilde{\mathcal T}$ partition function, see \eqref{polepositionS5}.} These quantities are in their Higgs branch localized form,\footnote{\label{footnoteHBLS3}The squashed three-sphere partition function of a theory $\tau$ can be computed using two different localization schemes. The usual ``Coulomb branch localization'' computes it as a matrix integral of the schematic form \cite{Kapustin:2009kz,Jafferis:2010un,Hama:2010av,Hama:2011ea}
\begin{equation*}
Z^{(\tau,S^3_{b})} = \int \text{d}\sigma\  Z_{\text{cl}}^{({\tau},S^3_{b})}(\sigma)\  Z_{\text{1-loop}}^{({\tau},S^3_{b})}(\sigma)\;,
\end{equation*}
while a ``Higgs branch localization'' computation brings it into the form \cite{Fujitsuka:2013fga,Benini:2013yva}
\begin{equation*}
Z^{(\tau,S^3_{b})}  = \sum_{\text{HV}} Z_{\text{cl}|\text{HV}}^{({\tau},S^3_{b})}\  Z_{\text{1-loop}|\text{HV}}^{({\tau},S^3_{b})}\ Z_{\text{vortex}|\text{HV}}^{(\tau,\mathbb R^2\times S^1)}(b)\  Z_{\text{vortex}|\text{HV}}^{(\tau, \mathbb R^2\times S^1)}(b^{-1})\;.
\end{equation*}
Here the sum runs over all Higgs vacua $\text{HV}$ and the subscript $|\text{HV}$ denotes that the quantity is evaluated in the Higgs vacuum HV. Furthermore, one needs to include two copies of the K-theoretic vortex partition function $Z_{\text{vortex}}^{\mathbb R^2\times S^1}$. The two expressions for $Z$ are related by closing the integration contours in the former and summing over the residues of the enclosed poles. In the main text the theory $\tau$ will always be SQCDA and hence we omit the superscripted label. Note that for SQCDA, the sum over vacua is a sum over partitions of the rank of the gauge group. See appendix \ref{HBLS2S3} for all the details.} hence the additional subscript indicating the Higgs branch vacuum, \ie, the partition $\vec n^{(\alpha)}$. Their explicit expressions can be found in appendix \ref{subapp: S3 partition function}. The Fayet-Iliopoulos parameter $\xi^{(\alpha)}_{\text{FI}}$, the adjoint mass $m_X^{(\alpha)}$, and the fundamental and antifundamental masses $m_I^{(\alpha)},\tilde m_I^{(\alpha)}$ entering the three-dimensional partition function on $S^{3}_{(\alpha)}$ are identified with the five-dimensional parameters as follows, with ${\lambda _{(\alpha)} } \equiv \sqrt {{\omega _{(\alpha)} }/({\omega _1}{\omega _2}{\omega _3})}$,
\begin{align}\label{parameteridentifications_S5_SQCD}
  &\xi _{{\text{FI}}}^{(\alpha)} =  \frac{{8{\pi ^2}{\lambda _{(\alpha)} }}}{{g_{{\text{YM}}}^2}},& & m_X^{(\alpha)} = i{\omega _{(\alpha)} }{\lambda _{(\alpha)} } \;, & \\
  &m^{(\alpha)}_I = {\lambda _{(\alpha)} }\left( { {M_I} + \frac{i}{2}(|\vec\omega | + {\omega _{(\alpha)} })} \right), & & \tilde m^{(\alpha)} _J = -i{\omega _{(\alpha)} }{\lambda _{(\alpha)} }  + \lambda _{(\alpha)} \left(\tilde M_J + \frac{i}{2}(|\vec\omega | + {\omega _{(\alpha)} })\right) \;.&\label{parameteridentifications2_S5_SQCD}
\end{align}
Note that the relation on the $U(1)$ mass $\frac{1}{N}\sum_{I=1}^N iM_I = \frac{|\vec \omega|}{2} + \frac{n^{(1)}}{N} \omega_1 + \frac{n^{(2)}}{N}\omega_2 $ translates into a relation on the $U(1)$ mass of the fundamental chiral multiplets. Finally, both the classical action and the one-loop determinant produce extra factors which cannot be factorized in terms of information depending only on $\vec n^{(1)}$ or $\vec n^{(2)}$,
\begin{equation}
 Z_\text{1-1oop,extra}^{\tilde{\mathcal{T}};{{\vec n}^{(1)}},{{\vec n}^{(2)}}} = Z_{{\text{vf,extra}}}^{{{\vec n}^{(1)}},{{\vec n}^{(2)}}}(M)\ Z_{{\text{afund,extra}}}^{{{\vec n}^{(1)}},{{\vec n}^{(2)}}}(\tilde M)\;, \qquad Z_\text{cl,extra}^{\vec n^{(1)}, \vec n^{(2)}} = (q_3\bar q_3)^{- \sum_{A=1}^N n^{(1)}_A n^{(2)}_A}\;,
\end{equation}
where $Z_\text{afund,extra}^{\vnl, \vnr}$ captures  the non-factorizable factors from the antifundamental one-loop determinant, while $Z_\text{vf,extra}^{\vnl, \vnr}$ captures those from the vector multiplet and fundamental hypermultiplet one-loop determinants, which can be found in \eqref{def:Z-afund-extra}-\eqref{def:Z-vf-extra}. These factors will cancel against factors produced by the instanton partition functions, which we consider next.

When employing the Higgsing prescription to compute the partition function in the presence of defects, the most interesting part of the computation is the result of the analysis and massaging of the instanton partition functions \eqref{IPFS5_SQCD} evaluated at the value \eqref{poleS3S3} for their gauge equivariant parameter. We find that each term in the sum over Young diagrams can be brought into an almost factorized form. As mentioned before, certain non-factorizable factors cancel against the extra factors in \eqref{S5cl1loop_atpole_SQCD}, but a simple non-factorizable factor remains. When recasting the final expression in the form of a 5d/3d/1d coupled system, it is precisely this latter factor that captures the contribution of the degrees of freedom living on the intersection $S^1_{(1\cap 2)}$ of the three-spheres on which the defects live. 

Let us start by analyzing the instanton partition functions $Z_{{\text{inst}}}^{(\widetilde{\mathcal T},\mathbb R^4\times S^1_{(2\cap 3)})}$ and $Z_{{\text{inst}}}^{(\widetilde{\mathcal T},\mathbb R^4\times S^1_{(1\cap 3)})}$. It is clear from \eqref{IPFR4S1_fund} that upon plugging in the gauge equivariant parameter \eqref{poleS3S3} in $Z_{{\text{inst}}}^{(\widetilde{\mathcal T},\mathbb R^4\times S^1_{(2\cap 3)})}$, the $N$-tuple of Young diagrams $\vec Y$ has zero contribution if any of the Young diagrams $Y_A$ has more than $n^{(2)}_A$ rows. Similarly, $Z_{{\text{inst}}}^{(\widetilde{\mathcal T},\mathbb R^4\times S^1_{(1\cap 3)})}$ does not receive contributions from $\vec Y$ if any of its members $Y_A$ has more than $n^{(1)}_A$ rows. Hence the sum over Young diagrams simplifies to a sum over all possible sequences of $n^{(\alpha)}$ non-decreasing integers. The summands of the instanton partition functions undergo many simplifications at the special value for the gauge equivariant parameter, and in fact one finds that they become precisely the K-theoretic vortex partition function for SQCDA upon using the parameter identifications \eqref{parameteridentifications_S5_SQCD} (see appendix \ref{subapp: reduction-to-vortex-partition-function} for more details):\footnote{This fact has for example also been observed in \cite{Bonelli:2011wx,Bonelli:2011fq,Nieri:2013vba,Aganagic:2013tta}, and can also be read off from the brane picture in figure \ref{fig:manynodequiver_higgsed}. Before Higgsing, the instantons of the extra $SU(N)$ gauge node are realized by D0-branes spanning in between the NS5-branes. After Higgsing, the D0-branes can still be present if they end on the D2$_\text{R}$ and D2$_\text{L}$-branes. If, say, $n^{\text{L}}=0$, they precisely turn into vortices of the two-dimensional theory living on the D2-branes.}
\begin{equation}
Z_{{\text{inst}}}^{(\widetilde{\mathcal T},\mathbb R^4\times S^1_{(2\cap 3)})} \rightarrow Z_{\text{vortex}|\vec{n}^{(2)}}^{\mathbb R^2\times S^1_{(2\cap 3)}}(b_{(2)}^{-1})\;, \qquad Z_{{\text{inst}}}^{(\widetilde{\mathcal T},\mathbb R^4\times S^1_{(1\cap 3)})} \rightarrow Z_{\text{vortex}|\vec{n}^{(1)}}^{\mathbb R^2\times S^1_{(1\cap 3)}}(b_{(1)}^{-1})\;,
\end{equation}
with the three dimensional squashing parameters defined as
\begin{equation}
  {b_{(1)}} \equiv \sqrt {{\omega _2}/{\omega _3}} , \qquad {b_{(2)}} \equiv \sqrt {{\omega _1}/{\omega _3}}, \qquad {b_{(3)}} \equiv \sqrt {{\omega _1}/{\omega _2}} \;.
\end{equation}

The third instanton partition function, $Z_{{\text{inst}}}^{(\widetilde{\mathcal T},\mathbb R^4\times S^1_{(1\cap 2)})}$, behaves more intricately when substituting the gauge covariant parameter of \eqref{IPFS5_SQCD}. From \eqref{IPFR4S1_fund} one immediately finds that $N$-tuples of Young diagrams $\vec Y$ have zero contribution if any of its constituting diagrams $Y_A$ contain the ``forbidden box'' with coordinates $\text{(column,row)}=(n_A^{(1)}+1, n_A^{(2)}+1)$. We split the remaining sum over $N$-tuples of Young diagrams into two, by defining the notion of \textit{large} $N$-tuples, as those $N$-tuples satisfying the requirement that all of its members $Y_A$ contain the box with coordinates $(n_A^{(1)}, n_A^{(2)})$, and calling all other $N$-tuples \textit{small}. Let us focus on the former sum first. 

\begin{figure}[t]
\centering
\includegraphics[width=.8\textwidth]{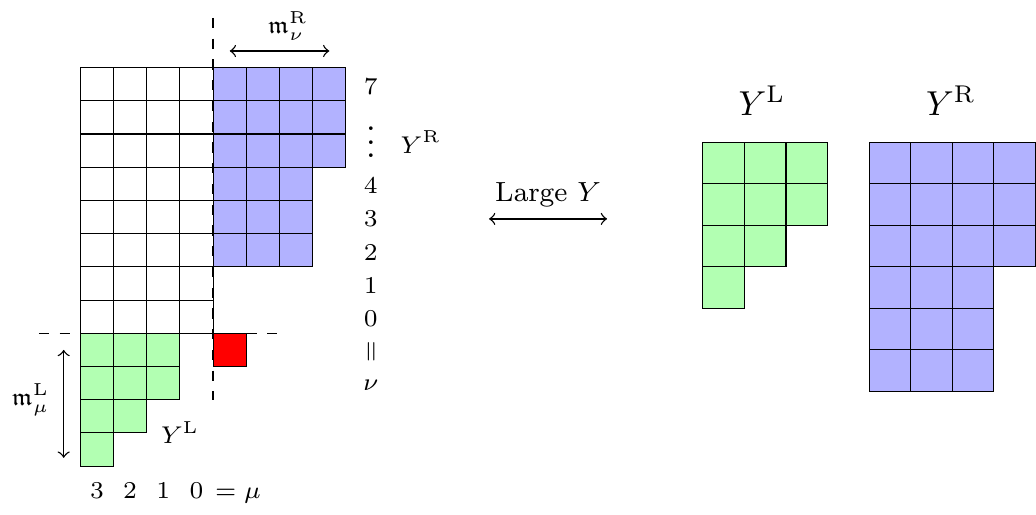}
\caption{\label{fig:Y} A constituent $Y$ of a large $N$-tuple of Young diagrams $\vec Y$ for $n^{(1)}=4, n^{(2)}=8$. The red box denotes the ``forbidden box'' with coordinates $(n^{(1)}+1,n^{(2)}+1).$ The green and blue areas denote $Y^\text{L}$ and $Y^\text{R}$ respectively. The definitions of $\mathfrak m_\mu^{\text L}$ and  $\mathfrak m_\nu^{\text R},$ see \eqref{definitionsms}, are also indicated.} 
\end{figure}

Given a large $N$-tuple $\vec Y$, we define $\vec Y^\text{L}$ and $\vec Y^\text{R}$ as the Young diagrams
\begin{equation}\label{defYLYR}
\begin{aligned}
&Y^\text{L}_{Ar} = Y_{Ar} - n^{(2)}_A &&\quad \text{for} \quad 1\leq r \leq n^{(1)}_A\;, \qquad \text{and}\; \qquad Y^\text{L}_{Ar} =0 \quad &&\text{for} \quad  n^{(1)}_A < r \\
&Y^\text{R}_{Ar} = Y_{A(n^{(1)}_A + r)} &&\quad \text{for} \quad 1 \le  r\;. &&
\end{aligned}
\end{equation}
Furthermore, we define the non-decreasing sequences of integers
\begin{equation}\label{definitionsms}
  \mathfrak{m}_{A\mu }^{\text{L}} \equiv Y_{A(n_A^{(1)} - \mu )}^{\text{L}},\quad\mu  = 0,...,n_A^{(1)} - 1,\qquad \qquad \mathfrak{m}_{A\nu }^{\text{R}} \equiv \tilde Y_{A(n_A^{(2)} - \nu )}^{\text{R}},\quad\nu  = 0,...,n_A^{(2)} - 1\;,
\end{equation}
where $\tilde Y^{\text{R}}_A$ denotes the transposed diagram of $Y^{\text{R}}_A$. Figure \ref{fig:Y} clarifies these definitions.
With these definitions in place, one can show (see appendix \ref{subapp:factorization-of-instanton-partition-function-large}) the following factorization of the summand of the instanton partition function for large tuples of Young diagrams $\vec Y$
\begin{multline}
q_3^{|\vec Y_{\text{large}}|}\   Z_{{\text{inst}}}^{(\widetilde{\mathcal T},{\mathbb{R}^4} \times {S^1_{(1\cap 2)}})} \left(\vec Y_{\text{large}}\right)  \rightarrow\  q_3^{|\mathfrak m^{\text L}|+|\mathfrak m^{\text R}|}\ Z_{\text{vortex}|\vec n^{(1)}}^{\mathbb{R}^2\times S^1_{(1\cap 2)}}(\mathfrak m^{\text L}|b_{(1)} )\  Z_{\text{vortex}|\vec n_2}^{\mathbb{R}^2\times S^1_{(1\cap 2)}}(\mathfrak m^{\text R}| b_{(2)} ) \\ \times  Z_{\text{intersection}}^{\text{large}|\vec n^{(1)},\vec n^{(2)}}(\mathfrak m^{\text L},\mathfrak m^{\text R})\ \Big(Z^{\widetilde{\mathcal T};\vec n^{(1)},\vec n^{(2)}}_{\text{cl,extra}}\ Z^{\widetilde{\mathcal T};\vec n^{(1)},\vec n^{(2)}}_{\text{1-loop,extra}}\Big)^{-1} \;.\label{factorizeLarge}
\end{multline}
Here we used $Z_{\text{vortex}|\vec n}^{\mathbb{R}^2\times S^1}(\mathfrak m|b )$ to denote the summand of the $U(n)$ SQCDA K-theoretic vortex partition function, \ie{},
\begin{equation}
Z_{\text{vortex}|\vec n}^{\mathbb{R}^2\times S^1}( b ) = \sum_{\substack{\mathfrak m_{A\mu}\geq 0\\\mathfrak m_{A\mu}\leq \mathfrak m_{A(\mu+1)}}} z_b^{|\mathfrak m|} Z_{\text{vortex}|\vec n}^{\mathbb{R}^2\times S^1}(\mathfrak m| b )  \;,
\end{equation}
where $|\mathfrak m|=\sum_A\sum_\mu \mathfrak m_{A\mu}.$ (See appendix \ref{subapp: S3 partition function} for concrete expressions.) The factor $Z_{\text{intersection}}^{\text{large}|\vec n^{(1)},\vec n^{(2)}}$ is given by
\begin{align}\label{intersectionfactorlarge}
&Z_{\text{intersection}}^{\text{large}|\vec n^{(1)},\vec n^{(2)}}(\mathfrak m^{\text L},\mathfrak m^{\text R})\nn\\ 
&\equiv  \;\prod_{A,B = 1}^N \prod_{\mu  = 0}^{n^{(1)}_A - 1} \prod_{\nu  = 0}^{n^{(2)}_B - 1} {\Big(2i\sinh \pi i \Big[i\frac{\beta}{2\pi}(M_A-M_B) + \epsilon_2 (\mathfrak m_{A\mu}^{\text{L}}+\nu) - \epsilon_1(\mathfrak m_{B\nu}^{\text{R}}+\mu) - \epsilon_1\Big]\Big)^{-1} }\nn\\
&\qquad\qquad\qquad\quad\ \ \  \times{\Big(2i\sinh \pi i \Big[i\frac{\beta}{2\pi}(M_A-M_B) + \epsilon_2 (\mathfrak m_{A\mu}^{\text{L}}+\nu) - \epsilon_1(\mathfrak m_{B\nu}^{\text{R}}+\mu) + \epsilon_2 \Big]\Big)^{-1} }\;.
\end{align}
As announced, the extra factors in the second line of \eqref{factorizeLarge} cancel against those in \eqref{S5cl1loop_atpole_SQCD}.

For small diagrams, we can still define $\vec Y^{\text{R}}$ as in the second line of \eqref{defYLYR}, but $\vec Y^{\text{L}}$ is not a proper $N$-tuple of Young diagrams due to the presence of negative entries. Nevertheless, we can define sets of non-decreasing integers as
\begin{equation}
\mathfrak m_{A\mu}^{\text{L}} \equiv Y_{A(n_A^{(1)}-\mu)}-n^{(2)}_A\;, \quad  \text{for} \quad 0\leq \mu \leq n_A^{(1)}-1\;, \qquad \mathfrak m_{A\nu}^{\text{R}}\equiv \tilde Y^{\text{R}}_{A(n_A^{(2)}-\nu)}\;, \quad  \text{for} \quad 0\leq \nu \leq n_A^{(2)}-1\;.\label{smallms}
\end{equation}
It is clear that $\mathfrak{m}_{A\mu}^\text{L} $ can take negative values. Then one can show (see appendix \ref{subapp:factorization-for-small-young-diagrams}) that
\begin{multline}
q_3^{|\vec Y_{\text{large}}|}\ Z_{{\text{inst}}}^{(\widetilde{\mathcal T},{\mathbb{R}^4} \times {S^1_{(1\cap 2)}})} \left(\vec Y_{\text{small}}\right)
  \rightarrow\ q_3^{|\mathfrak m^{\text L}|+|\mathfrak m^{\text R}|}\  Z_{\text{(semi-)vortex}|\vec n^{(1)}}^{\mathbb{R}^2\times S^1_{(1\cap 2)}}(\mathfrak m^{\text L}|b_{(1)} )\ Z_{\text{vortex}|\vec n_2}^{\mathbb{R}^2\times S^1_{(1\cap 2)}}(\mathfrak m^{\text R}| b_{(2)} ) \\ 
  \times Z_{\text{intersection}}^{\vec n^{(1)},\vec n^{(2)}}(\mathfrak m^{\text L},\mathfrak m^{\text R})\   \Big(Z^{\widetilde{\mathcal T};\vec n^{(1)},\vec n^{(2)}}_{\text{cl,extra}}\ Z^{\widetilde{\mathcal T};\vec n^{(1)},\vec n^{(2)}}_{\text{1-loop,extra}}\Big)^{-1} \;.\label{factorizeSmall}
\end{multline}
The intersection factor for generic (small) $N$-tuples of Young diagrams is a generalization of \eqref{intersectionfactorlarge} that can be found explicitly in \eqref{smallintersectionfactor}. The factor $Z_{\text{(semi-)vortex}|\vec n^{(1)}}^{\mathbb{R}^2\times S^1_{(1\cap 2)}}(\mathfrak m^{\text L}|b_{(1)} )$ is a somewhat complicated expression generalizing $Z_{\text{vortex}|\vec n^{(1)}}^{\mathbb{R}^2\times S^1_{(1\cap 2)}}$, which we present in \eqref{smallfactorization}.

Putting everything together, and noting that summing over all $N$-tuples of Young diagrams avoiding the forbidden box is equivalent to summing over all possible values of $\mathfrak m^{\text{L/R}}_{A\mu}$ , we find the following result for the Higgsed partition function 
\begin{align}\label{totalS5n1n2}
Z^{(\widetilde{\mathcal T},S^5_{\vec\omega})}\rightarrow Z_{\text{1-loop}}^{(\mathcal T,S^5_{\vec\omega})}\ &\Bigg(\ \sideset{}{^{\prime}}\sum_{\text{large }\vec Y} Z_{\vec n^{(1)}}(\mathfrak m^{\text{L}}| b_{(1)} )\;\;  Z_{\text{intersection}}^{\text{large}|\vec n^{(1)},\vec n^{(2)}}(\mathfrak m^{\text{L}},\mathfrak m^{\text{R}}) \;\;  Z_{\vec n^{(2)}}(\mathfrak m^{\text{R}}| b_{(2)} )\nn \\
&  \quad + \sideset{}{^{\prime}}\sum_{\text{small }\vec Y} \hat Z_{\vec n^{(1)}}(\mathfrak m^{\text{L}}| b_{(1)} )\;\;  Z_{\text{intersection}}^{\vec n^{(1)},\vec n^{(2)}}(\mathfrak m^{\text{L}},\mathfrak m^{\text{R}}) \;\;  Z_{\vec n^{(2)}}(\mathfrak m^{\text R}| b_{(2)} )\Bigg)
\end{align}
where 
\begin{align}\label{Zn1}
Z_{\vec n^{(1)}}(\mathfrak m^{\text{L}}| b_{(1)} ) &= \ Z^{S^3_{(1)}}_{\text{cl}|\vec n^{(1)}}\ Z^{S^3_{(1)}}_{\text{1-loop}|\vec n^{(1)}}\ q_3^{|\mathfrak m^{\text L}|}Z_{\text{vortex}|\vec n^{(1)}}^{\mathbb{R}^2\times S^1_{(1\cap 2)}}(\mathfrak m^{\text L}|b_{(1)} ) \ Z_{\text{vortex}|\vec n^{(1)}}^{\mathbb{R}^2\times S^1_{(1\cap 3)}}(b_{(1)}^{-1} )\;, 
\end{align}
and similarly for $Z_{\vec n^{(2)}}(\mathfrak m^{\text R}| b_{(2)} )$. The expression for $\hat Z_{n_1}(\mathfrak m^\text{L}|b_{(1)} )$ is obtained by replacing $Z_{\text{vortex}|n_1}^{\mathbb{R}^2\times S^1_{(1\cap 2)}}$ with $Z_{\text{(semi-)vortex}|n_1}^{\mathbb{R}^2\times S^1_{(1\cap 2)}}$. The prime on the sums over Young diagrams in \eqref{totalS5n1n2} indicates that only $N$-tuples of Young diagrams avoiding the ``forbidden box'' are included. To obtain the final result of the Higgsed partition function, we need to sum the right-hand side of \eqref{totalS5n1n2} over all partitions $\vec n^{(1)}$ of $n^{(1)}$ and $\vec n^{(2)}$ of $n^{(2)}$.

\subsubsection{Matrix model description and 5d/3d/1d coupled system}\label{subsubsection: matrixmodel5d/3d/1d}
Our next goal is to write down a matrix model integral that reproduces the $S^5_{\vec \omega}$-partition function of the theory $\mathcal T$ of $N^2$ free hypermultiplets in the presence of intersecting codimension two defects, \ie{}, a matrix integral that upon closing the integration contours appropriately reproduces the expression on the right-hand side of \eqref{totalS5n1n2}, summed over all partitions of $n^{(1)}$ and $n^{(2)}$, as its sum over residues of encircled poles.

A candidate matrix model is obtained relatively easily by analyzing the contribution of the large tuples of Young diagrams in \eqref{totalS5n1n2}. It reads
\begin{equation}\label{matrixmodelS3US3}
Z^{(\mathcal T,S^3_{(1)}\cup S^3_{(2)}\subset S^5_{\vec \omega})} = \frac{Z_{\text{1-loop}}^{(\mathcal T,S^5_{\vec\omega})}}{n^{(1)}!n^{(2)}!} \int_{\mathrm{JK}} \prod_{a=1}^{n^{(1)}} \text{d}\sigma^{(1)}_a \  \prod_{b=1}^{n^{(2)}} \text{d}\sigma^{(2)}_b \ Z^{S^3_{(1)}}(\sigma^{(1)}) \ Z_{\text{intersection}}(\sigma^{(1)},\sigma^{(2)})\   Z^{S^3_{(2)}}(\sigma^{(2)})\;,
\end{equation} 
where $Z^{S^3_{(1)}}(\sigma^{(1)})$ denotes the classical action times the one-loop determinant of the $S^3_{(1)}$ partition function of SQCDA, that is, of a three-dimensional $\mathcal N=2$ gauge theory with gauge group $U(n^{(1)})$, and $N$ fundamental, $N$ antifundamental and one adjoint chiral multiplet, and similarly for $Z^{S^3_{(2)}}(\sigma^{(2)})$.\footnote{See appendix \ref{subapp: S3 partition function} for concrete expressions for the integrand of the three-sphere partition function.}  The contribution from the intersection $S^1_{(1\cap 2)}$ reads
\begin{equation}
Z_{\text{intersection}}(\sigma^{(1)},\sigma^{(2)}) = \prod_{a  = 1}^{n^{(1)} } \prod_{b  = 1}^{n^{(2)} } \prod_{\pm} \left[2i\sinh \pi i \Big(\Delta_{ab}\pm\frac{1}{2}\big(b_{(1)}^2 +b_{(2)}^2\big)\Big)\right]^{-1} \;,
\end{equation}
with $\Delta _{ab} =-i b_{(2)}\sigma ^{(2)}_b +i b_{(1)}\sigma ^{(1)}_a$. Note that from \eqref{parameteridentifications_S5_SQCD} we deduce that the Fayet-Iliopoulos parameters $\xi_\text{FI}^{(1)}$ and $\xi_\text{FI}^{(2)}$ are both positive. The mass and other parameters on \emph{both} three-spheres satisfy relations which follow from the identifications in \eqref{parameteridentifications_S5_SQCD}-\eqref{parameteridentifications2_S5_SQCD}. Concretely, we find
\begin{equation}\label{parameterS3US3}
  \begin{aligned}
    &b_{(1)}\xi_{\text{FI}}^{(1)} = b_{(2)} \xi_{\text{FI}}^{(2)}\;, \qquad &&b_{(1)}\left(m^{(1)}_I + \frac{i}{2}b_{(1)}\right) =  b_{(2)}\left(m^{(2)}_I +\frac{i}{2} b_{(2)}\right)\;,\qquad && m_X^{(1)} = i\frac{b_{(2)}^2}{b_{(1)}}\;,\\
    &  &&b_{(1)}\left(\tilde m^{(1)}_J-\frac{i}{2}b_{(1)}  \right) =  b_{(2)}\left(\tilde m^{(2)}_J -\frac{i}{2}b_{(2)} \right)\;, \qquad && m_X^{(2)} = i\frac{b_{(1)}^2}{b_{(2)}}\;,
  \end{aligned}
\end{equation}
where $m^{(\alpha)}_I, \tilde m^{(\alpha)}_J$ and $m^{(\alpha)}_X$ are the fundamental, antifundamental and adjoint masses on the respective spheres. Moreover, the differences of the relations in \eqref{parameteridentifications2_S5_SQCD}, for fixed $\alpha$, relate the three-dimensional mass parameters on $S^3_{(\alpha)}$ to the five-dimensional mass parameters of the $N^2$ free hypermultiplets, \ie{}, to $M_{IJ}=M_I-\tilde M_J + i \frac{|\vec \omega|}{2}$:
\begin{equation}\label{MIJ_S5_relation}
M_{IJ} =  \lambda _{(\alpha )}^{ - 1}\left( {m_I^{(\alpha )} - \tilde m_J^{(\alpha )}} \right) - i{\omega _\alpha } + i\frac{{|\vec \omega |}}{2}\;.
\end{equation}

The matrix integral \eqref{matrixmodelS3US3} is evaluated using a Jeffrey-Kirwan-like residue prescription\cite{1993alg.geom..7001J}. We have derived it explicitly by demanding that the integral \eqref{matrixmodelS3US3} reproduces the result of the Higgsing computation (see below). The prescription is fully specified by the following charge assignments: the matter fields that contribute to $Z_{S^3_{(1)}}(\sigma^{(1)})$ and $Z_{S^3_{(2)}}(\sigma^{(2)})$ are assigned their standard charges under the maximal torus $U(1)^{n^{(1)}}\times U(1)^{n^{(2)}}$ of the total gauge group $U(n^{(1)})\times U(n^{(2)})$, while \textit{all} factors contributing to $Z_{\text{intersection}}(\sigma^{(1)},\sigma^{(2)})$ are assigned charges of the form $(0,\ldots, 0, +b_{(1)},0\ldots,0 \ ;\  0,\ldots, 0, -b_{(2)},0\ldots,0 )$. Furthermore, we pick the JK-vector $\eta = (\xi_{\text{FI}}^{(1)}; \xi_{\text{FI}}^{(2)})$, where we treat the Fayet-Iliopoulos parameters as an $n^{(1)}$-vector and $n^{(2)}$-vector respectively. Recall from \eqref{parameteridentifications_S5_SQCD} that both are positive.

\begin{figure}[t!]
  \centering
  \includegraphics[width=0.3\textwidth]{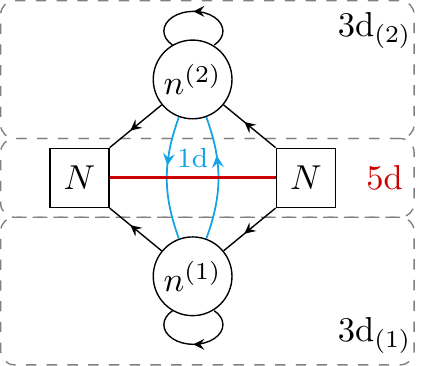}
  \caption{\label{fig:InsertSymmetrics5dfreeHM} Coupled 5d/3d/1d quiver gauge theory realizing intersecting M2-brane surface defects labeled by $n^{\text{R}}$- and $n^{\text{L}}$-fold symmetric representations in the five-dimensional theory of $N^2$ free hypermultiplets. The three-dimensional degrees of freedom are depicted in $\mathcal N=2$ quiver gauge notation, while the one-dimensional ones are denoted using one-dimensional $\mathcal N=2$ quiver notations, with solid lines representing chiral multiplets. Various superpotential couplings are turned on, as in figure \ref{fig:InsertSymmetrics} (see \cite{Gomis:2016ljm}). Applying the Higgsing prescription to SQCD precisely results in the partition function of this quiver gauge theory.}
\end{figure}

Before verifying that the matrix model \eqref{matrixmodelS3US3}, with the pole prescription just described, indeed faithfully reproduces the expression \eqref{totalS5n1n2} summed over all partitions $\vec n^{(1)},\vec n^{(2)}$, we remark that it takes precisely the form of the partition function of the 5d/3d/1d coupled system of figure \ref{fig:InsertSymmetrics5dfreeHM}, which is the trivial dimensional uplift of figure \ref{fig:InsertSymmetrics} specialized to the case of $N^2$ free hypermultiplets described by a two-flavor-node quiver. This statement can be verified by dimensionally uplifting the localization computation of \cite{Gomis:2016ljm}.
In some detail, $Z_{\text{1-loop}}^{(\mathcal T,S^5_{\vec\omega})}$ captures the contributions to the partition function of the five-dimensional degrees of freedom, \ie{}, of the theory $\mathcal T$ consisting of $N^2$ free hypermultiplets, while $Z^{S^3_{(\alpha)}}$ encodes those of the degrees of freedom living on $S^3_{(\alpha)}$, described by $U(n^{(\alpha)})$ SQCDA, for $\alpha=1,2$, and the factor $Z_{\text{intersection}}$ precisely equals the one-loop determinant of the one-dimensional bifundamental chiral multiplets living on the intersection $S^3_{(1)} \cap S^3_{(2)} = S^1_{(1\cap 2)}$. Moreover, the mass relations \eqref{MIJ_S5_relation}, which we find straightforwardly from the Higgsing prescription, are the consequences of cubic superpotential couplings in the 5d/3d/1d coupled system, which were analyzed in detail in \cite{Gomis:2016ljm}. The mass relations  among the (anti)fundamental chiral multiplet masses in \eqref{parameterS3US3} are in fact a solution of \eqref{MIJ_S5_relation} obtained by subtracting the equation for $\alpha=1$ and $\alpha=2$ and subsequently performing a separation of the indices $I,J$. The separation constants appearing in the resulting solutions can be shifted to arbitrary values by performing a change of variables in the three-dimensional integrals, up to constant prefactors stemming from the classical actions. The Higgsing prescription also fixes the classical actions and hence we find specific values for the separation constants. The adjoint masses in \eqref{parameterS3US3} are the consequence of a quartic superpotential. Also observe that our computation fixes the flavor charge of the one-dimensional chiral multiplets, which enter explicitly in $Z_{\text{intersection}}$, and for which no first-principles argument was provided in \cite{Gomis:2016ljm}.

The integrand of \eqref{matrixmodelS3US3} has poles in each of the three factors; the Jeffrey-Kirwan-like residue prescription is such that, among others, it picks out classes of poles, which we refer to as poles of type-$\hat \nu$. They read, for partitions $\vec n^{(1)}$ and $\vec n^{(2)}$ of $n^{(1)}$ and $n^{(2)}$ respectively, over all of which we sum, and for sequences of integers $\{\hat \nu_A\}$ where $\hat \nu_A \in \{-1, 0, \ldots, n_A^{(2)} - 1\}$,  
\begin{equation}\label{polesnuhat_main}
  \begin{aligned}
    & \text{poles of type-}\hat{\nu}: & \sigma _{A\mu} ^{(1)} = & \;m^{(1)}_A + \mu m_X^{(1)} - i\mathfrak{m}_{A\mu} ^{\text{L}}{b_{(1)}} - i\mathfrak{n}_{A\mu} ^{\text{L}}b_{(1)}^{ - 1} , \qquad && \mu = 0, \ldots, n^{(1)}_A - 1\\
    & ~ & \sigma _{B\nu} ^{(2)} = & \;{m_B^{(2)}} + \nu m_X^{(2)} - i\mathfrak{m}_{B\nu} ^{\text{R}}{b_{(2)}} - i\mathfrak{n}_{B\nu} ^{\text{R}}b_{(2)}^{ - 1} \;, \qquad &&\nu = 0, \ldots, n^{(2)}_B - 1\;.
  \end{aligned}
\end{equation}
where $\mathfrak{m}^{\text{L}}_\mu$, $\mathfrak{m}^{\text{R}}_\nu$, $\mathfrak{n}^{\text{L}}_\mu$, $\mathfrak{n}^{\text{R}}_\nu$ are non-decreasing sequences of integers, such that $\mathfrak{n}_{A\mu} ^{\text{L}},\mathfrak{n}_{B\nu} ^{\text{R}} \geqslant 0$ and (where $\hat \nu$ enters)
\begin{equation}\label{polesnuhat_main2}
\left\{ \begin{aligned}
   & \mathfrak{m}_{A\mu \ge 0} ^{\text{L}} \ge 0\qquad &&{\text{if }}\hat\nu_A  = -1 \\
   & \mathfrak{m}_{A\mu  \ge 1}^{\text{L}} \ge \mathfrak{m}_{A0}^{\text{L}} =  - \hat\nu_A - 1 \quad &&{\text{if }}\hat\nu_A  \ge 0 \hfill \\ 
  \end{aligned}  \right.\;,  \qquad \quad \mathfrak{m}_{0 \leqslant \nu  \leqslant \hat \nu_A}^{\text{R}} = 0\;,\qquad\mathfrak{m}_{\nu  \ge \hat \nu_A + 1  }^{\text{R}} \ge 0\;.
\end{equation}
Note that if all $\hat{\nu}_A = -1$ these poles are simply obtained by assigning to $\sigma^{(1)}$ a pole position of ${Z_{S_{(1)}^3}}$ and to $\sigma^{(2)}$ a pole position of ${Z_{S_{(2)}^3}}$, whose residues precisely reproduce the sum over large diagrams in \eqref{totalS5n1n2}. Precisely this fact motivated the candidate matrix model in \eqref{matrixmodelS3US3}. In appendix \ref{subapp:constructing-Young-diagrams}, we describe a simple algorithm to construct Young diagrams avoiding the ``forbidden box'' associated with poles of type-$\hat \nu$. Furthermore, we show in appendix \ref{subapp:residues-and-instanton-partition-function} that the sum over the corresponding residues precisely reproduce the sum over Young diagrams in \eqref{totalS5n1n2}. Finally, we show in appendix \ref{subapp:extra-poles-and-diagrams} that the residues of poles not of type-$\hat \nu$, but contained in the Jeffrey-Kirwan-like prescription, cancel among themselves by studying a simplified example. We thus conclude that the integral \eqref{matrixmodelS3US3} indeed faithfully reproduces the sum over Young diagrams in \eqref{totalS5n1n2}.

\subsection{Intersecting surface defects on \texorpdfstring{$S^4_{b}$}{S4b}}
Let us next study the partition function of $N^2$ free hypermultiplets on $S^4_{b}$ in the presence of intersecting codimension two defects wrapping the two-spheres $S^2_{\text{L/R}}$, the fixed loci of the $U(1)^{\text{L/R}}$ isometries (see footnote \ref{definitionsquashedfoursphere}). The intersection of $S^2_{\text{L}}$ with $S^2_{\text{R}}$ consists of two points. The analysis largely parallels the one in the previous subsection, so we will be more brief. 

\subsubsection{\texorpdfstring{$S^4_{b}$}{S4b} partition function of \texorpdfstring{$\widetilde{\mathcal T}$}{\widetilde T}}
The theory $\widetilde{\mathcal T}$ is an $\mathcal N=2$ supersymmetric gauge theory with gauge group $SU(N)$ and $N$ fundamental and $N$ antifundamental hypermultiplets. Its squashed four-sphere partition function is computed by the matrix integral \eqref{S4bpartitionfunction} (or \eqref{S4bpartitionfunctionsplit}), 
\begin{equation}\label{S4b_SQCD_partitionfunction}
Z^{(\widetilde{\mathcal T},S^4_{b})}(M,\tilde M) =  \int \text{d}\Sigma \ Z_{\text{cl}}^{(\widetilde{\mathcal T},S^4_{b})}(\Sigma)\ Z_{\text{1-loop}}^{(\widetilde{\mathcal T},S^4_{b})}(\Sigma, M,\tilde M)\ |Z_{\text{inst.}}^{(\widetilde{\mathcal T},\mathbb R^4)}(q,\Sigma ,M^\epsilon ,\tilde M^\epsilon )|^2 \;.
\end{equation}
The classical action is given by 
\begin{equation}\label{classactionSQCD}
Z_{\text{cl}}^{(\widetilde{\mathcal T},S^4_{b})}(\Sigma) = \exp\left[-\frac{8\pi^2}{g_{\text{YM}}^2} \Tr\Sigma^{2} \right]
\end{equation}
and the one-loop factor reads
\begin{equation}\label{oneloopSQCD}
Z_{\text{1-loop}}^{(\widetilde{\mathcal T},S_b^4)}(\Sigma, M,\tilde M) = Z_{{\text{vect}}}^{S_b^4}(\Sigma )\ Z_{\text{fund}}^{S_b^4}(\Sigma ,M)\ Z_{\text{afund}}^{S_b^4}(\Sigma ,\tilde M)\;,
\end{equation}
where
\begin{align}
Z_{\text{fund}}^{S_b^4}(\Sigma ,M) = &\; \prod_{I = 1}^N {\prod_{A = 1}^N {\frac{1}{{{\Upsilon _b}(i{\Sigma _A}-i{M_I} + Q/2)}}} } \;,\qquad &Z_{{\text{vect}}}^{S_b^4}(\Sigma ) = &\; \prod_{\substack{A,B = 1\\A \ne B}}^N {{\Upsilon _b}(i{\Sigma _A}-i{\Sigma _B})}\;,\nn\\
Z_{\text{afund}}^{S_b^4}(\Sigma ,\tilde M) = &\; \prod_{J = 1}^N {\prod_{A = 1}^N {\frac{1}{{{\Upsilon _b}(-i{{\Sigma}_A} + i{{\tilde M}_J} + Q/2)}}} }\;.& \label{S4b_one_loop_SQCD}
\end{align}
We have denoted the masses associated with the $U(N)$ flavor symmetry of the $N$ fundamental hypermultiplets as $M_I$ and those of the $N$ antifundamental hypermultiplets as $\tilde M_J$.  We also denote $Q = b + b^{-1}$.

The instanton partition functions can be written as a sum over $N$-tuples of Young diagrams as
\begin{equation}\label{SQCDIPF}
Z_{\text{inst.}}^{(\widetilde{\mathcal T},\mathbb R^4)}(q,\Sigma ,M^\epsilon ,\tilde M^\epsilon )=\sum_{\vec Y} {{q^{|\vec Y|}}\ z_{{\text{vect}}}^{\mathbb R^4}(\vec Y,\Sigma,\epsilon_1,\epsilon_2 )\ z_{{\text{fund}}}^{\mathbb R^4}(\vec Y,\Sigma ,M^\epsilon,\epsilon_1,\epsilon_2)\ z_{{\text{afund}}}^{\mathbb R^4}(\vec Y,\Sigma ,\tilde M^\epsilon,\epsilon_1,\epsilon_2)} \;.
\end{equation}
The various factors in the summand are defined in \eqref{zvectIPF} and \eqref{z(a)fundIPF} in appendix \ref{appendix:IPF-factorization}. The $\Omega$-deformation parameters are identified as $\epsilon_1=b$ and $\epsilon_2=b^{-1}$, the superscript $^\epsilon$ denotes the usual shift of hypermultiplet masses \cite{Okuda:2010ke}
\begin{equation}
  {M^\epsilon } \equiv M - \frac{i}{2}({\epsilon _1} + {\epsilon _2})\;,\qquad   {\tilde M^\epsilon } \equiv \tilde M - \frac{i}{2}({\epsilon _1} + {\epsilon _2})\;,
\end{equation}
and the modulus squared simply entails sending $q = \exp(2\pi i \tau)\rightarrow \bar q$, with $\tau = \frac{\vartheta}{2\pi} + \frac{4\pi i}{g_{\text{YM}}^2}$.

\subsubsection{Implementing the Higgsing prescription}
The Higgsing prescription instructs us to consider the poles of the fundamental one-loop factor given by
\begin{equation}\label{poleS2S2}
  i\Sigma_A = i M_{\sigma(A)} -  n_A^{\text{L}} b -  n_A^{\text{R}} b^{-1} - \frac{b+b^{-1}}{2} \qquad \text{for} \qquad  A=1,\ldots,N\;,
\end{equation}
with $\sigma$ a permutation of $N$ elements, which we choose to be the identity. At the end of the computation, we should sum over all partitions $\vec n^{\text{L/R}}$ of $n^{\text{L/R}}$, \ie{}, $n^{\text{L/R}}= \sum_A n_A^{\text{L/R}}$.

The fact that the two two-spheres intersect at two disjoint points, namely their north poles and south poles, adds another layer of complication compared to the analysis in the previous subsection. Even so, when evaluating the residue at \eqref{poleS2S2}, the analysis of the classical action and one-loop determinants is straightforward. Both can be brought into a factorized form in terms of pieces depending only on information on either two-sphere, using the shift formula \eqref{recursion-upsilon} for the latter, up to extra factors which will cancel against certain non-factorizable factors coming from the instanton partition functions. Explicitly,
\begin{equation}\label{S4cl1loop_atpole_SQCD}
Z_{\text{cl}}^{(\widetilde{\mathcal T},S^4_{b})}\ Z_{\text{1-loop}}^{(\widetilde{\mathcal T},S^4_{b})} \rightarrow  Z_{\text{1-loop}}^{(\mathcal T,S^4_{b})}\ \Big(Z^{S^2_{\text{L}}}_{\text{cl}|\vec n^{\text{L}}}\ Z^{S^2_{\text{L}}}_{\text{1-loop}|\vec n^{\text{L}}}\Big)\ \Big(Z^{S^2_{\text{R}}}_{\text{cl}|\vec n^{\text{R}}}\ Z^{S^2_{\text{R}}}_{\text{1-loop}|\vec n^{\text{R}}}\Big) \Big(Z^{\widetilde{\mathcal T};\vec n^{\text{L}},\vec n^{\text{R}}}_{\text{cl,extra}}\ Z^{\widetilde{\mathcal T};\vec n^{\text{L}},\vec n^{\text{R}}}_{\text{1-loop,extra}}\Big)^{2}\;,
\end{equation}
where $Z_{\text{1-loop}}^{({\mathcal T},S^4_{b})}$ is the one-loop determinant of $N^2$ hypermultiplets, which constitute the infrared theory $\mathcal T$, and have masses $M_{IJ} = M_I - \tilde M_J + i\frac{Q}{2}$. Furthermore, $Z^{S^2_{\text{L/R}}}_{\text{\ldots}|\vec n^{\text{L/R}}}$ denote factors in the Higgs branch localized two-dimensional $\mathcal N=(2,2)$ SQCDA two-sphere partition function (see footnote \ref{footnoteHBLS3} for the equivalent three-sphere discussion, and appendix \ref{subapp: S2 partition function} for explicit expressions). The two-dimensional FI-parameter $\xi_{\text{FI}}$, fundamental masses $m_I$, antifundamental masses $\tilde m_J$ and adjoint masses $m_X$ are related to the four-dimensional parameters as
\begin{align}\label{realtionsS4_free}
  &\xi _{{\text{FI}}}^{\text{L}} = \xi _{{\text{FI}}}^{\text{R}} = \frac{{4\pi }}{{g_{{\text{YM}}}^2}}\;,& &m_I^\text{L} = b M_I + \frac{i}{2} + i b^2\;, & &\tilde m^\text{L}_J = b \tilde M_J + \frac{i}{2}\;, &&  m_X^\text{L} = ib^2\\\label{realtionsS4_free2}
  &{\vartheta ^{{\text{L/R}}}} = \vartheta \;, & &m_I^\text{R} = b^{-1} M_I + \frac{i}{2} + i b^{-2}\;, & &\tilde m^\text{R}_J = b^{-1} \tilde M_J + \frac{i}{2}\;, &&  m_X^\text{R} = ib^{-2}\;.
\end{align}
Finally the extra factors are
\begin{equation}
  Z_\text{1-loop,extra}^{\vnl, \vnr} = Z_\text{vf,extra}^{\vnl, \vnr}(M)\ Z_\text{afund,extra}^{\vnl, \vnr}(\tilde M), \qquad Z_\text{cl,extra} = (q\bar q)^{-\sum_{A = 1}^N n^\text{L}_An^\text{R}_A}\;,
\end{equation}
where $Z^{\vnl, \vnr}_\text{vf,extra}$ and $Z^{\vnl, \vnr}_\text{afund,extra}$ are as before the non-factorizable pieces produced by applying the shift formulae to the vector and (anti)fundamental one-loop determinant and can be found in \eqref{def:Z-afund-extra}-\eqref{def:Z-vf-extra}.

The massaging of each of the two instanton partition functions, which now both describe instantons located at intersection points, is completely similar to the one we performed above. First, the sum over $N$-tuples of Young diagrams $\vec Y$ can be restricted to a sum over tuples whose constituents $Y_A$ all avoid the ``forbidden'' box at $(n^{\text{L}}_A+1,n^{\text{R}}_A+1)$. Second, the left-over sum can be decomposed into sums over large and small diagrams, and moreover their summands can almost be factorized in terms of the summands of vortex partition functions, after canceling some overall factors with the extra factors from the classical action and one-loop determinants in \eqref{S4cl1loop_atpole_SQCD}. The remaining non-factorizable factor is an intersection factor,
\begin{multline}
Z_{\text{intersection}}^{\text{large}|\vec n^{(1)},\vec n^{(2)}}(\mathfrak m^{\text L},\mathfrak m^{\text R})=  \;\prod_{A,B = 1}^N \prod_{\mu  = 0}^{n^{(1)}_A - 1} \prod_{\nu  = 0}^{n^{(2)}_B - 1} {\Big(i(M_A-M_B) + \epsilon_2 (\mathfrak m_{A\mu}^{\text{L}}+\nu) - \epsilon_1(\mathfrak m_{B\nu}^{\text{R}}+\mu) - \epsilon_1\Big)^{-1} }\\
\times{\Big( i(M_A-M_B) + \epsilon_2 (\mathfrak m_{A\mu}^{\text{L}}+\nu) - \epsilon_1(\mathfrak m_{B\nu}^{\text{R}}+\mu) + \epsilon_2 \Big)^{-1} }\;. \label{intersection_factor_HB_R4}
\end{multline}
for large diagrams, and \eqref{smallintersectionfactor} for generic diagrams. The full expression for the residue at the pole location \eqref{poleS2S2} thus involves the product of the two massaged instanton partition functions, together with the leftover classical action and one-loop determinant factors,
\begin{align}\label{totalS4n1n2}
Z^{(\widetilde{\mathcal T},S^4_b)} \rightarrow &\ Z_{\text{1-loop}}^{(\mathcal T,S^4_b)}\ \Big(Z^{S^2_{\text{L}}}_{\text{cl}|\vec n^{\text{L}}}\ Z^{S^2_{\text{L}}}_{\text{1-loop}|\vec n^{\text{L}}}\Big)\ \Big(Z^{S^2_{\text{R}}}_{\text{cl}|\vec n^{\text{R}}}\ Z^{S^2_{\text{R}}}_{\text{1-loop}|\vec n^{\text{R}}}\Big) \nn\\
&\times \Bigg|\ \sideset{}{^{\prime}}\sum_{\text{large }\vec Y} q^{|\mathfrak m^{\text{L}}|+|\mathfrak m^{\text{R}}|} Z_{\text{vortex}|\vec n^{\text{L}}}^{\mathbb{R}^2}(\mathfrak m^{\text L} ) \;\;  Z_{\text{intersection}}^{\text{large}|\vec n^{\text{L}},\vec n^{\text{R}}}(\mathfrak m^{\text{L}},\mathfrak m^{\text{R}}) \;\;  Z_{\text{vortex}|\vec n^{\text{R}}}^{\mathbb{R}^2}(\mathfrak m^{\text R} )  \\
&\qquad + \sideset{}{^{\prime}}\sum_{\text{small }\vec Y}  q^{|\mathfrak m^{\text{L}}|+|\mathfrak m^{\text{R}}|} Z_{\text{semi-vortex}|\vec n^{\text{L}}}^{\mathbb{R}^2}(\mathfrak m^{\text L} ) \;\;  Z_{\text{intersection}}^{\vec n^{\text{L}},\vec n^{\text{R}}}(\mathfrak m^{\text{L}},\mathfrak m^{\text{R}}) \;\;  Z_{\text{vortex}|\vec n^{\text{R}}}^{\mathbb{R}^2}(\mathfrak m^{\text R} )  \Bigg|^2\;.\nn
\end{align}
The final result for the Higgsed partition function is obtained by summing the right-hand side of this expression over all partitions $\vec n^{\text{L/R}}$ of $n^{\text{L/R}}$.

\subsubsection{Matrix model description and 4d/2d/0d coupled system}
As in the previous subsection, the contribution of large tuples in both instanton partition functions suggests the following matrix integral
\begin{multline}
  Z^{(\mathcal{T},S^2_\text{L} \cup S^2_\text{R} \subset S^4_b)} = Z^{({\mathcal T},S^4_b)}_{\text{1-loop}} \frac{1}{n^{\text{L}}!n^{\text{R}}!}\sum_{B^{\text{R}}\in\mathbb Z^{n^\text{R}}}\sum_{B^{\text{L}}\in\mathbb Z^{n^\text{L}}} \int_{\mathrm{JK}} \prod_{a=1}^{n^\text{R}} \frac{d\sigma^{\text{R}}_a}{2\pi} \  \prod_{c=1}^{n^\text{L}} \frac{d\sigma^{\text{L}}_c}{2\pi} \ Z^{S^2_{\text{R}}}(\sigma^{\text{R}},B^{\text{R}}) \ Z^{S^2_{\text{L}}}(\sigma^{\text{L}},B^{\text{L}})\\
  \times\ \prod_\pm Z_{\text{intersection}}^\pm(\sigma^{\text{L}}, B^{\text{L}},\sigma^{\text{R}},B^{\text{R}})   \label{matrixmodelS2US2} \;,
\end{multline}
where $Z^{S^2_{\text{R}}}(\sigma^{\text{R}},B^{\text{R}})$ denotes the summand/integrand of the $S^2_{\text{R}}$ partition function for SQCDA with gauge group $U(n^{\text{R}})$,  and similarly for $Z^{S^2_{\text{L}}}(\sigma^{\text{L}},B^{\text{L}})$.\footnote{Concrete expressions can be found in appendix \ref{subapp: S2 partition function}.} The intersection factors read
\begin{equation}
Z_{\text{intersection}}^\pm(\sigma^{\text{L}}, B^{\text{L}},\sigma^{\text{R}},B^{\text{R}})  = \prod_{a  = 1}^{n^\text{R} } \prod_{c  = 1}^{n^\text{L} } \left[ \left(\Delta_{ac}^\pm + \frac{b+b^{-1}}{2}\right)\left(\Delta_{ac}^\pm - \frac{b+b^{-1}}{2}\right)\right]^{-1} \;,  \label{intersection-factor-CBL-S2}
\end{equation}
with $\Delta_{ac}^\pm =b^{-1}\left(i\sigma^{\text{R}}_a \pm \frac{B^{\text{R}}_a}{2}\right)- b \left(i\sigma^{\text{L}}_c \pm \frac{B^{\text{L}}_c}{2}\right)$ and where $b$ is the four-sphere squashing parameter. The factor labeled by the plus sign arises from the intersection point at the north pole, and the other factor from the south pole. The mass and other parameters on both two-spheres satisfy relations, which can be derived from \eqref{realtionsS4_free}-\eqref{realtionsS4_free2}, 
\begin{equation}\label{paramidentificationsS2US2}
\begin{aligned}
  &\xi_{\text{FI}}^{\text{L}} = \xi_{\text{FI}}^{\text{R}}\;, \qquad &&b^{-1} \left(m^{\text{L}}_I + \frac{i}{2} \right)  = b\left(m^{\text{R}}_I + \frac{i}{2}\right)\;,\qquad && m_X^\text{L} = ib^2\;,\\
  &\vartheta^{\text{L}}  = \vartheta^{\text{R}} \;,  &&b^{-1}\left(\tilde m^{\text{L}}_J - \frac{i}{2} \right) =  b\left(\tilde m^{\text{R}}_J-\frac{i}{2} \right)\;, \qquad && m_X^{\text{R}} = ib^{-2}\;,
  \end{aligned}
\end{equation}
while the hypermultiplet masses $M_{IJ} = M_I - \tilde M_J + i\frac{Q}{2}$ are related to the two-dimensional mass parameters as
\begin{equation}\label{4dmassfreeHM}
i{b^{ - 1}} = \left[ {{M_{IJ}} + \frac{i}{2}(b + {b^{ - 1}})} \right] - {b^{ - 1}}(m_I^{\text{L}} - \tilde m_J^{\text{L}}) \;, \qquad i{b} = \left[ {{M_{IJ}} + \frac{i}{2}(b + {b^{ - 1}})} \right] - {b}(m_I^{\text{R}} - \tilde m_J^{\text{R}})\;.
\end{equation}

The residue prescription used to evaluate the integrals in \eqref{matrixmodelS2US2} is completely similar to Jeffrey-Kirwan-like prescription introduced in the previous subsection: the matter fields contributing to $Z^{S^2_{\text{R/L}}}$ are assigned their natural charges under the Cartan subgroup $U(1)^{n^\text{R}}\times U(1)^{n^\text{L}}$ of the total gauge group, while all factors of the intersection factors are assigned charges of the form $(0,\ldots, 0, b^{-1},0\ldots,0 \ ;\  0,\ldots, 0, -b,0\ldots,0 )$. The JK-vector is again given in terms of the FI-parameters, $\eta = (\xi_{\text{FI}}^{\text{R}},\xi_{\text{FI}}^{\text{L}})$. We have derived this prescription by demanding that the matrix integral reproduces the result of the Higgsing computation.

\begin{figure}[t!]
  \centering
  \includegraphics[width=0.3\textwidth]{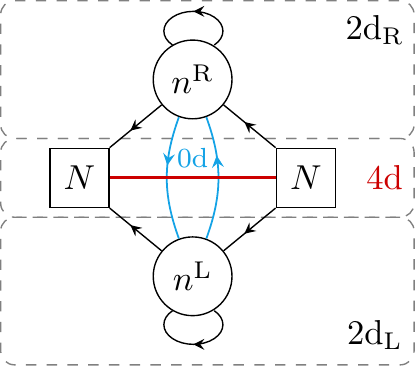}
  \caption{\label{fig:InsertSymmetrics4dfreeHM} Coupled 4d/2d/0d quiver gauge theory realizing intersecting M2-brane surface defects labeled by $n^{\text{R}}$- and $n^{\text{L}}$-fold symmetric representations in the four-dimensional theory of $N^2$ free hypermultiplets. Various superpotential couplings are turned on and are given in detail in \cite{Gomis:2016ljm}. The Higgsing prescription applied to SQCD precisely reproduces the partition function of this coupled system.}
\end{figure}

It was shown in \cite{Gomis:2016ljm} that the partition function of the 4d/2d/0d coupled system of figure \ref{fig:InsertSymmetrics} for the case of $N^2$ free hypermultiplets described by a two-flavor-node quiver, reproduced in figure \ref{fig:InsertSymmetrics4dfreeHM} for convenience, precisely equals the matrix integral \eqref{matrixmodelS2US2}. 
In particular, $\prod_\pm Z_{\text{intersection}}^\pm$ computes the one-loop determinant of the zero-dimensional bifundamental chiral multiplets at the two intersection points of the two-spheres $S^2_{\text{R}}$ and $S^2_{\text{L}}$. In the first-principles localization computation of \cite{Gomis:2016ljm}, the relations \eqref{4dmassfreeHM} are consequences of cubic superpotential couplings. Up to separation constants, their solutions can be found to be the mass relations in \eqref{paramidentificationsS2US2}. As explained in the previous subsection, the Higgsing computation fixes the separation constants to specific values. Note that our computations fixes the flavor symmetry charges of the zero-dimensional fields and provides a derivation of the residue prescription.

The proof that the matrix integral reproduces the result of the Higgsing computation follows the same logic as the one in the previous subsection, but is substantially more involved due to the fact that two copies of the intersection factor are present. We present some of the details in appendix \ref{appendix:extra-pole-2d}.

\section{Intersecting surface defects in interacting theories}\label{section: interacting theories}
In the previous section, we have computed the expectation value of intersecting surface defects in four-/five-dimensional theories $\mathcal T$ of free hypermultiplets placed on the four-/five-sphere. In this section, we consider intersecting surface defects inserted in interacting theories. More precisely, we focus on $\mathcal T$ being an $\mathcal N=2$ supersymmetric theory with gauge group $SU(N)$ and $N$ fundamental and $N$ anti-fundamental hypermultiplets, \ie{}, $\mathcal N=2$ SQCD. 

The partition function of SQCD on the four-sphere has appeared in our earlier computations, see \eqref{S4b_SQCD_partitionfunction}. In particular, it involves the contribution of instantons located at the north pole and south pole of the four-sphere. When decorating the computation with intersecting surface defects, which precisely have these points as their intersection locus, we should expect the instanton counting to be modified non-trivially. By performing the Higgsing procedure on a theory $\widetilde{\mathcal T}$ described by the $\mathcal N=2$ quiver 
\begin{center}
  \includegraphics[width=.3\textwidth]{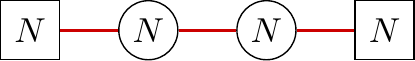}\;,
\end{center}
we will be able to derive a precise description of the modified ADHM integral by casting both the Higgsed partition function as well as its instanton contributions in a matrix integral form.

\paragraph{The $S^4_b$-partition function of $\widetilde{\mathcal T}$} is given by
\begin{multline}
Z^{(\widetilde{\mathcal T},S^4_{b})}(M,\tilde M,\hat M) =  \int \prod_{A,B = 1}^N {d{\Sigma _A}\ d{\Sigma'_B}}\  Z_{\text{cl}}^{(\widetilde{\mathcal T},S^4_{b})}(\Sigma ,\Sigma')\ Z_{\text{1-loop}}^{(\widetilde{\mathcal T},S^4_{b})}(\Sigma ,\Sigma', M,\tilde M,\hat M) \\
\times \ {\left| {Z_{\text{inst.}}^{(\widetilde{\mathcal T},\mathbb R^4)}(q,q',\Sigma ,\Sigma',M^\epsilon,\tilde M^\epsilon,\hat M^\epsilon)} \right|}^2 \;,
\end{multline}
where $M_I$ and $\tilde M_J$ denote the masses associated to the $U(N)$ flavor symmetry of the $N$ fundamental and antifundamental hypermultiplets respectively, while $\hat M$ is the mass associated to the $U(1)$ flavor symmetry of the bifundamental hypermultiplet. The classical action reads
\begin{equation}
Z_{\text{cl}}^{(\widetilde{\mathcal T},S^4_{b})}(\Sigma,\Sigma') = \exp\left[-\frac{8\pi^2}{g_{\text{YM}}^2} \Tr\Sigma^{ 2}  -\frac{8\pi^2}{g_{\text{YM}}^{\prime 2}} \Tr\Sigma^{\prime 2} \right]\;,
\end{equation}
while the one-loop determinant is given by
\begin{equation}
Z_{\text{1-loop}}^{(\widetilde{\mathcal T},S^4_{b})}(\Sigma ,\Sigma', M,\tilde M,\hat M) = Z_{{\text{vect}}}^{S_b^4}(\Sigma )\ Z_{{\text{vect}}}^{S_b^4}(\Sigma' )\ Z_{\text{fund}}^{S_b^4}(\Sigma,M)\ Z_{\text{afund}}^{S_b^4}(\Sigma' ,\tilde M)\ Z_{\text{bifund}}^{S_b^4}(\Sigma ,\Sigma',\hat M)\;,
\end{equation}
where all factors were defined in \eqref{S4b_one_loop_SQCD} but
\begin{equation}
Z_{\text{bifund}}^{S_b^4}(\Sigma ,\Sigma',\hat M) = \prod_{A = 1}^N {\prod_{B = 1}^N {\frac{1}{{{\Upsilon _b}(i\Sigma _B'-i{\Sigma _A} + i\hat M + Q/2)}}} } \;.
\end{equation}

The instanton partition function is given by a double sum over $N$-tuples of Young diagrams
\begin{multline}\label{IPFtwonode}
{Z_{\text{inst.}}^{(\widetilde{\mathcal T},\mathbb R^4)}(q,q',\Sigma ,\Sigma',M^\epsilon,\tilde M^\epsilon,\hat M^\epsilon)} =  \sum_{\vec Y,\vec Y'} {q^{|\vec Y|}}{{q'}^{|\vec Y'|}}\ z_{{\text{vect}}}^{{\mathbb{R}^4}}(\vec Y,\Sigma )\ z_{{\text{vect}}}^{{\mathbb{R}^4}}(\vec Y',\Sigma ')\  z_{{\text{fund}}}^{{\mathbb{R}^4}}(\vec Y,\Sigma ,M^\epsilon)\\
\times z_{{\text{afund}}}^{{\mathbb{R}^4}}(\vec Y',\Sigma' , \tilde M^\epsilon)\ z_{{\text{bifund}}}^{{\mathbb{R}^4}}(\vec Y,\vec Y',\Sigma ,\Sigma ',\hat M^\epsilon) \;. 
\end{multline}
The contributions of the various multiplets can be found in appendix \ref{appendix:IPF-factorization}. The superscripts $^\epsilon$ again denote the usual shift \cite{Okuda:2010ke}
\begin{equation}
M^\epsilon  = M-\frac{i}{2}({\epsilon _1} + {\epsilon _2})\;, \qquad \tilde M^\epsilon  = \tilde M-\frac{i}{2}({\epsilon _1} + {\epsilon _2})\;, \qquad \hat M^\epsilon  = \hat M-\frac{i}{2}({\epsilon _1} + {\epsilon _2})\;.
\end{equation}

\paragraph{Implementing the Higgsing prescription} once again amounts to considering the poles of the fundamental one-loop factor given by
\begin{equation}\label{poleS2S2_bis}
i\Sigma_A = i M_{\sigma(A)} -  n_A^{\text{L}} b -  n_A^{\text{R}} b^{-1} - \frac{b+b^{-1}}{2} \qquad \text{for} \qquad  A=1,\ldots,N\;,
\end{equation}
with $\sigma$ a permutation of $N$ elements, which we choose to be the identity. Here $\vec n^{\text{L/R}}$ is a partition of $n^{\text{L/R}}$, and we will sum over all.

It is straightforward to compute the residues of the one-loop determinant at \eqref{poleS2S2_bis}:
\begin{equation}\label{S4cl1loop_atpole_twonode}
Z_{\text{cl}}^{(\widetilde{\mathcal T},S^4_{b})}\ Z_{\text{1-loop}}^{(\widetilde{\mathcal T},S^4_{b})} \rightarrow  Z_{\text{cl}}^{({\mathcal T},S^4_{b})}Z_{\text{1-loop}}^{(\mathcal T,S^4_{b})}\ \Big(Z^{S^2_{\text{L}}}_{\text{cl}|\vec n^{\text{L}}}\ Z^{S^2_{\text{L}}}_{\text{1-loop}|\vec n^{\text{L}}}\Big)\ \Big(Z^{S^2_{\text{R}}}_{\text{cl}|\vec n^{\text{R}}}\ Z^{S^2_{\text{R}}}_{\text{1-loop}|\vec n^{\text{R}}}\Big) \Big(Z^{\widetilde{\mathcal T};\vec n^{\text{L}},\vec n^{\text{R}}}_{\text{cl,extra}}\ Z^{\widetilde{\mathcal T};\vec n^{\text{L}},\vec n^{\text{R}}}_{\text{1-loop,extra}}\Big)^{2}\;,
\end{equation}
Here $Z_{\text{cl}}^{({\mathcal T},S^4_{b})}Z_{\text{1-loop}}^{(\mathcal T,S^4_{b})}$ are the classical action and one-loop determinant of the theory $\mathcal T$, \ie{}, of four-dimensional $\mathcal N=2$ supersymmetric SQCD. Their expression can be found in \eqref{classactionSQCD} and \eqref{oneloopSQCD} respectively.\footnote{In the previous section the theory $\widetilde {\mathcal T}$ was SQCD.} The antifundamental masses of $\mathcal T$ are simply given by $\tilde M_J$, but the fundamental masses take the values 
\begin{equation}
M'_A = M_A - \hat M + iQ/2
\end{equation}
in terms of the fundamental and bifundamental masses of the quiver theory $\widetilde{\mathcal T}$. As before, $Z^{S^2_{\text{L/R}}}_{\text{\ldots}|\vec n^{\text{L/R}}}$ denote factors in the Higgs branch localized SQCDA two-sphere partition function. The two-dimensional FI-parameters $\xi^\text{L/R}_{\text{FI}}$, fundamental masses $m^\text{L/R}_I$, antifundamental masses $\tilde m^\text{L/R}_J$ and adjoint masses $m^\text{L/R}_X$ are now related to the four-dimensional parameters of theory $\widetilde{\mathcal T}$ as
\begin{align}\label{paramsS4_1}
  &\xi _{{\text{FI}}}^{\text{L}} = \frac{{4\pi }}{{g_{{\text{YM}}}^2}}\;,& &m_I^{\text{L}} = b({M_I} + iQ/2) + \frac{i}{2}{b^2}\;, & &\tilde m_J^\text{L} = b(\Sigma '_J + \hat M) + \frac{i}{2}\;, &&  m_X^{\text{L}} = ib^2\\
  &\xi _{{\text{FI}}}^{\text{R}} = \frac{{4\pi }}{{g_{{\text{YM}}}^2}}\;,& &m_I^\text{R} = b^{ - 1}(M_I + iQ/2) + \frac{i}{2}b^{ - 2}\;, & &\tilde m_J^\text{R} = b^{-1}(\Sigma '_J + \hat M) + \frac{i}{2}\;, &&  m_X^{\text{R}} = ib^{-2}\;,\label{paramsS4_2}
\end{align}
together with $\vartheta^\text{L/R} = \theta $. Note that the two-dimensional masses depend on the four-dimensional gauge parameter. The explicit expressions for the extra one-loop factors, which now receives contributions from the fundamental hypermultiplet, vector multiplet and bifundamental hypermultiplet one-loop determinant, can be found in \eqref{def:Z-vf-extra}-\eqref{def:Z-bifund-extra}. Again, $Z_\text{cl,extra}^{\vnl, \vnr} = (q\bar q)^{-\sum_A \nl_A \nr_A}$.

When substituting the gauge equivariant parameter \eqref{poleS2S2_bis} in the instanton partition functions \eqref{IPFtwonode}, the only non-vanishing contributions arise from $N$-tuples $\vec Y$ avoiding the ``forbidden box'' and arbitrary $N$-tuples $\vec Y'$. As before, we can split the sum over the former into one over large and one over small tuples. As we have learned in the previous section, the analysis of the large tuples is sufficient to derive the matrix model integral describing the infrared system, \ie{}, the theory $\mathcal T$ with intersecting defects inserted. We thus focus only on such large tuples. We find
\begin{align}
&q^{|\vec Y_{\text{large}}|} q^{\prime |\vec Y'|}\ Z_{{\text{inst}}}^{(\widetilde{\mathcal T},{\mathbb{R}^4} \times {S^1_{(1\cap 2)}})} (\vec Y',\vec Y_{\text{large}}) \nn \\
&\rightarrow\ q^{\prime |\vec Y'|}\ z_{{\text{vect}}}^{{\mathbb{R}^4}}(\vec Y',\Sigma ')\  z_{{\text{afund}}}^{{\mathbb{R}^4}}(\vec Y',\Sigma' ,\tilde M^\epsilon)\ z_{{\text{fund}}}^{{\mathbb{R}^4}}(\vec Y',\Sigma' ,M^{\prime\epsilon}) \ z^{\mathbb{R}^2_{\text{L}}}_\text{defect}(\vec Y', \Sigma' , \mathfrak{m}^\text{L})\ z^{\mathbb{R}^2_{\text{R}}}_\text{defect}(\vec {\tilde Y}',\Sigma' , \mathfrak{m}^\text{R}) \nn\\
&\phantom{\rightarrow\ }\times q^{|\mathfrak m_\text{L}| + |\mathfrak m_\text{R}|}\ Z_{\text{vortex}|\vec n^{\text{L}}}^{\mathbb{R}^2}(\mathfrak m^{\text L} ) \  Z_{\text{intersection}}^{\text{large}|\vec n^{\text{L}},\vec n^{\text{R}}}(\mathfrak m^{\text{L}},\mathfrak m^{\text{R}}) \ Z_{\text{vortex}|\vec n^{\text{R}}}^{\mathbb{R}^2}(\mathfrak m^{\text R} )  \ \Big( Z^{\widetilde{\mathcal T};\vec n^{(1)},\vec n^{(2)}}_{\text{cl,extra}}\ Z^{\widetilde{\mathcal T};\vec n^{(1)},\vec n^{(2)}}_{\text{1-loop,extra}} \Big)^{-1}\label{S4inst_atpole_twonode}\;,
\end{align}
The intersection factor was already given in \eqref{intersection_factor_HB_R4}. The expression for the factors $z^{\mathbb{R}^2_{\text{L/R}}}_\text{defect}$ can be found in \eqref{zLdef_HB}. They clearly correspond to new ingredients in the instanton partition function of $\mathcal T$, arising due to the presence of the defects on the local $\mathbb R^2_{\text{L/R}}$. Momentarily, we will study the modified ADHM data and its corresponding ADHM integral computing this modified instanton partition function. Recall that to obtain the final expression for the Higgsed partition function, we need to sum over all partitions of $n^{\text{L/R}}$.

\paragraph{A matrix model integral} describing the $S^4_b$ partition function of $SU(N)$ SQCD in the presence of intersecting surface defects supported on $S^2_\text{L, R}$ can be inferred from \eqref{S4cl1loop_atpole_twonode} and \eqref{S4inst_atpole_twonode} to be 
\begin{small}
\begin{align} 
  &Z^{(\mathcal{T},S^2_\text{L} \cup S^2_\text{R} \subset S^4_b)}=\frac{1}{n^{\text{L}}!n^{\text{R}}!}\sum_{B^{\text{R}}\in\mathbb Z^{n^\text{R}}}\sum_{B^{\text{L}}\in\mathbb Z^{n^\text{L}}} \int d\Sigma ' \int_{\mathrm{JK}} \prod_{a=1}^{n^\text{R}} \frac{d\sigma^{\text{R}}_a}{2\pi} \  \prod_{b=1}^{n^\text{L}} \frac{d\sigma^{\text{L}}_b}{2\pi}  \;Z^{({\mathcal T},S^4_b)}_{\text{cl}}(\Sigma')\ Z^{({\mathcal T},S^4_b)}_{\text{1-loop}}(\Sigma',M',\tilde M) \nn\\
  &\times  Z^{S^2_{\text{R}}}(\sigma^{\text{R}},B^{\text{R}};\Sigma') \ Z^{S^2_{\text{L}}}(\sigma^{\text{L}},B^{\text{L}};\Sigma')\ Z_{\text{intersection}}^+(\sigma^{\text{L}}, B^{\text{L}},\sigma^{\text{R}},B^{\text{R}}) \ Z_{\text{intersection}}^-(\sigma^{\text{L}}, B^{\text{L}},\sigma^{\text{R}},B^{\text{R}})    \label{def:SQCDA-defect-matrix-model} \\
  & \times {\left| {\sum_{\vec Y'} {{q'}^{|\vec Y'|}} 
  z_{{\text{vect}}}^{{\mathbb{R}^4}}(\vec Y',\Sigma ')\,  z_{{\text{afund}}}^{{\mathbb{R}^4}}(\vec Y',\Sigma' ,\tilde M^\epsilon)\, z_{{\text{fund}}}^{{\mathbb{R}^4}}(\vec Y',\Sigma' ,M^{\prime\epsilon})
  \, z_{{\text{defect}}}^{\mathbb{R}_{\text{L}}^2}(\vec Y', \Sigma', \sigma^\text{L}, B^\text{L})\, z_{{\text{defect}}}^{\mathbb{R}_{\text{R}}^2}(\vec Y', \Sigma', \sigma^\text{R}, B^\text{R}) } \right|^2}.\nn
\end{align}
\end{small}%
Here the factors in the first lines are the classical action and one-loop determinant of $\mathcal T$, \ie{}, four-dimensional SQCD, and the factors in the second line are the $S^2_{\text{L/R}}$ partition functions for SQCDA as well as the intersection factors \eqref{intersection-factor-CBL-S2}. The last line contains two copies of the instanton partition function, computed in the presence of the locally planar intersecting surface defects. 

The mass parameters on the two two-spheres are related as in \eqref{paramidentificationsS2US2}, while the parameters of the four-dimensional theory $\mathcal T$ are related to the two-dimensional ones as
\begin{equation}\label{4dmassSQCD}
i{b^{ - 1}} = \left[ {{M_{I}'-\Sigma'_J} + \frac{i}{2}(b + {b^{ - 1}})} \right] - {b^{ - 1}}(m_I^{\text{L}} - \tilde m_J^{\text{L}}) \;, \qquad i{b} = \left[ {{M_{I}'-\Sigma'_J} + \frac{i}{2}(b + {b^{ - 1}})} \right] - {b}(m_I^{\text{R}} - \tilde m_J^{\text{R}})\;.
\end{equation}
Note that when performing the integral over the four-dimensional gauge parameter $\Sigma'$, one should use
\begin{equation}
\tilde m_J^\text{L} = b\Sigma_J' + \tilde m^{\text{L}}_{U(1)} \;, \qquad \tilde m_J^\text{R}= b^{-1} \Sigma_J' + \tilde m^{\text{R}}_{U(1)}\;,
\end{equation}
where $\tilde m^{\text{L/R}}_{U(1)} = \frac{1}{N}\sum_{K=1}^N \tilde m_K^\text{L/R}$. These follow directly from \eqref{4dmassSQCD}. In the two-dimensional one-loop determinants we have made explicit this $\Sigma'$-dependence. Note that by performing the change of variables $\sigma^{\text{L/R}}_I \rightarrow  \sigma^{\text{L/R}}_I + \tilde m^{\text{L/R}}_{U(1)}$ in the two-dimensional integrals, one effectively changes the $U(1)$ masses as $\tilde m^{\text{L/R}}_{U(1)}\rightarrow 0$ and $m^{\text{L/R}}_{U(1)}\rightarrow m^{\text{L/R}}_{U(1)} -\tilde m^{\text{L/R}}_{U(1)}$ in the matrix integral, up to an overall constant factor originating from the two-dimensional classical actions.\footnote{Note that in terms of the effective variables, the relation \eqref{4dmassSQCD} remains unaffected, but \eqref{paramidentificationsS2US2} is modified as
\begin{equation}
b^{-1} \left(m^{\text{L}}_I + \frac{i}{2} \right)  = b\left(m^{\text{R}}_I + \frac{i}{2}\right) + c\;, \qquad b^{-1}\left(\tilde m^{\text{L}}_J - \frac{i}{2} \right) =  b\left(\tilde m^{\text{R}}_J-\frac{i}{2} \right) + c\;,
\end{equation} 
with $c = b^{-1} \tilde m^{\text{L}}_{U(1)}-b \tilde m^{\text{R}}_{U(1)}+\frac{i}{2}(b-b^{-1})$.} Henceforth, we choose to work with this effective new integral.

The contribution of the locally planar surface defect, supported on the local $\mathbb R^2_{\text{L}}$, to the north pole copy of the instanton partition function is given by
\begin{equation}\label{zdefect}
z_{{\text{defect}}}^{\mathbb{R}_\text{L}^2}(\vec Y',\Sigma',\sigma ^{\text{L}}, {B^{\text{L}}}) = \prod_{a = 1}^{{n^{\text{L}}}} \prod_{B=1}^N \prod_{r=1}^{W_{Y'_B}} \prod_{s=1}^{Y'_{Br}} \frac{-\epsilon_2 (i\sigma_a + B_a/2 - i \tilde m_B ) + r\epsilon_1 + s \epsilon_2}{-\epsilon_2 (i\sigma_a + B_a/2 - i \tilde m_B ) + (r-1)\epsilon_1 + s \epsilon_2}\;,
\end{equation}
where $W_{Y_B'}$ denotes the width of the Young diagram $Y_B'$. Similarly, $z_{{\text{defect}}}^{\mathbb{R}_\text{R}^2}$ is obtained by swapping $\text{L}\leftrightarrow \text{R}$, $\epsilon_1 \leftrightarrow \epsilon_2$ and $Y_B \leftrightarrow \tilde Y_B$. The combination $i \sigma + \frac{1}{2}B$ is valid for the north pole contributions; to get the south pole counterpart one replaces it with $i \sigma - \frac{1}{2}B$.

One can verify that if we perform the integrations over $\sigma^\text{L/R}$ and the sums over $B^\text{L/R}$ using the same Jeffrey-Kirwan-like residue prescription as discussed in the previous section, the matrix model \eqref{def:SQCDA-defect-matrix-model} reproduces the result obtained from the Higgsing prescription.

\begin{figure}[t!]
  \centering
  \includegraphics[width=0.4\textwidth]{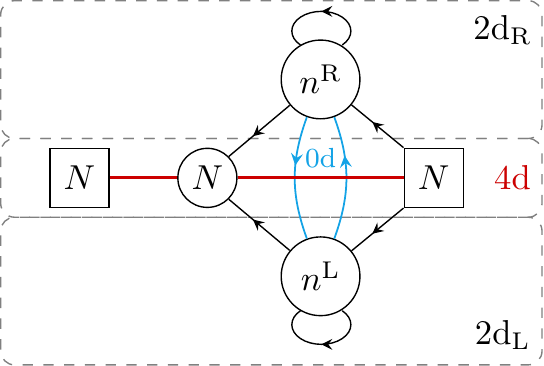}
  \caption{\label{fig:InsertSymmetricsSQCD} Coupled 4d/2d/0d quiver gauge theory realizing the insertion, in four-dimensional $\mathcal N=2$ SQCD, of intersecting M2-brane surface defects labeled by symmetric representations of rank $n^{\text{R}}$ and $n^{\text{L}}$ respectively . Various superpotential couplings are turned, in direct analogy to the ones given in detail in \cite{Gomis:2016ljm}. The Higgsing prescription applied to a linear quiver gauge theory with two gauge nodes reproduces the partition function of this coupled system.}
\end{figure}

\paragraph{The 4d/2d/0d coupled system} whose partition function is computed by \eqref{def:SQCDA-defect-matrix-model} is depicted in figure \ref{fig:InsertSymmetricsSQCD}. 
The first line of \eqref{def:SQCDA-defect-matrix-model} captures the classical action and one-loop determinant of the four-dimensional theory, while the second line captures the contributions of the two-dimensional degrees of freedom residing on the intersecting two-spheres as well as the one-loop determinants of the zero-dimensional bifundamental chiral multiplets at their intersection points. The most salient new feature of this coupled system is the fact that part of the two-dimensional flavor symmetry is gauged by the four-dimensional gauge symmetry. This fact is reflected in the relations in \eqref{4dmassSQCD}, relating the two-dimensional mass parameters $\tilde m$ to the gauge parameter $\Sigma'$, which are the consequence of the usual cubic superpotential couplings. As mentioned above, when computing the squashed four-sphere partition function of the coupled system, the instanton counting is modified non-trivially due to the presence of the intersecting surface defects. The argument of the modulus squared in the last line of \eqref{def:SQCDA-defect-matrix-model} provides a concrete expression for the modified instanton partition function. In the next section, we turn to a more detailed analysis of the degrees of freedom which give rise to this instanton partition function.

\section{Instanton partition function and intersecting surface defects}\label{sec:IPFwDef}
\begin{figure}[t!]
  \centering
  \includegraphics[width=0.8\textwidth]{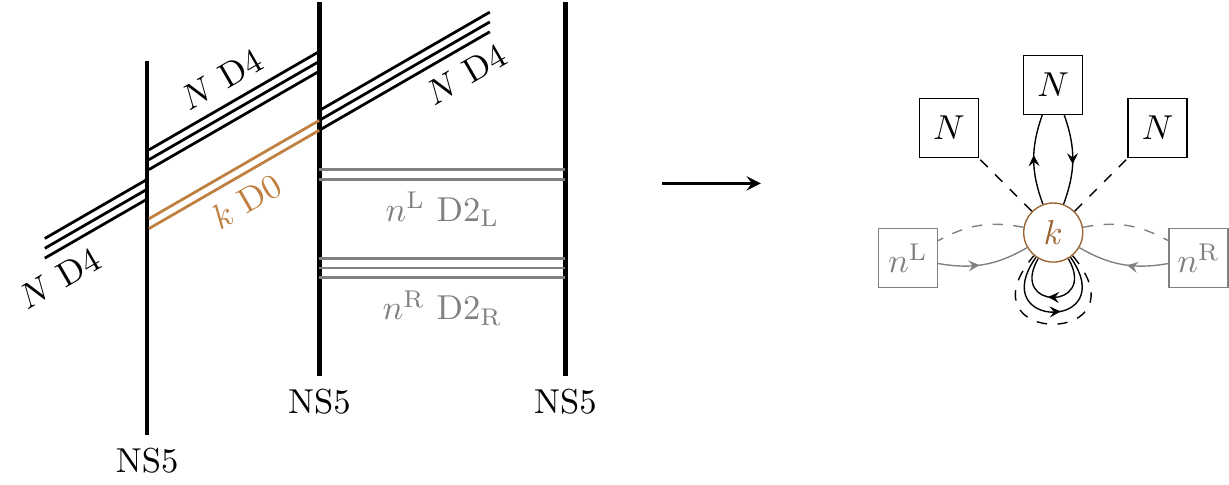}
  \caption{\label{fig:SQCD_Brane_Instanton} The left part of the figure depicts the brane configuration realizing $k$-instantons in $\mathcal N=2$ SQCD, in the presence of intersecting surface defects, of M2-type and labeled by symmetric representations, represented by the gray branes.\protect\footnotemark{} The right part of the figure shows the quiver description of the worldvolume theory of the D0-branes. As the system preserves two-dimensional $\mathcal N=(0,2)$ supersymmetry dimensionally reduced to zero dimensions, the quiver is drawn using $\mathcal N=(0,2)$ notations, with full lines representing chiral multiplets and dashed lines Fermi multiplets. In the absence of the defects, the preserved supersymmetry is $\mathcal N=(0,4)$. We thus learn that one needs to turn a J-type superpotential $J_\Lambda$ for the adjoint Fermi multiplet $\Lambda$ consisting of the sum of the adjoint bilinears of the scalars of the two pairs of chiral multiplets. The charges in table \ref{table:ADHMstrings} are also compatible with quadratic E- or J-type superpotentials for the Fermi multiplets charged under $U(n^{\text{L/R}})$.
  }
\end{figure}

Let us start by considering the familiar brane realization of a $k$-instanton in $SU(N)$ $\mathcal N=2$ SQCD as an additional stack of $k$ D0-branes as depicted in the the left part of figure \ref{fig:SQCD_Brane_Instanton}, ignoring the gray branes for the time being.
\footnotetext{The brane directions are as in footnote \ref{branedirections}.}
The supersymmetry preserved by the worldvolume theory of the D0-branes is the dimensional reduction to zero dimensions of two-dimensional $\mathcal N=(0,4)$ supersymmetry. Its matter content can be straightforwardly read off by quantizing the open strings stretching between the D0-brane and the various D4-branes, as well as between the D0-branes themelves, see \cite{Douglas:1995bn,Douglas:1996uz}. We summarize it in table \ref{table:ADHMstrings}, and have depicted the resulting quiver gauge theory in the right part of figure \ref{fig:SQCD_Brane_Instanton} (omitting the gray quiver nodes and links). The partition function of this zero-dimensional theory computes the (non-perturbative) $k$-instanton partition function, which we denote as $Z_k^{\mathbb R^4}$.
\renewcommand{\arraystretch}{1.3}
\begin{table}[ht]
\begin{center}
\begin{tabular}{c||c|c|c|c|c|c|c|c|c|c|c|c}
strings & D0-D4$_1$ & \multicolumn{2}{|c|}{D0-D4$_2$} & D0-D4$_3$ & \multicolumn{4}{|c|}{D0-D0} & \multicolumn{2}{|c}{{\color{gray} D0-D2$_\text{R}$}} & \multicolumn{2}{|c}{{\color{gray} D0-D2$_\text{L}$}} \\
\hline\hline
$\mathcal N=(0,4)$ & FM & \multicolumn{2}{|c|}{HM} & FM & \multicolumn{2}{|c|}{VM} & \multicolumn{2}{|c|}{HM} & \multicolumn{2}{|c}{{\color{gray}\tiny (not preserved)}} & \multicolumn{2}{|c}{{\color{gray}\tiny (not preserved)}}\\
\hline\hline
$\mathcal N=(0,2)$ & FM & CM & CM & FM & VM & FM & CM & CM & {\color{gray}CM} & {\color{gray}FM} & {\color{gray}CM}&{\color{gray}FM}\\
\hline
$J$ & 0 &  $\frac{1}{2}$ & $\frac{1}{2}$ & 0 & 0&1 & $\frac{1}{2}$ & $\phantom{-}\frac{1}{2}$ & {\color{gray}0} & {\color{gray}$\frac{1}{2}$} & {\color{gray}$\phantom{-}$0} & {\color{gray}$\frac{1}{2}$}\\
\hline
$J_l$ & 0& 0 & 0 & 0 & 0 & 0& $\frac{1}{2}$ & $-\frac{1}{2}$ & {\color{gray}$\frac{1}{2}$} & {\color{gray}0} & {\color{gray} $-\frac{1}{2}$} & {\color{gray}0}\\
\end{tabular}
\caption{Massless excitations of strings stretching between the branes indicated in the first row organized in multiplets of the dimensional reduction of two-dimensional $\mathcal N=(0,4)$  and $\mathcal N=(0,2)$ supersymmetry to zero dimensions in the second and third row respectively. Here VM denotes vectormultiplet, HM hypermultiplet, FM Fermi multiplet and CM chiral multiplet. Note that the system including the D2-branes only preserves $\mathcal N=(0,2)$, hence we leave the $\mathcal N=(0,4)$ entries corresponding to D0-D2 strings open. The last two rows list the charges of the mutliplets under the flavor symmetry charges $J$ and $J_l$. \label{table:ADHMstrings}}
\end{center}
\end{table}

In some more detail, the instanton partition function of a four-dimensional $\mathcal N=2$ supersymmetric theory is computed by a localization computation on $\mathbb R^4_{\epsilon_1, \epsilon_2}$, \ie{}, in the $\Omega$-background parametrized by $\epsilon_1,\epsilon_2$ \cite{Nekrasov:2002qd,Nekrasov:2003rj}. The localizing supercharge $\mathcal Q$ squares to
\begin{equation}\label{QsqIPF}
\mathcal Q^2 = (\epsilon_1 + \epsilon_2)(J_r + \mathcal R) + (\epsilon_1 - \epsilon_2)J_l + i\Sigma\cdot G + iM\cdot F\;,
\end{equation}
where $J_l,J_r$ are the Cartan generators of the $SU(2)_l\times SU(2)_r \simeq SO(4)$ rotational symmetries of $\mathbb R^4$,\footnote{The $\Omega$-deformation breaks the rotational symmetry to $SO(2)_1\times SO(2)_2$. In terms of their Cartan generators $J_1, J_2$ one has $J_l = \frac{1}{2}(J_1-J_2)$ and $J_r = \frac{1}{2}(J_1+J_2)$.} while $\mathcal R$ is the $SU(2)_{\mathcal R}$ generator. We define $J=J_r+\mathcal{R}$. Furthermore, $G$ denotes the collection of Cartan generators of the gauge symmetry and $F$ those of the flavor symmetry; $\phi$ and $M$ are their respective equivariant parameters. The localization locus consists of point-instantons located at the origin. The integration over the $k$-instanton moduli space is captured by the non-perturbative $k$-instanton partition function, which equals the partition function of the zero-dimensional ADHM model, read off as the worldvolume theory on the D0-branes, computed by localization with respect to the induced supercharge, that is, with respect to the supercharge in its $\mathcal N=(0,2)$ supersymmetry (sub)algebra satisfying the same square as in \eqref{QsqIPF} (up to gauge transformations). From the $\mathcal N=(0,2)$ zero-dimensional point of view, the charges $J=J_r + \mathcal R$ and $J_l$ appear as flavor charges, as do both $G$ and $F$. $\mathcal Q^2$ additionally includes $\phi\cdot G_{0d}$.

Dimensionally reducing the localization results of \cite{Hori:2014tda} and in particular \cite{Hwang:2014uwa}, it is now straightforward to compute $Z_k^{\mathbb R^4}$ as
\begin{equation}\label{Zkintegral}
Z_k^{\mathbb R^4}= \int_{\text{JK}} \prod_{I = 1}^k {d{\phi _I}\ {Z_{{\text{D0-D0}}}}(\phi)\  Z_{\text{D0-D4}_1} (\phi,\tilde M)\ {Z_{{\text{D0-D4}_2}}}(\phi ,\Sigma ')}\  Z_{\text{D0-D4}_3} (\phi,M)\;,
\end{equation}
where
\begin{align}
{Z_{{\text{D0-D0}}}}(\phi ) = &\; \prod_{I,J = 1}^k {\frac{{({\phi _{IJ}})'({\phi _{IJ}} + {\epsilon _1} + {\epsilon _2})}}{{({\phi _{IJ}} + {\epsilon _1})({\phi _{IJ}} + {\epsilon _2})}}} \;, \\
Z_{\text{D0-D4}_1} (\phi,\tilde M) = &\; \prod_{I = 1}^k {\prod_{A = 1}^N {({\phi _I} - i{\tilde M_A})} } \;, \qquad Z_{\text{D0-D4}_3} (\phi, M) = \; \prod_{I = 1}^k {\prod_{A = 1}^N {({\phi _I} - i{ M_A})} }\;,\\
{Z_{{\text{D0-D4}_2}}}(\phi ,\Sigma ') = & \; \prod_{I = 1}^k {\prod_{A = 1}^N {\frac{1}{{({\phi _I} - i\Sigma _A' + \frac{1}{2}({\epsilon _1} + {\epsilon _2}))( - {\phi _I} + i\Sigma _A' + \frac{1}{2}({\epsilon _1} + {\epsilon _2}))}}} } \;,
\end{align}
where ${\phi _{IJ}} = \phi_I - \phi_J$, and the prime on $({\phi_{IJ}})'$ indicates to omit the factors with $I=J$. Here we denoted the equivariant parameters for the various $SU(N)$ symmetries as in the previous section.

The integral \eqref{Zkintegral} is computed using the Jeffrey-Kirwan residue prescription. We choose to select the contributions of negatively charged fields, and thus collect the residues of the poles defined by solving the equations
\begin{equation}
  {\phi _I} = i\Sigma_A' + \frac{1}{2}({\epsilon _1} + {\epsilon _2}) , \qquad {\phi _I} = \phi _J + \epsilon_1 , \qquad  {\phi _I} = {\phi _J} + {\epsilon _2} \;.
\end{equation}
The contributing poles are labeled by $N$-tuples of Young diagrams $\vec Y = \{Y_A\}$ such that $\sum_A |Y_A| = k$,
\begin{equation}
  {\phi _I} = i{\Sigma _A'} - \frac{1}{2}({\epsilon _1} + {\epsilon _2}) + r{\epsilon _1} + s{\epsilon _2}\;, \qquad (r,s) \in Y_A\;.
\end{equation}
It is easy to convince oneself that summing over the residues precisely reproduces the $q^k$ term of the SQCD instanton partition function given in \eqref{SQCDIPF}.

Let us now re-introduce the intersecting surface defects in the setup.\footnote{See also \cite{Gaiotto:2014ina} for an analysis of the equivariant integral of a five-dimensional theory in the presence of three-dimensional chiral multiplets.} The brane configuration was already depicted in the left part of figure \ref{fig:SQCD_Brane_Instanton}, now also considering the gray branes. Upon inserting the defects, the $\mathcal N=(0,4)$ symmetry, dimensionally reduced to zero dimensions, carried by the D0-branes is broken to the dimensional reduction of $\mathcal N=(0,2)$. We have used precisely this subalgebra in the previous paragraphs already. The zero-dimensional model is enriched by the modes arising from the quantization of the open strings stretching between the D0 and D2$_{\text{L}}$ and D2$_{\text{R}}$-branes. They each contribute an additional $\mathcal N=(0,2)$ Fermi and chiral multiplet,\footnote{The brane system consisting of a stack of D0-branes and one stack of D2-branes, each ending on an NS5-brane, preserves on the D0-brane the dimensional reduction to zero dimensions of $\mathcal N=(2,2)$ supersymmetry. The open string modes thus organize themselves in an $\mathcal N=(2,2)$ chiral multiplet.} and the final ADHM quiver theory is depicted in the right part of figure \ref{fig:SQCD_Brane_Instanton}. The additional multiplets carry charges under $J$ and $J_l$ as in table \ref{table:ADHMstrings}. It is then straightforward to include their contributions to the ADHM matrix model. It is given by
\begin{multline}
  Z_\text{D0-D2}(\phi ,\Sigma ',\sigma ,B) \\
  \equiv \prod_{I = 1}^K \prod_{A = 1}^N \Bigg[ \prod_{a  = 0}^{n_A^{\text{L}} - 1} \frac{{\phi _I} - \epsilon_2(i\sigma_a^{\text{L}} + \frac{B_a^{\text{L}}}{2})+\frac{1}{2}(\epsilon_1+\epsilon_2)}{{\phi _I} - \epsilon_2(i\sigma_a^{\text{L}} + \frac{B_a^{\text{L}}}{2})-\frac{1}{2}(\epsilon_1-\epsilon_2)} \prod_{a  = 0}^{n_A^{\text{R}} - 1} \frac{{\phi _I} - \epsilon_1(i\sigma_a^{\text{R}} + \frac{B_a^{\text{R}}}{2})+\frac{1}{2}(\epsilon_1+\epsilon_2)}{{\phi _I} - \epsilon_1(i\sigma_a^{\text{R}} + \frac{B_a^{\text{R}}}{2})+\frac{1}{2}(\epsilon_1-\epsilon_2)} \Bigg]\;.
\end{multline}
where we used $\sigma_a^{\text{L/R}} -i \frac{B_a^{\text{L/R}}}{2}$ as the gauge equivariant parameter of the $U(n^{\text{L/R}})$ symmetry, as this is the combination that enters in our computations on $S^4_b$ at the north pole. (The south pole contribution would be obtained by changing the sign in front of $B$.)\footnote{Note that we are using the effective description obtained by performing a change of variables in the two-dimensional integrals and omitting some irrelevant constant prefactor as explained below equation \eqref{def:SQCDA-defect-matrix-model}.}

Noting that our JK-prescription does not select the poles of the above factor, it is straightforward to see that the matrix integral 
\begin{equation}\label{Zkintegraldefects}
Z_k^{\mathbb R^2_{\text{L}} \cup \mathbb R^2_{\text{R}} \subset \mathbb R^4}= \int_{\text{JK}} \prod_{I = 1}^k {d{\phi _I}\, {Z_{{\text{D0-D0}}}}(\phi)\,  Z_{\text{D0-D4}_1} (\phi,\tilde M)\, {Z_{{\text{D0-D4}_2}}}(\phi ,\Sigma ')}\,  Z_{\text{D0-D4}_3} (\phi,M)\, Z_\text{D0-D2}(\phi ,\Sigma ',\sigma ,B)
\end{equation}
precisely reproduces the modified instanton partition function as it appeared in the last line of \eqref{def:SQCDA-defect-matrix-model}.

\section{Discussion}\label{sec:conclusions}
\begin{figure}[t]
  \centering
  \includegraphics[width=\textwidth]{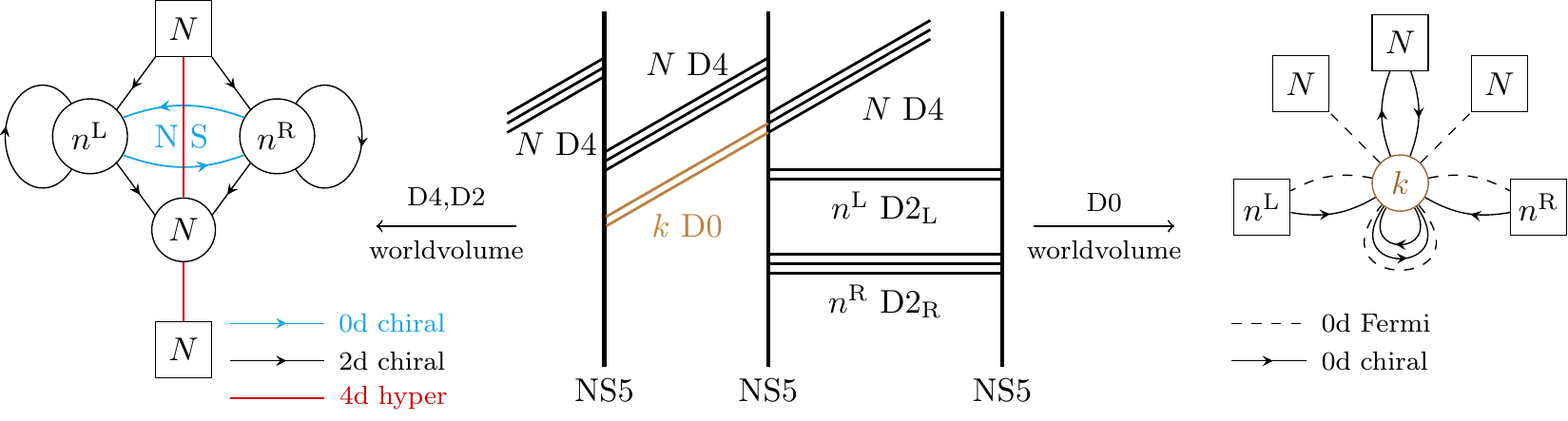}
  \caption{\label{fig:ADHM_conclusions} The type IIA brane-configuration in the middle describes the 4d/2d/0d coupled system on the left as the worldvolume theory of the D4/D2$_\text{L}$/D2$_\text{R}$-branes, see also figure \ref{fig:InsertSymmetricsSQCD}. The worldvolume theory of the $k$ D0-branes is shown on the right, see figure \ref{fig:SQCD_Brane_Instanton} for more details. Its partition function computes the $k$-instanton partition function of the 4d/2d/0d coupled system.}
\end{figure}

In this paper, we have extended the study of intersecting codimension two defects, initiated in \cite{Gomis:2016ljm}, to interacting four-dimensional theories. We have employed the Higgsing prescription of \cite{Gaiotto:2012xa,Gaiotto:2014ina} to compute the vacuum expectation value of intersecting M2-brane defects, labeled by $n^{\text{L}}$ and $n^{\text{R}}$-fold symmetric representations respectively, inserted in four-dimensional $\mathcal N=2$ SQCD. Subsequently we cast the result in the form of a partition function of a coupled 4d/2d/0d system, see \eqref{def:SQCDA-defect-matrix-model}, which takes the schematic form\footnote{Before tackling this computation, we also considered the theory of $N^2$ free hypermultiplets. Also for this case, we cast the result in the form of a partition function of a coupled system.}
\begin{equation}\label{SQCD_conclusion}
Z^{(\mathcal{T},S^2_\text{L} \cup S^2_\text{R} \subset S^4_b)} = \SumInt \ Z_{\text{pert}}^{(\mathcal T,S^4_b)}\ Z_{\text{pert}}^{(\tau^{\text{L}},S^2_\text{L})} \ Z_{\text{pert}}^{(\tau^{\text{R}},S^2_\text{R})} \   Z^{+}_{\text{intersection}}\ Z^{-}_{\text{intersection}}\ \left|Z_{\text{inst}}^{(\mathcal T, \mathbb R^2_{\text{L}}\cup \mathbb R^2_{\text{R}} \subset \mathbb R^4)}\right|^2\;.
\end{equation}
The leftmost subfigure of figure \ref{fig:ADHM_conclusions} depicts the 4d/2d/0d coupled system under consideration. The theory $\mathcal T$ is four-dimensional $\mathcal N=2$ SQCD and $\tau^{\text{L/R}}$ are identified as two-dimensional $\mathcal N=(2,2)$ $U(n^{\text{L/R}})$ SQCDA. 
Our computation provides an explicit formula for the instanton partition function in the presence of the above-mentioned intersecting defects, $Z_{\text{inst}}^{(\mathcal T, \mathbb R^2_{\text{L}}\cup \mathbb R^2_{\text{R}} \subset \mathbb R^4)}$, appearing in \eqref{SQCD_conclusion}, see equation \eqref{def:SQCDA-defect-matrix-model}. We also found the ADHM model whose partition function computes the $k$-instanton contribution to $Z_{\text{inst}}^{(\mathcal T, \mathbb R^2_{\text{L}}\cup \mathbb R^2_{\text{R}} \subset \mathbb R^4)}$, see the rightmost subfigure in figure \ref{fig:ADHM_conclusions}. This model can be read off from a D-brane construction, as also indicated in the figure.

\begin{figure}[t]
  \centering
  \includegraphics[width=\textwidth]{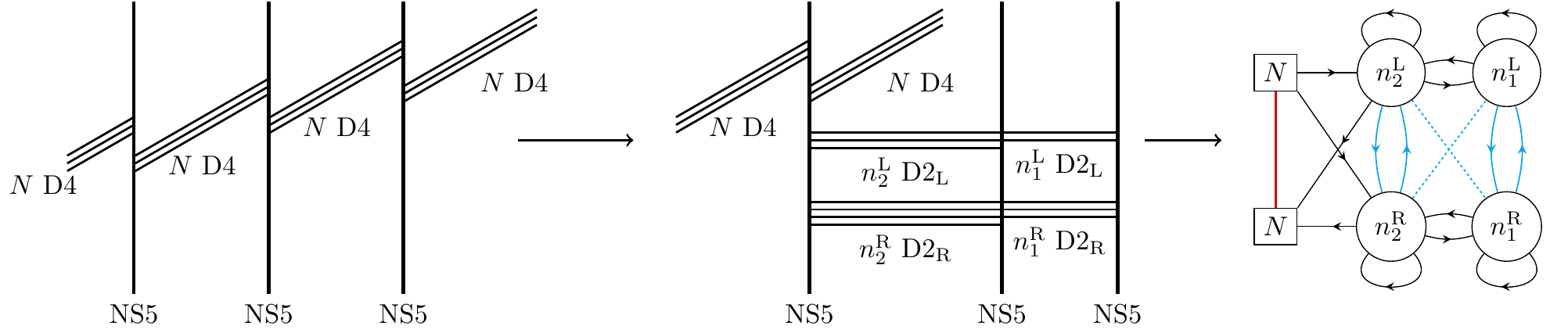}
  \caption{\label{figure:consecutive-Higgsing-brane-construction} One starts with the theory $\mathcal T$ of $N^2$ free hypermultiplets and successively gauges in two more theories of $N^2$ free hypermultiplets. The brane realization of the resulting theory $\widetilde {\mathcal T}$ is shown in the leftmost figure. One can then apply the Higgsing prescription twice, corresponding to pulling the two rightmost NS5-branes away from the main stack, and stretching $(n^\text{L}_2 - n^\text{L}_1, n^\text{R}_2 - n^\text{R}_1)$ (D2$_\text{L}$, D2$_\text{R}$) branes and ($n^\text{L}_1, n^\text{R}_1$) (D2$_\text{L}$, D2$_\text{R}$) branes respectively in between them and the flavor D4- branes. The two-dimensional part of the system is in its Higgs phase, and can be brought into its Coulomb phase by aligning the two displaced NS5-branes, as shown in the middle figure. The corresponding 4d/2d/0d coupled system can be read off easily, and is shown in the last figure. This system was also considered in \cite{Gomis:2016ljm}.}
\end{figure}

Starting from a theory $\mathcal T$ whose flavor symmetry contains an $SU(N)$ factor, one can gauge in successively multiple, say $p$, theories of $N^2$ free hypermultiplets. The resulting theory $\widetilde {\mathcal T}$ has $p$ additional $U(1)$ factors in its flavor symmetry group compared to the original theory $\mathcal T$. It is clear that one can apply the Higgsing prescription consecutively to each of these starting from the outermost one along the quiver. The associated type IIA brane-realization is a simple generalization of the one we have discussed in section \ref{subsec:brane realization}. We depict the case $p=2$ for $\mathcal T$ the theory of $N^2$ free hypermultiplets in figure \ref{figure:consecutive-Higgsing-brane-construction}.
The corresponding 4d/2d/0d coupled system can be read off from the brane picture and is given in the rightmost subfigure in figure \ref{figure:consecutive-Higgsing-brane-construction}. We conjecture that the M-theory interpretation of this procedure corresponds to the insertion of multiple intersecting M2-branes ending on the main stack of M5-branes, describing theory $\mathcal T$, all labeled by symmetric representations.  

\begin{figure}[t!]
  \centering
  \includegraphics[width=\textwidth]{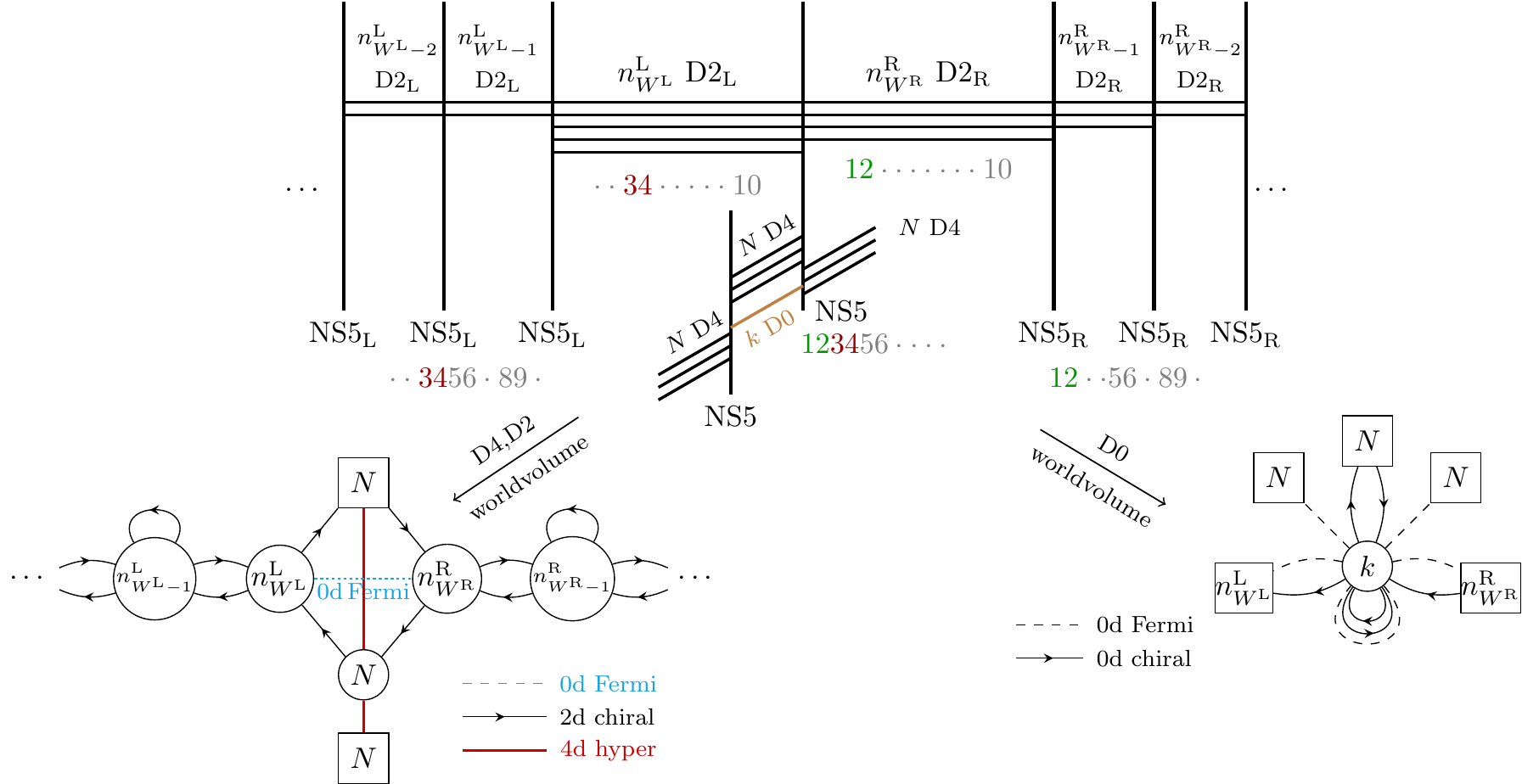}
  \caption{\label{figure:spacetime-ADHM-quivers-antisymmetric} The type IIA brane realization of general intersecting M2-brane defects labeled by two irreducible representation $(\mathcal{R}^\text{L}, \mathcal{R}^\text{R})$ inserted in SQCD, as well as its corresponding 4d/2d/0d coupled system and ADHM model are depicted.}
\end{figure}

General intersecting M2-brane defects labeled by two generic irreducible representations $(\mathcal{R}^\text{L}, \mathcal{R}^\text{R})$ of $SU(N)$ can also be described by 4d/2d/0d coupled systems \cite{Gomis:2016ljm}; when the four-dimensional theory is $\mathcal N=2$ SQCD, we have depicted an example in the bottom left of figure \ref{figure:spacetime-ADHM-quivers-antisymmetric}. The two-dimensional degrees of freedom are described by quiver gauge theories which encode the representation $\mathcal R$ through their gauge group ranks \cite{Gomis:2014eya}. The coupled system involves zero-dimensional Fermi multiplets, transforming in the bifundamental representation of the innermost two-dimensional gauge groups, as degrees of freedom living at the intersection points. Such 4d/2d/0d coupled system can be engineered as the worldvolume theory of the D4/D2$_\text{L}$/D2$_\text{R}$-branes in the type IIA system shown in figure \ref{figure:spacetime-ADHM-quivers-antisymmetric}.

When attempting to use this coupled system to compute the vacuum expectation value of general intersecting M2-brane defects, one needs as an input the instanton partition function in the presence of the defects. We propose that the structure of its $k$-instanton ADHM model can also in this case be read off from the D0-brane worldvolume theory in the type IIA system. The resulting quiver theory, which has two-dimensional $\mathcal{N} = (0,2)$ supersymmetry reduced to zero dimensions, is also included in figure \ref{figure:spacetime-ADHM-quivers-antisymmetric}. It is almost the same as the one in figure \ref{fig:ADHM_conclusions}, up to the orientation of an arrow. This new ADHM model however leads to a dramatically different ADHM integration: extra poles coming from the factor $Z_\text{D0-D2}$ will be selected by the JK-prescription, and the result of the ADHM integral is a double sum over $N$-tuples of Young diagrams and, separately, $n^\text{L}$-tuples of Young diagrams, together having $k$ boxes in total. It would be very interesting to study in more detail these new ADHM integrals.

When $\mathcal{R}^\text{L/R}$ are both symmetric representations, the descriptions of figures \ref{figure:spacetime-ADHM-quivers-antisymmetric} and \ref{fig:ADHM_conclusions} are both valid. In \cite{Gomis:2016ljm}, the equality of the resulting squashed four-sphere partition functions was verified for four-dimensional theories without gauge fields, and for defects labeled by fundamental representations. It would be of interest to study the duality between the two descriptions in interacting theories.

A construction for general intersecting M2-brane defects in terms of a renormalization group flow from a larger theory $\widetilde{\mathcal T}$ triggered by some position-dependent Higgs branch vacuum expectation value is currently unknown. Presumably it requires $\widetilde{\mathcal T}$ to be a non-Lagrangian theory of class $\mathcal S$. Reversing the logic, one might hope to recover information about the partition function of the UV non-Lagrangian theory $\widetilde {\mathcal{T}}$ by investigating the partition function of all intersecting M2-brane defects, for which we have nice quiver description, and to which the UV theory can flow. Note that Higgs branch localized expressions of partition functions are a simple example of this aspiration \cite{Pan:2015hza}. 

It was conjectured in \cite{Gomis:2016ljm} that the partition function of $SU(N)$ SQCD in the presence of general intersecting M2-brane defects can be identified with Liouville/Toda five-point functions. In particular, identifying $x' = qz$, $x = z$, with $|q|, |z| < 1$, along with other parameter identifications, one expects
\begin{align}
  Z^{(\mathcal T, S_{\text{L}}^2 \cup S_{\text{R}}^2 \subset S_b^4)} (q,z,M',\tilde M) = \mathcal{A}(x, x') \left\langle {{{\hat V}_{{\alpha _0}}}(0)\ \hat V_{-b \Omega_{\mathcal{R}^\text{L}} - b^{-1} \Omega_{\mathcal{R}^\text{R}} }(x')\ {{\hat V}_\beta }(x)\ {{\hat V}_{{\alpha _1}}}(1)\ {{\hat V}_{{\alpha _\infty }}}(\infty )} \right\rangle \;, \label{AGT-correspondence-defect}
\end{align}
where $\mathcal{A}(x, x') \equiv A|x'{|^{2{\gamma _0}}}|1 - x'{|^{2{\gamma _1}}}|x{|^{2{\gamma _2}}}|1 - x{|^{2{\gamma _3}}}|x - x'{|^{2\gamma _4 }}$, for some $\gamma_i$. Furthermore $\alpha_0$, $\alpha_\infty$ are generic, while $\beta$, $\alpha_1$ are semi-degenerate momenta determined in terms of the masses of the gauge theory. Let us perform a few checks of this statement, leaving a more thorough analysis for the future.

Consider the simple case with $\mathcal{R}^\text{L}=\mathbf 1$ and $\mathcal{R}^\text{R} = \operatorname{symm}^{n^\text{R}} \square$. In the OPE limit $1 > |q| > |z| \to 0$, the leading terms in $z$ read, up to the factor $\mathcal{A}$,
\begin{equation}
\sum\limits_{\mathfrak{t}} |z|^{2\Delta ({\alpha _0} - b^{-1}\mathfrak{t}) - 2\Delta ({\alpha _0}) - 2\Delta ( - {n^{\text{R}}}b^{-1}h_1)}\hat C_{{\alpha _0}, - {n^{\text{R}}}b^{-1}h_1}^{{\alpha _0} - b^{-1}\mathfrak{t}}\left\langle {{{\hat V}_{{\alpha _0} - b^{-1}\mathfrak{t}}}(0){{\hat V}_\beta }(x){{\hat V}_{{\alpha _1}}}(1){{\hat V}_{{\alpha _\infty }}}(\infty )} \right\rangle  \;.
  \label{AGT-with-defect-small-z}
\end{equation}
Here $\mathfrak{t} = \sum_{A = 1}^N n^\text{R}_A h_A$ and $h_A$ are the weights of the fundamental representation of $SU(N)$. The set of natural numbers $\vnr$ is any partition of $\nr$, and the sum over $\mathfrak{t}$ means summing over all such partitions. On the gauge theory side, in the $z \to 0$ limit, one can close the contour of the integration over $\sigma^\text{R}$ in the partition function $Z^{(\mathcal{T}, S^2_\text{R} \subset S^4_b)}$, as in \eqref{def:SQCDA-defect-matrix-model} with $\nl = 0$, and obtain the Higgs branch localized expression as a sum over the two-dimensional Higgs vacua labeled by partitions $\vec n^\text{R}$. This sum is mapped to the sum over $\mathfrak{t}$ in \eqref{AGT-with-defect-small-z}. The leading terms in $z$ simply come from the zero-vortex sector. The four-dimensional matrix integral in each leading term, which now depends on the Higgs vacuum $\vnr$, is simply an $S^4_b$-partition function with shifted fundamental masses $M_A^{{n^{\text{R}}}} \equiv M'_A - \hat M + i(b + {b^{ - 1}})/2 + in_A^{\text{R}}b^{-1}$, thanks to
\begin{equation}
  z^{\mathbb{R}^4}_\text{fund} (\vec Y',{{M'}^\varepsilon }) \ z_\text{defect}^{\mathbb{R}_{\text{R}}^2 \subset \mathbb{R}^4}(\vec Y',\Sigma ',i{\sigma ^{\text{R}}}, {B^{\text{R}}}) \to {z^{\mathbb{R}^4}_{{\text{fund}}}}(\vec Y',{({M^{{n^{\text{R}}}}})^\varepsilon })\;,
\end{equation}
where $\to$ indicates the evaluation at the Higgs vacuum $\vnr$. This $S^4_b$-partition function is mapped to the four-point function in \eqref{AGT-with-defect-small-z}. In particular, the four-dimensional one-loop determinant together with the two-dimensional one-loop determinant evaluated at the Higgs vacuum $\vnr$ is precisely equal to the structure constants $\hat C_{{\alpha _0}, - {n^{\text{R}}}b^{-1}h_1}^{{\alpha _0} - b^{-1}\mathfrak{t}} \hat C_{\alpha_0 - b^{-1} \mathfrak{t} , \beta}^{\alpha} \hat C_{\alpha, \alpha_1, \alpha_\infty}$, up to some uninteresting constants.

\begin{figure}[t]
\centering
  \includegraphics[width=0.8\textwidth]{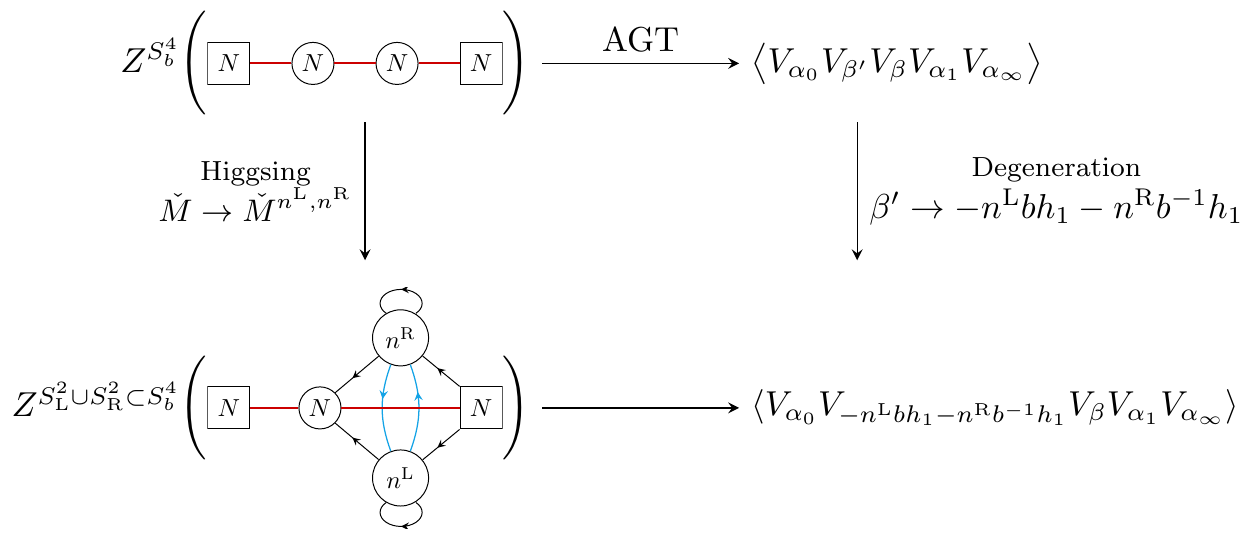}
  \caption{\label{degeneration-of-AGT} A commuting diagram showing the relation between the Higgsing prescription and degenerating semi-degenerate momentum.}
\end{figure}

In fact, the statement \eqref{AGT-correspondence-defect} in the case of $\mathcal{R}^\text{L/R}$ being both symmetric, can be viewed as a degeneration of the well-established AGT conjecture without surface defects, by considering the commuting diagram in figure \ref{degeneration-of-AGT} \cite{Bonelli:2011wx,Bonelli:2011fq,Nieri:2013vba}.
We remind ourselves that $i\check M^{\nl, \nr} = (b{n^{\text{L}}} + {b^{ - 1}}{n^{\text{R}}})/N + (b + {b^{ - 1}})/2$, and in the AGT correspondence $\beta'  = \left[ {N(b + {b^{ - 1}})/2 - \sum_{A = 1}^N {i{M_A}} } \right]{h_1}$. The Higgsing prescription sends the $U(1)$ mass $\check M \to \check M^{\nl, \nr}$, which is equivalent to degenerating the semi-degenerate momentum $\beta' \to - \nl b h_1 - \nr b^{-1} h_1$.

The correspondence \eqref{AGT-correspondence-defect} is a great tool to discover and understand new dualities of the 4d/2d/0d coupled system. The Liouville/Toda correlation functions enjoy various symmetries, including, but not limited to, the invariance under conjugation and Weyl reflection of the momenta, and conformal and crossing symmetries. It would be interesting to translate these CFT symmetries to dualities on the gauge theory side, especially when the intersecting defects are coupled to four-dimensional interacting theories.

\section*{Acknowledgments}
The authors would like to thank Giulio Bonelli, Jaume Gomis, Bruno Le Floch, Fabrizio Nieri, Daniel Park, Massimiliano Ronzani, Alessandro Tanzini and Maxim Zabzine for helpful conversations and useful suggestions. We are also grateful to Bruno Le Floch for comments on a draft of this paper. Y.P. is supported in part by Vetenskapsr{\aa}det under grant {\#}2014- 5517, by the STINT grant and by the grant ``Geometry and Physics'' from the Knut and Alice Wallenberg foundation. The work of W.P. is supported in part by the DOE grant DOE-SC0010008. 

\appendix

\section{Special functions}\label{app:Special functions}
In this appendix we briefly recall the definitions and some useful properties of the special functions which play an essential role in this paper.

\subsection{Factorials}\label{subapp: generalized factorials}
In analyzing vortex and instanton partition functions, one often encounters products of the form $\prod_{k = 0}^{m - 1} f_{\epsilon_1, \epsilon_2}(z + k),$ for some function $f_{\epsilon_1, \epsilon_2}$. Due to its frequent occurrence and to streamline the discussion we introduce the factorial 
\begin{equation}\label{generalizedfactorial}
  (z)_m^f \equiv \prod_{k = 0}^{m - 1} f_{\epsilon_1, \epsilon_2}(z + k)\;.
\end{equation}
As a trivial example, it includes the standard Pochhammer symbol $(x)_m \equiv \prod_{k=0}^{m-1}(x+k)$, for $f_{\epsilon_1, \epsilon_2}=\operatorname{id}$. We often abbreviate $f_{\epsilon_1, \epsilon_2}$ as $f$ when no confusion is expected.

\subsection{Double- and triple-sine functions}
\paragraph{Double-sine function $s_b(z)$:} the double-sine function $s_b(z)$ is defined as the regularized version of the product
\begin{equation}
s_b(z ) \equiv \prod_{m,n \geqslant 0} {\frac{{(m + 1/2)b + (n + 1/2){b^{ - 1}} - iz}}{{(m + 1/2)b + (n + 1/2){b^{ - 1}} + iz}}} \;.
\end{equation}
It is a common practice to define $Q \equiv b + b^{-1}$. The function $s_b(z + iQ/2)$ has poles at $z = + im_{\ge 0}b + in_{\ge 0}{b^{ - 1}}$ and zeros at $z = - im_{\ge 1}b - in_{\ge 1}{b^{ - 1}}$. The double sine function satisfies the following recursion relation
\begin{equation}
  \begin{split}
    {s_b}\left( {z + \frac{{iQ}}{2} + imb + in{b^{ - 1}}} \right) = \frac{{{{\left( { - 1} \right)}^{mn}}{s_b}\left( {z + \frac{{iQ}}{2}} \right)}}{{\prod_{{k} = 1}^m {2i\sinh \pi b(z + i{k}b)} \prod_{{k} = 1}^n {2i\sinh \pi {b^{ - 1}}(z + i{k}b^{-1})} }}\;,
  \end{split}
  \label{recursion-double-sine}
\end{equation}
as well as the symmetry properties
\begin{equation}
  {s_b}\left( z \right){s_b}\left( { - z} \right) = 1\;,\qquad {s_b}\left( {z + \frac{{iQ}}{2}} \right){s_b}\left( { - z + \frac{{iQ}}{2}} \right) = \frac{1}{{2\sinh (\pi bz)\ 2\sinh (\pi {b^{ - 1}}z)}}\;.
\end{equation}

An alternative definition of the double-sine function is given by
\begin{equation}
S_2( z|\omega _1,\omega _2 ) \equiv \prod_{m,n \geq 0} \frac{m\omega _1 + n\omega _2 + z}{(m + 1)\omega _1 + (n + 1)\omega _2 - z} \;.
\end{equation}
The definitions of $s_b(z)$ and $S_2(z|\omega_1, \omega_2)$ are related by $b = \sqrt{\omega_1/\omega_2}$ and
\begin{equation}
S_2\left( z|\omega _1,\omega _2 \right) = s_b\left(  - i\frac{Q}{2} + i\frac{z}{\sqrt {\omega _1\omega _2} } \right),\qquad {S_2}\left( {\frac{{{\omega _1} + {\omega _2}}}{2} - iz\Big|{\omega _1},{\omega _2}} \right) = {s_b}\left( \frac{z}{\sqrt {{\omega _1}{\omega _2}}}  \right)\;.
\end{equation}

\paragraph{Triple-sine function $S_3(z|\vec \omega)$:} the triple-sine function $S_3(z|\vec \omega)$ for any triplet $\vec \omega = (\omega_1, \omega_2, \omega_3)$ is defined as the regularization of the product
\begin{equation}
S_3\left( {z|\vec \omega } \right) = \prod_{{n_1, n_2, n_3} = 0}^{ + \infty } {\left( {z + \vec n \cdot \vec \omega } \right)\left( {(\vec n + 1) \cdot \vec \omega  - z} \right)} \;.
\end{equation}
It has no poles, and its zeros are located at $z = (\vec n + 1) \cdot \vec \omega$ or $z =  - \vec n \cdot \vec \omega$ for $n_\alpha \ge 0 $. It satisfies a convenient recursion relation (where ${S_1}(z|\omega ) \equiv 2\sinh (\pi z/\omega )$)
\begin{small}
\begin{align}
&S_3\left( {z + \sum_{\alpha=1}^3{n_\alpha}{\omega _\alpha}|\omega_1,\omega_2,\omega_3 } \right) \nn\\
    = &\; {\left( { - 1} \right)^{{n_1}{n_2}{n_3}}}{S_3}\left( {z|\omega_1,\omega_2,\omega_3 } \right)\ \prod_{{k_1} = 0}^{{n_1} - 1} {\frac{1}{{{S_2}\left( {z + {k_1}{\omega _1}|{\omega _2},{\omega _3}} \right)}}} \prod_{{k_2} = 0}^{{n_2} - 1} {\frac{1}{{{S_2}\left( {z + {k_2}{\omega _2}|{\omega _1},{\omega _3}} \right)}}} \prod_{{k_3} = 0}^{{n_3} - 1} {\frac{1}{{{S_2}\left( {z + {k_3}{\omega _3}|{\omega _1},{\omega _2}} \right)}}}  \nn\\
    & \times\prod_{{k_2} = 0}^{{n_2} - 1} {\prod_{{k_3} = 0}^{{n_3} - 1} {{S_1}\left( {z + {k_2}{\omega _2} + {k_3}{\omega _3}|{\omega _1}} \right)} } \prod_{{k_1} = 0}^{{n_1} - 1} {\prod_{{k _2} = 0}^{{n_2} - 1} {{S_1}\left( {z + {k_1}{\omega _1} + {k_2}{\omega _2}|{\omega _3}} \right)} } \prod_{{k_1} = 0}^{{n_1} - 1} {\prod_{{k_3} = 0}^{{n_3} - 1} {{S_1}\left( {z + {k_1}{\omega _1} + {k_3}{\omega _3}|{\omega _2}} \right)} } \;,
    \label{recursion-triple-sine}
\end{align}
\end{small}%
and has the following symmetry property
\begin{equation}
S_3\left( z|\vec \omega  \right) = S_3\left( \omega_1+\omega_2+\omega_3 - z|\vec \omega  \right)\;.
\end{equation}

\subsection{\texorpdfstring{$\Upsilon _b$}{Yb} functions}
The function $\Upsilon_b(z) $ is defined as
\begin{equation}
\Upsilon_b(z) = \prod_{m,n\geq 0} (mb+nb^{-1} + z) ((m+1)b + (n+1)b^{-1} - z) \;.
\end{equation}
It has no poles, but zeros located at $z=-mb-nb^{-1}$ and $z=(m+1)b+(n+1)b^{-1},$ for $m,n\geq 0$, and satisfies the recursion relation for $m,n \in \mathbb{Z}$
\begin{multline}
\Upsilon _b(z - mb - n{b^{ - 1}}) = \; {( - 1)^{\left| {mn} \right|}}{\Upsilon _b}(z)\ \frac{{\prod_{r = 0}^{ - m - 1} {\prod_{s = 0}^{ - n - 1} {{{(z + rb + s{b^{ - 1}})}^2}} \prod_{r = 1}^m {\prod_{s = 1}^n {{{(z - rb - s{b^{ - 1}})}^2}} } } }}{{\prod_{r = 0}^{ - m - 1} {\prod_{s = 1}^n {{{(z + rb - s{b^{ - 1}})}^2}} } \prod_{r = 1}^m {\prod_{s = 0}^{ - n - 1} {{{(z - rb + s{b^{ - 1}})}^2}} } }}  \\
\times \frac{{\prod_{r = 0}^{ - m - 1} {\gamma (b(z + rb))} \prod_{s = 0}^{ - n - 1} {\gamma ({b^{ - 1}}(z + s{b^{ - 1}}))} }}{{\prod_{r = 1}^m {\gamma (b(z - rb))} \prod_{s = 1}^n {\gamma ({b^{ - 1}}(z - s{b^{ - 1}}))} }}\ 
 \frac{{\prod_{r = 1}^m {{b^{ - 1 + 2(z - rb)b}}} \prod_{s = 1}^n {{b^{1 - 2(z - s{b^{ - 1}}){b^{ - 1}}}}} }}{{\prod_{r = 0}^{ - m - 1} {{b^{ - 1 + 2(z + rb)b}}} \prod_{s = 0}^{ - n - 1} {{b^{1 - 2(z + s{b^{ - 1}}){b^{ - 1}}}}} }}  \;.
  \label{recursion-upsilon}
\end{multline}
Here we defined $Q\equiv b + b^{-1}$, and for each product, when the lower limit is strictly larger than the upper limit, the product reduces to 1. Some other useful properties include ${\Upsilon _b}(Q - z) = {\Upsilon _b}(z)$ and ${\Upsilon _b}(Q/2) = 1$.

\section{The \texorpdfstring{$S^2$}{S2} and \texorpdfstring{$S^3_b$}{S3b} SQCDA partition function}\label{HBLS2S3}
In this appendix we briefly recall the two-dimensional $\mathcal N=(2,2)$ supersymmetric and three-dimensional  $\mathcal N=2$ supersymmetric sphere partition function of a $U(n_\text{c})$ gauge theory with $n_\text{f}$ fundamental, $n_{\text{af}}$ antifundamental, and one adjoint chiral multiplet, which we henceforth call SQCDA, and present its Higgs branch localized form.

Here and in the next appendices, we will use the notation $(i, \mu)$ in substitution of the original color index $a \in \{1, ..., n_\text{c}\}$, where $i = 1, ..., n_\text{f}$, $\mu = 0, ..., k_i - 1$, with $\sum _i k_i = n_\text{c}$. Each partition $\vec k = \{k_i\}$ corresponds to a Higgs vacuum of the theory. Therefore, for arbitrary functions $\Phi$ of arbitrary sequences of $n_\text{c}$ variables $x_a$, we have 
\begin{equation}
  \prod_{a = 1}^{{n_\text{c}}} {\Phi ({x _a})}  \to \prod_{i = 1}^{{n_\text{f}}} {\prod_{\mu  = 0}^{{k_i} - 1} {\Phi ({x _{i\mu }})} } \equiv \prod_{(i,\mu)}{\Phi ({x _{i\mu }})}  \;, \qquad \sum_{a = 1}^{{n_\text{c}}} {\Phi ({x _a})}  \to \sum_{i = 1}^{{n_\text{f}}} {\sum_{\mu  = 0}^{{k_i} - 1} {\Phi ({x _{i\mu }})} } \equiv \sum_{(i,\mu)}{\Phi ({x _{i\mu }})}\;.
\end{equation}

\subsection{The \texorpdfstring{$S^2$}{S2} SQCDA partition function}\label{subapp: S2 partition function}
The two-sphere partition function of an $\mathcal N=(2,2)$ supersymmetric gauge theory with gauge group $U(n_\text{c})$ and with $n_\text{f}$ fundamental chiral multiplets with masses $m_j$, $n_\text{af}$ antifundamental chiral multiplets with masses $\tilde m_t$ and an adjoint chiral multiplet with mass $m_X$ is computed by \cite{Benini:2012ui,Doroud:2012xw}
\begin{equation}
  \begin{aligned}
    Z^{{S^2}}_{\text{SQCDA}} = \;\frac{1}{{{n_\text{c}}!}}\sum_{B \in {\mathbb{Z}^{{n_\text{c}}}}}   \int & \prod_{a = 1}^{{n_\text{c}}} \left[{\frac{d{\sigma _a}}{2\pi}} {e^{ - 4\pi i \xi_\text{FI} \sigma_a  - i (\vartheta - n_\text{c} - 1) B_a}}\right]\prod_{a > b} {\left[ {{{({\sigma _a} - {\sigma _b})}^2} + \frac{{{{({B_a} - {B_b})}^2}}}{4}} \right]}  \\
    & \times \prod_{j = 1}^{{n_\text{f}}} {\prod_{a = 1}^{{n_\text{c}}} {\frac{{\Gamma ( - i{\sigma _a} - \frac{{{B_a}}}{2} + i{m_j})}}{{\Gamma (1 + i{\sigma _a} - \frac{{{B_a}}}{2} - i{m_j})}}} } \prod_{t = 1}^{{n_\text{af}}} {\prod_{a = 1}^{{n_\text{c}}} {\frac{{\Gamma ( + i{\sigma _a} + \frac{{{B_a}}}{2} - i{{\tilde m}_t})}}{{\Gamma (1 - i{\sigma _a} + \frac{{{B_a}}}{2} + i{{\tilde m}_t})}}} }  \\
    & \times \prod_{a,b = 1}^{{n_\text{c}}} {\frac{{\Gamma ( - i{\sigma _a} + i{\sigma _b} - \frac{{{B_a} - {B_b}}}{2} + i{m_X})}}{{\Gamma (1 + i{\sigma _a} - i{\sigma _b} - \frac{{{B_a} - {B_b}}}{2} - i{m_X})}}} \;.
  \end{aligned}
  \label{S2CBlocalization}
\end{equation}
Here we have set the $r$-charges to zero. They can be reinstated by analytically continuing the masses. In comparison to higher-dimensional sphere partition functions, the two-sphere partition function also involves a sum over magnetic fluxes $B.$ We have also included the two-dimensional $\vartheta$-angle in the classical action. The one-loop determinants for the chiral multiplets are expressed in terms of the standard Gamma-function. Finally, note that we have turned off background fluxes for the flavor symmetries, as they will play no role in this paper.

Assuming that $n_\text{f}>n_\text{af}$ or $n_\text{f}=n_\text{af}$ with $\xi _\text{FI} >0,$ we can close the integration contours of \eqref{S2CBlocalization} in the lower half-plane. The integrand of the $\sigma$-integrations in \eqref{S2CBlocalization} is an infinite sum over $B \in \mathbb{Z}^{n_\text{c}}$. One can show that the sum of the residues of different permutations of the color label give the same result, canceling the $n_\text{c}!$ in \eqref{S2CBlocalization}. For each summand with a given $B_{a}$, the relevant poles are then given by
\begin{equation}
i{\sigma _{j\mu }} = i{m_j} + \mu {m_X} + {N_{ j\mu}} + \frac{{|{B_{j\mu }}|}}{2}\;, \qquad \text{for all}\quad   0 \le {N_{j0}} \le {N_{j1}} \le ... \le {N_{j({k_j} - 1)}} \in \mathbb{N}\;,
  \label{old-pole-equations-original-2d}
\end{equation}
where $\mu = 0, ..., k_j - 1$, and $\vec k$ is a partition of $n_\text{c}$: $\sum_{j = 1}^{n_\text{f}} k_j = n_\text{c}$, labeling the Higgs vacua. These poles can be rewritten into a more useful form, by introducing ${\mathfrak{m}_{j\mu }} = {N_{j\mu }} + \frac{{{B_{j\mu }}}}{2} + \frac{{|{B_{j\mu }}|}}{2}$, ${\mathfrak{n}_{j\mu }} = {N_{j\mu }} + \frac{{|{B_{j\mu }}|}}{2} - \frac{{{B_{j\mu }}}}{2}$. We collectively denote these poles as $\sigma_{\mathfrak{m}, \mathfrak{n}}$,
\begin{equation}
  \text{poles of type } \sigma_{\mathfrak{m}, \mathfrak{n}}:\qquad  i{\sigma _{j\mu }} + \frac{{{B_{j\mu }}}}{2} = i{m_j} + \mu i{m_X} + {\mathfrak{m}_{j\mu }}\;, \qquad i{\sigma _{j\mu }} - \frac{{{B_{j\mu }}}}{2} = i{m_j} + \mu i{m_X} + {\mathfrak{n}_{j\mu }}\;.
  \label{old-pole-equations-2d}
\end{equation}
Note that $\mathfrak{m}, \mathfrak{n}$ are sequences of non-decreasing natural numbers, such that $0 \le {\mathfrak{m}_{j0}} \le {\mathfrak{m}_{j1}} \le ... \le {\mathfrak{m}_{j({k_j} - 1)}}$, $0 \le {\mathfrak{n}_{j0}} \le {\mathfrak{n}_{j1}} \le ... \le {\mathfrak{n}_{j({k_j} - 1)}}$, $\forall j = 1, ..., n_\text{f}$. Alternatively, one can solve $N$ and $B$ in terms of $\mathfrak{m}, \mathfrak{n}$ by definition.

The sum over residues can be brought into the form \cite{Gomis:2014eya}
\begin{equation}
  Z^{S^2}_\text{SQCDA} = \sum\limits_k  Z^{S^2}_{\text{cl}|\vec k}\;Z_{\text{1-loop}|\vec k}^{{S^2}}\;\sum\limits_\mathfrak{m} {{{\hat z}^{\left| \mathfrak{m} \right|}}} Z_{\text{vortex}|\vec k}^{{\mathbb{R}^2}}(\mathfrak{m})\;\sum\limits_\mathfrak{n} {{{\bar {\hat z}}^{\left| \mathfrak{n} \right|}}} Z_{{\text{vortex}|\vec k}}^{{\mathbb{R}^2}}(\mathfrak{n})\;,
  \label{SQCDA-partition-function-2d}
\end{equation}
where $\hat z$ is defined as $\hat z=e^{ - 2\pi \xi _{\text{FI}} + i\vartheta^\prime },$ with $\vartheta^\prime = \vartheta + \pi(n_\text{f}-n_\text{af}).$ Furthermore, we introduced $|\mathfrak{m}| \equiv \sum_{j\mu} \mathfrak{m}_{j\mu},$ and similarly for $|\mathfrak n|.$ The classical and one-loop factors read, with $m_{jl} \equiv m_j - m_l$ and $\tilde m_{jt} \equiv m_j - \tilde m_t$,
\begin{align}
  Z_{\text{cl}|\vec k} = &\; \exp \left[ {- 4\pi i{\xi _{{\text{FI}}}}\sum_{j = 1}^{{n_\text{f}}} {\left( {{k_j}{m_j} + \frac{{{m_X}}}{2}({k_j} - 1){k_j}} \right)} } \right]\\
  Z_{\text{1-loop}|\vec k}^{{S^2}} = & \;\frac{{\prod_{j = 1}^{{n_\text{f}}} {\prod_{l = 1}^{{n_\text{f}}} {\prod_{\mu  = 0}^{{k_j} - 1} {\gamma (- i{m_{jl}} + {n_l}i{m_X} - \mu i{m_X})} } } }}{{\prod_{j = 1}^{{n_\text{f}}} {\prod_{t = 1}^{{n_\text{af}}} {\prod_{\mu  = 0}^{{k_j} - 1} {\gamma (1 - i{\tilde m_{jt}} - \mu i{m_X})} } } }} \label{def:fund-one-loop-HBL-S2}\;,
\end{align}
where $\gamma(x)=\frac{\Gamma(x)}{\Gamma(1-x)}.$ The summand $Z_{\text{vortex}|\vec k}^{{\mathbb{R}^2}}(\mathfrak m)$ of the vortex partition function is given by
\begin{align}
  \begin{split}
      {Z^{\mathbb{R}^2}_{{\text{vortex}}| \vec k}}(\mathfrak{m}) = &\; \prod_{t = 1}^{{n_\text{af}}} {\prod_{j = 1}^{{n_\text{f}}} {\prod_{\mu  = 0}^{{k_j} - 1} {{{(1 - i({m_j - \tilde m_t}) - \mu i{m_X} - {\mathfrak{m}_{j\mu }})}_{{\mathfrak{m}_{j\mu }}}}} } }    \\
      & \times \prod_{j,l = 1}^{{n_\text{f}}} \Bigg[ \prod_{\mu  = 0}^{{k_j} - 1} {\prod_{\nu  = 0}^{{k_l} - 1} {\frac{1}{{{{(1 - i{m_{jl}} - (\mu  - \nu )i{m_X} + {\mathfrak{m}_{l\nu }} - {\mathfrak{m}_{j\mu }})}_{{\mathfrak{m}_{j\mu }} - {\mathfrak{m}_{j\mu  - 1}}}}}}} }   \\
      & \qquad \qquad \times\prod_{\mu  = 0}^{{k_j} - 1} {\frac{{{{(1 + i{m_{jl}} - ({k_l} - \mu )i{m_X} + {\mathfrak{m}_{j\mu }} - {\mathfrak{m}_{l({k_l} - 1)}})}_{{\mathfrak{m}_{l({k_l} - 1)}}}}}}{{{{(1 + i{m_{jl}} - ({k_l} - \mu )i{m_X})}_{{\mathfrak{m}_{j\mu }}}}}}} \Bigg] \;,
  \end{split}
  \label{vortex-partition-function-2d}
\end{align}
and $\sum\limits_\mathfrak{m}$ denotes a sum over all possible non-decreasing sequences of natural numbers $\mathfrak{m}_j$.

\subsection{The \texorpdfstring{$S^3_b$}{S3b} SQCDA partition function}\label{subapp: S3 partition function}
The squashed three-sphere is defined by its embedding in $\mathbb C^2$ as
\begin{equation}
\omega_1^2 |z_1|^2 + \omega_2^2 |z_2|^2 = 1\;,
\end{equation}
and the parameter $b$ is given by $b = \sqrt{\omega_2/{\omega_1}}.$ The partition function of an $\mathcal N=2$ supersymmetric gauge theory with gauge group $U(n_\text{c})$ and with $n_\text{f}$ fundamental chiral multiplets with masses $m_j$, $n_\text{af}$ antifundamental chiral multiplets with masses $\tilde m_t$ and an adjoint chiral multiplet with mass $m_X$ is computed by \cite{Kapustin:2009kz,Jafferis:2010un,Hama:2010av,Hama:2011ea}
\begin{multline}
Z^{S_b^3} = \frac{1}{{n_\text{c}!}}\int {\left( {\prod_{a = 1}^{n_\text{c}} d {\sigma _a}} \right)} \;{e^{ - 2\pi i\xi_\text{FI}\Tr\sigma }}\prod_{a > b} 2 \sinh (\pi b({\sigma _a} - {\sigma _b}))\;2\sinh (\pi {b^{ - 1}}({\sigma _a} - {\sigma _b})) \\ \times \frac{{\prod_{t = 1}^{{n_\text{af}}} {\prod_{a = 1}^{{n_\text{c}}} {{s_b}\left( {\frac{{iQ}}{2} + {\sigma _a} - {{\tilde m}_t}} \right)} } }}{{\prod_{j = 1}^{{n_\text{f}}} {\prod_{a = 1}^{{n_\text{c}}} {{s_b}\left( { - \frac{{iQ}}{2} + {\sigma _a} - {m_j}} \right)} } }}\prod_{a,b = 1}^{{n_\text{c}}} {{s_b}\left( {\frac{{iQ}}{2} - {\sigma _a} + {\sigma _b} + {m_X}} \right)} \;,
\label{matrix-model-3d}
\end{multline}
where $Q=b+b^{-1}$ and the matter one-loop determinants are expressed in terms of the double-sine function $s_b$. We have taken the Chern-Simons level to be zero. 

When $n_\text{f} > n_\text{af}$ or $n_\text{f} = n_\text{af}$ and the FI-parameter $\xi_\text{FI} > 0$,\footnote{Note that if we had turned on a Chern-Simons level, the convergence criterion would have been slightly more subtle than in two dimensions, as was explained in \cite{Benini:2013yva}.} we again consider poles of type $\sigma_{\mathfrak{m}, \mathfrak{n}}$ in the lower half-plane, labeled by ascending sequences of natural numbers $0 \le {\mathfrak{m}_{j0}} \le {\mathfrak{m}_{j1}} \le ... \le {\mathfrak{m}_{j({k_j} - 1)}}$, $\forall j = 1, ..., n_\text{f}$,
\begin{equation}
\text{poles of type } \sigma_{\mathfrak{m}, \mathfrak{n}}:\quad {\sigma _{i\mu }} = {m_j} + \mu {m_X} - i{\mathfrak{m}_{j\mu }}b - i{\mathfrak{n}_{j\mu }}{b^{ - 1}}\;, \;\;\mu = 0, ..., k_j - 1,\;\; \mathfrak{m}_{j\mu}, \mathfrak{n}_{j \mu} \in \mathbb{N}\;.
  \label{old-pole-equations-3d}
\end{equation}
Summing over the residues, one obtains the Higgs branch localized $S^3_{b}$-partition function \cite{Fujitsuka:2013fga,Benini:2013yva}, with Higgs vacua specified by a partition $\vec k$,
\begin{equation}
Z^{S_b^3}_{{\text{SQCDA}}} = \sum_{k} Z^{S_b^3}_{\text{cl}|\vec k}\ Z^{S_b^3}_{\text{1-loop}|\vec k}\ \left[ {\sum_\mathfrak{m} {z_b^{\left| \mathfrak{m} \right|}{Z^{\mathbb{R}^2\times S^1}_{{\text{vortex}|\vec k}}}\left( {\mathfrak{m}|b} \right)} } \right]\ \left[ {\sum_\mathfrak{n} {z_{{b^{ - 1}}}^{\left| \mathfrak{n} \right|}{Z^{\mathbb{R}^2\times S^1}_{{\text{vortex}| \vec k}}}\left( {\mathfrak{n}|{b^{ - 1}}} \right)} } \right]  \;,
  \label{SQCDA-partition-function-3d}
\end{equation}
where the classical and 1-loop factors, with $m_{ij} \equiv m_i - m_j$, are given by
\begin{align}
  {Z^{S_b^3}_{\text{cl}| \vec k}} \equiv &\; \exp \left[ { 2\pi i{\xi _{{\text{FI}}}}\left( {\sum\nolimits_{j = 1}^{{n_\text{f}}} {{m_j}{k_j}}  + \frac{{{m_X}}}{2}\sum\nolimits_{j = 1}^{{n_\text{f}}} {({k_j} - 1){k_j}} } \right)} \right]\\
  Z_{\text{1-loop}|\vec k}^{S_b^3} \equiv &\; \prod_{j = 1}^{{n_{\text{f}}}} {\prod_{(l,\mu )} {{s_b}\left( {\frac{{iQ}}{2} - {m_{lj}} + ({k_j} - \mu ){m_X}} \right)} } \prod_{t = 1}^{{n_{{\text{af}}}}} {\prod_{(j,\mu )} {{s_b}(\frac{{iQ}}{2} + {m_j} - {{\tilde m}_t} + \mu {m_X})} } \;.  \label{def: one-loop-HBL-S3}
\end{align}
The summand $Z_{\text{vortex}| \vec k}^{{\mathbb{R}^2} \times {S^1}}\left( {\mathfrak{m}|b} \right)$ of the vortex partition function is given by
\begin{align}
   \begin{split}
     {{Z^{\mathbb{R}^2\times S^1}_{{\text{vortex}}| \vec k}}\left( {\mathfrak{m}|b} \right)} \equiv &\; \prod_{t = 1}^{{n_{{\text{af}}}}} {\prod_{j = 1}^{{n_{\text{f}}}} {\prod_{\mu  = 0}^{{k_j} - 1} {( 1 - i{b^{ - 1}}({m_j} - {{\tilde m}_t}) - i\mu {b^{ - 1}}{m_X} - {\mathfrak{m}_{j\mu }} ) _{{\mathfrak{m}_{j\mu }}}^f} } }  \\
     & \times \prod_{j,l = 1}^{{n_\text{f}}} {\Bigg[\; {\prod_{\mu  = 0}^{{k_j} - 1} {\prod_{\nu  = 0}^{{k_l} - 1} {\frac{1}{{( 1 - i{b^{ - 1}}{m_{jl}} - i(\mu  - \nu ){b^{ - 1}}{m_X} + {\mathfrak{m}_{l\nu }} - {\mathfrak{m}_{j\mu }} ) _{{\mathfrak{m}_{j\mu }} - {\mathfrak m_{j,\mu  - 1}}}^f}}} } }}   \\
     &\qquad \qquad  { \times \prod_{\mu  = 0}^{{k_j} - 1} {\frac{{( 1 + i{b^{ - 1}}{m_{jl}} - i({k_l} - \mu ){b^{ - 1}}{m_X} + {\mathfrak{m}_{j\mu }} - {\mathfrak{m}_{l,{k_l} - 1}} ) _{{\mathfrak{m}_{l,{k_l} - 1}}}^f}}{{( 1 + i{b^{ - 1}}{m_{jl}} - i({k_l} - \mu ){b^{ - 1}}{m_X} ) _{{\mathfrak{m}_{j\mu }}}^f}}} } \Bigg]\;.
     \label{vortex-partition-function-3d}
   \end{split}
\end{align}
Here we used the function $(x)^f_m \equiv \prod_{k=0}^{m-1}f(x+k)$ with $f(x) = 2i \sinh \pi i b^2x$. See appendix \ref{app:Special functions}. Expression \eqref{vortex-partition-function-3d} is summed over all possible sequences of non-decreasing natural numbers $0 \le {\mathfrak{m}_{j0}} \le {\mathfrak{m}_{j1}} \le ... \le {\mathfrak{m}_{j({k_j} - 1)}}$, with weighting factor given in terms of
\begin{equation}
  z_{b^{\pm 1}} \equiv e^{- 2\pi {\xi _{{\text{FI}}}}b^{\pm 1}}\qquad |\mathfrak{m}| \equiv \sum\nolimits^{n_\text{f}}_{j = 1} \sum\nolimits_{\mu = 0}^{k_j - 1} \mathfrak{m}_{j \mu}\;.
\end{equation}

\subsection{Forest-tree representation}\label{foresttree}
The poles \eqref{old-pole-equations-2d} and \eqref{old-pole-equations-3d} admit a useful graphical representation in terms of forests of trees. Such representation will turn out to be useful in later appendices, so we introduce it here already for the simple case of SQCDA \cite{Hwang:2015wna}. We will consider the example of $S^3_b$; the case of $S^2$ is completely similar.

When $n_\text{f} \ge n_\text{af}$ and the FI-parameter $\xi_\text{FI} > 0$, the Jeffrey-Kirwan residue prescription, mentioned below \eqref{MIJ_S5_relation}, selects as poles the solutions to the equations
\begin{equation}\label{JKpolesS3}
  \begin{aligned}
    {\sigma _a} = & \;{m_{{i_a}}} - i{\mathfrak{m}_a}b - i{\mathfrak{n}_a}{b^{ - 1}} &  & {\mathfrak{m}_a},{\mathfrak{n}_a} \geqslant 0 \\
    {\sigma _a} = & \;{\sigma _b} + {m_X} - i\Delta {\mathfrak{m}_{ab}}b - i\Delta {\mathfrak{n}_{ab}}{b^{ - 1}} & & \Delta {\mathfrak{m}_{ab}},\Delta {\mathfrak{n}_{ab}} \geqslant 0, \;\;a\ne b\;.
  \end{aligned}
\end{equation}
where for each label $a$ the component $\sigma_a$ appears exactly once on the left-hand side, and $i_a \in \{1, ..., n_\text{f}\}$.  Note that \eqref{JKpolesS3} contains more poles than those described by \eqref{old-pole-equations-3d}.

\begin{figure}[t]
\centering
\includegraphics[width=.8\textwidth]{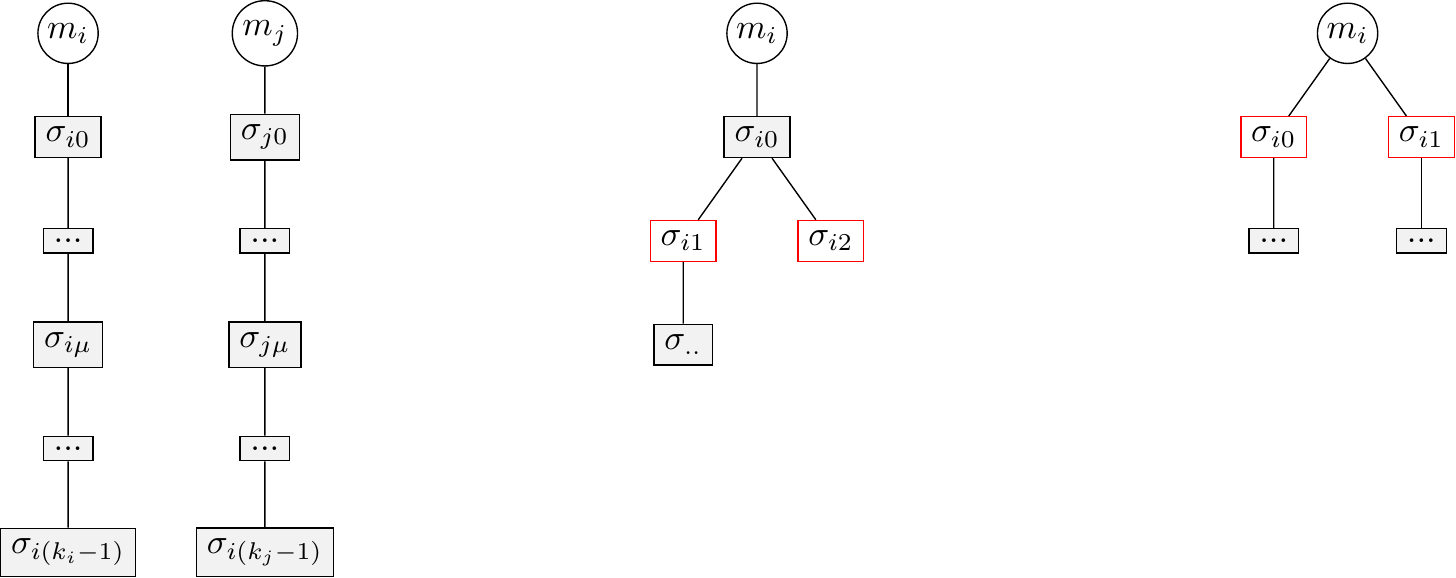}
\caption{Examples of forests of trees. The figure on the left shows two branch-less trees, associated with masses $m_i$ and $m_{j (\ne i)}$ respectively. Forests consisting of such branch-less trees will give non-zero contributions to the SQCDA partition function. The figure in the middle and on the right show trees with branches, or two trees associated to the same mass $m_i$; a forest that contains such trees does not contribute to the partition function by symmetry arguments.}
  \label{figure-forest-tree-intro}
\end{figure}

The poles constructed by solving $n_\text{c}$ of the equations in \eqref{JKpolesS3} for the $n_\text{c}$ components $\sigma_a$ can be represented by forests of trees by drawing nodes for all components $\sigma_a$ and all masses $m_i$ and connecting the nodes associated with the first symbol on the right-hand side of \eqref{JKpolesS3} (\ie{}, a component of $\sigma$ or a mass $m$) to that associated with the component on the left-hand side with a line, for all $n_\text{c}$ equations that were used. Note that trees consisting of a single mass node, can be omitted from the forest. As a result, each component $\sigma_a$ is linked to a fundamental mass $m_{i_a}$ (which occurs as the root node of the tree containing the node of $\sigma_a$), and the interrelations between components $\sigma_a$ form the structure of the forest of trees. Figure \ref{figure-forest-tree-intro} demonstrates a few examples. When no confusion is expected, we will sometimes omit the mass node at the root of the tree.

Using the symmetries of the one-loop determinants, one can show that, after summing over all possible poles, namely over all possible forest diagrams, only those forests whose trees are all \emph{branch-less} and where each fundamental mass is only linked to (at most) one branch-less tree, will contribute. The rest of the diagrams cancel among themselves.

In the residue computation, we encountered partitions $\vec k$ of the rank $n_\text{c}$ of the gauge group. Each entry $k_j$ is precisely the length of the length of the tree (or number of descendant nodes under mass $m_j$)

\section{Factorization of instanton partition function}\label{appendix:IPF-factorization}
In this appendix, we analyze the factorization of the summand of the instanton partition function, evaluated at special values of its gauge equivariant parameter, into the product of the summands of two (semi-)vortex partition functions. We can simultaneously consider the four-dimensional and five-dimensional instanton partition function by using the notation $(x)_m^f$ (see appendix \ref{app:Special functions}), where $f(x) = f_{\epsilon_1, \epsilon_2}(x)$ is some odd function that might depend on the $\Omega$-deformation parameters. Replacing $f$ by
\begin{equation}\label{fin4dand5d}
  \text{4d}: f(x)=\epsilon_2 x, \qquad \qquad  \text{5d}: f(x)=2 i \sinh(\pi i \epsilon_2 x)\;,
\end{equation}
the following results apply to the familiar instanton partition function respectively on $\mathbb{R}^4_{\epsilon_1, \epsilon_2}$ and $\mathbb{R}_{\epsilon_1, \epsilon_2}^4\times S^1_\beta$. In appendix \ref{appendix:extra-pole-3d}, \ref{appendix:extra-pole-2d}, we will discuss the relation between the factorization results in this appendix, and the poles and residues of the matrix models that describe gauge theories in the presence of intersecting defects.

\subsection{The instanton partition function}
We start with a four-/five-dimensional supersymmetric quiver gauge theory with gauge group $SU(N) \times SU(N)$\footnote{Instanton counting is typically performed for $U(N)$ gauge groups. We will not be careful about the distinction. In fact, removing the $U(1)$ factors is expected to just amount to some overall factor $(1-q)^{\#}$, as in \cite{Alday:2009aq}.}, with $N$ fundamental hypermultiplets, $N$ anti-fundamental hypermultiplets and one bi-fundamental hypermultiplet, with masses $M_I$, $\tilde M_J$ and $\hat M$ respectively. Let $\Sigma$ and $\Sigma'$ denote the Cartan-valued constant scalars of the two vector multiplets. The instanton partition function can be written as a sum over $N$-tuples of Young diagrams $\vec Y, \vec Y'$ and the individual contributions to each summand read
\begin{align}
&{z_{{\text{vect}}}}(\vec Y;\Sigma) \equiv \prod\limits_{A,B = 1}^{N_C} \prod\limits_{r,s = 1}^\infty \frac{{(  i\epsilon _2^{ - 1}\Sigma _{AB} - \mathfrak{b}^2(s - r + 1) - {Y_{Bs}})_{{Y_{Ar}}}^f}}{{(i\epsilon _2^{ - 1}\Sigma _{AB} - \mathfrak{b}^2(s - r + 1) - {Y_{Bs}})_{{Y_{Bs}}}^f}}\ \frac{{( i\epsilon _2^{ - 1}\Sigma _{AB} - \mathfrak{b}^2(s - r) - {Y_{Bs}})_{{Y_{Bs}}}^f}}{{( i\epsilon _2^{ - 1}\Sigma _{AB} - \mathfrak{b}^2(s - r) - {Y_{Bs}})_{{Y_{Ar}}}^f}}\;, \label{zvectIPF}\\
&{z_{{\text{(a)fund}}}}(\vec Y,\Sigma, \mu^\epsilon) \equiv \prod\limits_{A = 1}^N {\prod\limits_{I = 1}^N {\prod\limits_{r = 1}^\infty  {(i\epsilon _2^{ - 1}({\Sigma _A} - {\mu^\epsilon_I}) + \mathfrak{b}^2r + 1)_{{Y_{Ar}}}^f} } }  \;,\label{z(a)fundIPF}\\
&{z_{{\text{bifund}}}}(\vec Y,\vec Y',\Sigma ,\Sigma ',\hat M^\epsilon) \equiv \prod\limits_{A,B = 1}^N \prod\limits_{r,s = 1}^\infty  {\left[ {\frac{{(-i\epsilon _2^{ - 1}({\Sigma '_B} - {\Sigma _A} + \hat M^\epsilon) - {\mathfrak{b}^2}(s - r + 1) - {Y'_{Bs}})_{{Y'_{Bs}}}^f}}{{(-i\epsilon _2^{ - 1}({\Sigma '_B} - {\Sigma _A} + \hat M^\epsilon) - {\mathfrak{b}^2}(s - r + 1) - {Y'_{Bs}})_{{Y_{Ar}}}^f}}} \right.}  \nn \\
  &\qquad\qquad\qquad\qquad\qquad\qquad\qquad\qquad\ \times \left. {\frac{{(-i\epsilon _2^{ - 1}({\Sigma '_B} - {\Sigma _A} + \hat M^\epsilon) - {\mathfrak{b}^2}(s - r) - {Y'_{Bs}})_{{Y_{Ar}}}^f}}{{(-i\epsilon _2^{ - 1}({\Sigma '_B} - {\Sigma _A} + \hat M^\epsilon) - {\mathfrak{b}^2}(s - r) - {Y'_{Bs}})_{{Y'_{Bs}}}^f}}} \right]\;.\label{zbifundIPF}
\end{align}
Here $\Sigma _{AB} = \Sigma_A - \Sigma_B$ and $\mathfrak{b}^2 \equiv \epsilon_1 / \epsilon_2$. The full instanton partition function is thus\footnote{\label{simplifytoSQCD}On the one hand, the simpler case of $SU(N)$ SQCD, which we used in section \ref{section:free-hyper}, can be easily extracted from this expression, by setting all $Y'_A$ to empty Young diagrams and identifying the antifundamental mass as $\tilde M_A = \Sigma'_A + \hat M - i\epsilon_1 - i \epsilon_2$. On the other hand, it can also easily be generalized to linear $SU(N)$ quivers.} 
\begin{equation}\label{twogaugenodeIPF}
{Z_{{\text{inst}}}} \equiv \sum_{\vec Y,\vec Y'} {{q^{|\vec Y|}}{{q'}^{|\vec Y'|}}{z_{{\text{vect}}}}(\vec Y,\Sigma )\ {z_{{\text{vect}}}}(\vec Y',\Sigma ')\ {z_{{\text{afund}}}}(\vec Y',\Sigma ')\ {z_{{\text{bifund}}}}(\vec Y,\vec Y',\Sigma ,\Sigma ')\ {z_\text{fund}}(\vec Y,\Sigma )} \;,
\end{equation}
where we omitted the mass dependence.

We are interested in the instanton partition function evaluated at special values for its gauge equivariant parameter, 
\begin{equation}
  \Sigma _A \to \Sigma _A^{\vnl, \vnr} \equiv {M^\epsilon_A} + i(n_A^{\text{L}} + 1){\epsilon _1} + i(n_A^{\text{R}} + 1)\epsilon _2\;,
  \label{def:Sigma-special-value}
\end{equation}
for integers $n_A^{\text{L}/\text{R}} \ge 0$. Here $M$ denotes the mass of the fundamental hypermultiplets. We denote the collection of natural numbers simply by $\vec n^\text{L/R} \equiv \{n^\text{L/R}_A\}$, and their sums as $n^\text{L/R} \equiv \sum_{A=1}^N n^\text{L/R}_A$. Remarkably, when evaluated at these special values, the instanton partition function simplifies and exhibits useful factorization properties.

The most significant simplification comes from the evaluation of $z_\text{fund}$: if any Young diagram $Y_A$ of the $N$-tuple $\vec Y$ contains a box (the ``forbidden box'') at position $(n^\text{L}_A + 1, n^\text{R}_A + 1)$, then $z_\text{fund}(\vec Y, \Sigma)$ evaluates to zero. Hence, the sum over all $\vec Y$ is effectively restricted to those tuples all of whose members avoid the ``forbidden box''.\footnote{Such diagrams are sometimes referred to as hook Young diagrams.}

\subsection{Reduction to vortex partition function of SQCD instanton partition function}\label{subapp: reduction-to-vortex-partition-function}
Let us consider the SQCD instanton partition function and look at the case where $n^\text{R} = 0$. The forbidden boxes sit at $(n^\text{L}_A + 1, 1)$, implying that each $Y_A$ in a contributing tuple $\vec Y$ must have width $W_{Y_A} \leq n^\text{L}_A$.

Let $z^{\vnl, \vnr}_\text{vf}(\vec Y)$ denote the product ${z_{{\text{vect}}}}(\vec Y,\Sigma^{\vnl, \vnr} )\ {z_\text{fund}}(\vec Y, \Sigma^{\vnl, \vnr} ,M^\epsilon)$. It simplifies in the case $\vec n^\text{R} = \vec 0$ to
\begin{align}
&z_{{\text{vf}}}^{\vec n^\text{L},\vec 0}(\vec Y) = {z_{{\text{vect}}}}(\vec Y,\Sigma^{\vnl, \vec 0} )\ {z_\text{fund}}(\vec Y, \Sigma^{\vnl,\vec 0} ,M^\epsilon) \nn\\
&= (-1)^{N|\vec Y|}\prod\limits_{A,B = 1}^N \Bigg[  \frac{1}{{\prod\nolimits_{r = 1}^{n_A^\text{L}} {\prod\nolimits_{s = 1}^{n_B^\text{L}} {(1 - i\epsilon _2^{ - 1}{M_{AB}} + {\mathfrak{b}^2}(s - r + n_A^\text{L} - n_B^\text{L}) - {Y_{Ar}} + {Y_{Bs}})_{{Y_{Ar}} - {Y_{Ar + 1}}}^f} } }} \nn \\
  &\qquad\qquad\qquad\qquad \quad  \times \frac{{\prod\nolimits_{s = 1}^{n_B^\text{L}} {(1 + {\mathfrak{b}^2}s - i\epsilon _2^{ - 1}{M_{AB}} + {\mathfrak{b}^2}(n_A^\text{L} - n_B^\text{L}) + {Y_{Bs}} - {Y_{A1}})_{{Y_{A1}}}^f} }}{{\prod\nolimits_{s = 1}^{n_B^\text{L}} {(1 + {\mathfrak{b}^2}s - i\epsilon _2^{ - 1}{M_{AB}} + {\mathfrak{b}^2}(n_A^\text{L} - n_B^\text{L}))_{{Y_{Bs}}}^f} }} \Bigg]\;.
\end{align}
Multiplying in also $z_\text{afund}(\vec Y, \Sigma^{\vnl, \vec 0} ,\tilde M^\epsilon)$, we can identify the resulting product with a summand of a two-/three-dimensional SQCDA vortex partition function. We identify the number of colors and flavors as $n_\text{c} = n^\text{L}$, $n_\text{f} = N$, and $n_\text{af} = N$. The integer partitions are identified as $\{n_A^\text{L}\} \leftrightarrow \{k_i\}$, and finally ${\mathfrak{m}_{A\mu }} = {Y_{A({n^\text{L}_A} - \mu )}}$. Then we recover \eqref{vortex-partition-function-2d}, \eqref{vortex-partition-function-3d} in an obvious way, if one also sets
\begin{align}
    & \text{2d: } & f(x) \equiv & \; \epsilon_2 x, \\
    & \text{3d: } & f(x) \equiv &\; 2i\sinh \pi i{\epsilon _2}(x) & b_\text{3d} \equiv \sqrt {{\epsilon _2}}\;,
\end{align}
and identifies the masses as
\begin{align}
  & \text{2d } \xi_\text{FI} > 0: & {m_X} \equiv &\; i{\epsilon _1}/{\epsilon _2} \;, &  m_{AB} \equiv &\; \epsilon_2^{-1}{M_{AB}}\;, && {m_A} - {\tilde m_J} \equiv \epsilon _2^{ - 1}{\tilde M_{A,J}} + m_X\;, \label{2dparamsident}\\
  & \text{3d } \xi_\text{FI} > 0: & {m_X} \equiv & \; i{\epsilon _1}/\sqrt {{\epsilon _2}}\;, & m_{AB} \equiv &\; \epsilon _2^{-1/2}{M_{AB}}\;,  &&  {m_A} - {\tilde m_J} \equiv  \epsilon _2^{-1/2}{\tilde M_{A,J}} + m_X \;,\label{3dparamsident}
\end{align}
where $M_{AB} = M_A - M_B$ and $\tilde M_{AB} = M_A - \tilde M_B$.

\subsection{Factorization of instanton partition function for large \texorpdfstring{$N$}{N}-tuples of Young diagrams}\label{subapp:factorization-of-instanton-partition-function-large}

\begin{figure}[t]
  \centering
 \includegraphics[width=\textwidth]{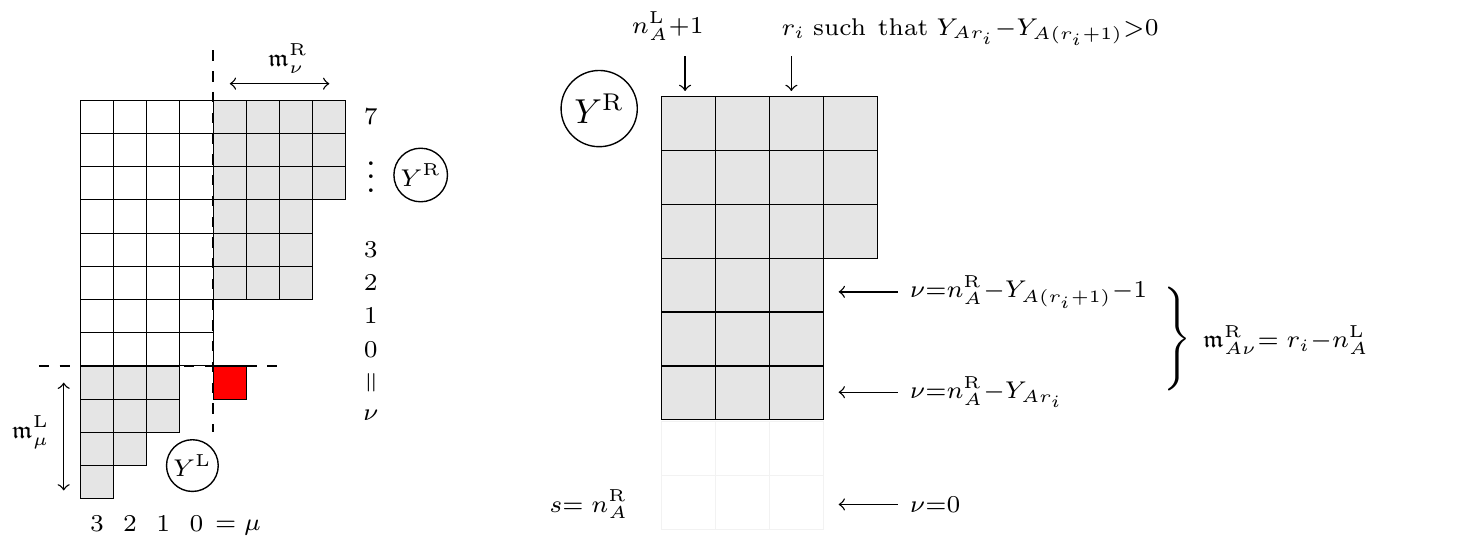}
  \caption{ The left figure demonstrates the decomposition of a large Young diagram $Y_A$ into $Y^\text{L}_A$ and $Y^\text{R}_A.$ The latter are filled in gray, while the ``forbidden box'' is colored red. The figure on the right demonstrates some convenient relations between $\mathfrak{m}^{\text{R}}_{A\mu}$ and $Y^\text{R}_{Ar}$.} \label{Young-diagram}
\end{figure}

Given the set of natural numbers $\vec n^\text{L}$, $\vec n^\text{R}$, we have defined in the main text the notion of large $N$-tuples of Young diagrams, see above equation \eqref{defYLYR}. For such large $N$-tuples we introduced subdiagrams $Y_A^{\text{L}}$ and $Y_A^\text{R}$ in \eqref{defYLYR}, and finally sequences of non-decreasing integers $\mathfrak{m}_{A\mu}^{\text{L}}$ and $\mathfrak{m}_{A\nu}^{\text{R}}$ in \eqref{definitionsms}. In figure \ref{Young-diagram} we remind the reader of these definitions.

Now we are ready to state the factorization of the various factors in the (summand of) the two-gauge-node instanton partition function of \eqref{twogaugenodeIPF}, associated to large Young diagrams, when evaluated on $\Sigma _A^{\vnl, \vnr}$ defined in \eqref{def:Sigma-special-value}. Introducing the shorthand notations $z_\text{afund}^{\vnl, \vnr}(\vec Y, \tilde M^\varepsilon)=z_\text{afund}(\vec Y,\Sigma^{\vnl, \vnr}, \tilde M^\epsilon)$, $z_\text{vf}^{\vec n^\text{L},\vec n^\text{R}}(\vec Y) = z_\text{vect}(\vec Y;\Sigma^{\vnl, \vnr})\ z_\text{fund}(\vec Y,\Sigma^{\vnl, \vnr}, M^\epsilon)$ and $z_{{\text{bifund}}}^{\vnl, \vnr}(\vec Y, \vec Y',\hat M^\epsilon)=z_\text{bifund}(\vec Y,\vec Y',\Sigma^{\vnl, \vnr} ,\Sigma ',\hat M^\epsilon)$, it is straightforward to show that
\begin{align}
z_\text{afund}^{\vnl, \vnr}(\vec Y, \tilde M^\varepsilon) = &\; z_\text{afund}^{\vnl,\vec 0}({{\vec Y}^{\text{L}}},\tilde M^\varepsilon)\ z_\text{afund}^{\vec 0,\vnr}(\vec Y^{\text{R}},\tilde M^\varepsilon )\ (Z_\text{afund,extra}^{\vnl,\vnr} (\tilde M))^{-1}\\
z_\text{vf}^{\vnl, \vnr}(\vec Y) = &\; ( - 1)^{N \vnl \cdot \vnr} \ z_{{\text{vf}}}^{{\vec n^{\text{L}}},\vec 0}({{\vec Y}^{\text{L}}})\ z_{{\text{vf}}}^{\vec 0,\vec n^{\text{R}}}({{\vec Y}^{\text{R}}})\ Z_{{\text{vf,intersection}}}^{\text{large}|\vnl, \vnr}({\mathfrak{m}^{\text{L}}},{\mathfrak{m}^{\text{R}}})\ (Z_\text{vf,extra}^{\vnl, \vnr})^{-1} \label{vf-IPF-factorization-large} \\
z_\text{bifund}^{\vnl, \vnr}(\vec Y, \vec Y',\hat M^\epsilon) = &\; ( - 1)^{N \vnl \cdot \vnr}\ z_{{\text{bifund}}}^{{\vec n^{\text{L}}},\vec 0}({{\vec Y}^{\text{L}}}, \vec Y',\hat M^\epsilon)\ z_{{\text{bifund}}}^{\vec 0,\vec n^{\text{R}}}({{\vec Y}^{\text{R}}}, \vec Y',\hat M^\epsilon)\nn \\
&\quad \times  Z_\text{bifund,intersection} (\vec Y') \ (Z_\text{bifund,extra}^{\vnl, \vnr})^{-1} \label{bifund-IPF-factorization-large}\;.
\end{align}
The product of the latter two can be simplified further to
\begin{multline}\label{factorizationsvfbifund}
z_\text{vf}^{\vnl, \vnr}(\vec Y) \ z_\text{bifund}^{\vnl, \vnr}(\vec Y, \vec Y',\hat M^\epsilon) = {Z_{\text{vortex}|\vnl}}({\mathfrak{m}^{\text{L}}})\ {Z_{{\text{vortex}}|\vnr}}({\mathfrak{m}^{\text{R}}})\ {z_\text{fund}}(\vec Y',\Sigma ',(M')^\epsilon) \\
  \times (Z_\text{vf,extra}^{\vnl, \vnr}\ Z_\text{bifund,extra}^{\vnl, \vnr})^{-1}\ Z_{{\text{vf,intersection}}}^{\text{large}|\vnl, \vnr}({\mathfrak{m}^{\text{L}}},{\mathfrak{m}^{\text{R}}})\   z_{\text{defect}}^{\text{L}}(Y', \mathfrak{m}^\text{L})\  z_{\text{defect}}^{\text{R}}(\tilde Y', \mathfrak{m}^\text{R})\;.
\end{multline}

Let us spell out in detail the various factors and quantities appearing in these factorization results. First of all, new masses of fundamental hypermultiplets, which we denoted as $M'$, appear. They are given by $M'_I = M_I - \hat M + i(\epsilon_1 + \epsilon_2)/2$, and their shifted versions are as usual $(M')^\epsilon = M' - i(\epsilon_1 + \epsilon_2)/2$. We also used the dot product $\vnl \cdot \vnr \equiv \sum_{A = 1}^N {n_A^{\text{L}}} n_A^{\text{R}}$. Next, as in the previous appendix, $Z_{\text{vortex}|\vec n^{\text{L/R}}}$ denotes the vortex partition function of $U(\sum n_A^{\text{L/R}})$ SQCDA with $n_\text{f} = n_\text{af} = N$, whose explicit expressions on $\mathbb{R}^2$ and $\mathbb{R}^2 \times S^1$ can be found in appendix \ref{subapp: S2 partition function} and \ref{subapp: S3 partition function}. The fundamental and adjoint masses are identified as in \eqref{2dparamsident}-\eqref{3dparamsident}, while the antifundamental masses are given by
\begin{align}
  & \text{2d} \ \xi_\text{FI} > 0: & & m_A - {{\tilde m}_J} = \varepsilon _2^{ - 1}({M_A} - {\Sigma '_J} - \hat M) + m_X\\
  & \text{3d} \ \xi_\text{FI} > 0: & &  m_A - {{\tilde m}_J} = \varepsilon _2^{ - 1/2}({M_A} - {\Sigma '_J} - \hat M) + m_X\;.
\end{align}

The factors labeled with `intersection' are given by
\begin{align}
Z_{{\text{vf,intersection}}}^{\text{large}|\vnl, \vnr}({\mathfrak{m}^{\text{L}}},{\mathfrak{m}^{\text{R}}}) \equiv & \;\prod_{A,b = 1}^N {\prod_{\mu  = 0}^{{n^\text{L}_A} - 1} {\prod_{\nu  = 0}^{{n^\text{R}_B} - 1} \frac{1}{{ {f({\Delta _{\text{C}}}(\mathfrak{m}) - {\mathfrak{b}^2})f({\Delta _{\text{C}}}(\mathfrak{m}) + 1)} } } }}\;, \label{intersection-factor-fund-large-diagram-HBL}\\
Z_\text{bifund,intersection}(\vec Y') \equiv &\; \prod_{A,B=1}^N  {\prod_{r = 1}^{W_{Y'_B}} {\prod_{s = 1}^{Y'_{Br}} \frac{1}{f( - i\epsilon _2^{ - 1}({\Sigma '_B} - {M_A} + \hat M) - {\mathfrak{b}^2}r - s)} } } \;,
  \label{intersection-factor-bifund-large-diagram-HBL}
\end{align}
with $\Delta_\text{C}(\mathfrak{m}) \equiv i \epsilon_2^{-1} (M_A - M_B) + (\mathfrak{m}^\text{L}_{A\mu} + \nu) - \mathfrak{b}^2 (\mathfrak{m}^\text{R}_{B\nu} + \mu)$. The factor $z_\text{defect}^{\text{L}}$  is defined as
\begin{align}
z_\text{defect}^{\text{L}}(Y',{\mathfrak{m}^{\text{L}}}) = \prod_{A,B = 1}^N {\prod_{\mu  = 0}^{n_A^{\text{L}} - 1} {\prod\limits_{s = 1}^{{W_{{Y'_B}}}} {\frac{{( - i\varepsilon _2^{ - 1}({\Sigma '_B} - (M'_A)^\epsilon ) + (\mu  + 1 + s){\mathfrak{b}^2} + \mathfrak{m}_{A\mu }^{\text{L}} - {Y'_{Bs}})_{{Y'_{Bs}}}^f}}{{( - i\varepsilon _2^{ - 1}({\Sigma '_B} - (M'_A)^\epsilon) + (\mu  + s){\mathfrak{b}^2} + \mathfrak{m}_{A\mu }^{\text{L}} - {Y'_{Bs}})_{{Y'_{Bs}}}^f}}} } } \;,\label{zLdef_HB}
\end{align}
and $z_\text{defect}^{\text{R}} (\tilde Y', \mathfrak{m}^\text{R})$ is the same expression but with $(\nl, \mathfrak{m}^\text{L}, \epsilon_2, Y') \leftrightarrow (\nr, \mathfrak{m}^\text{R}, \epsilon_1, \tilde Y')$.

Finally, the factors labeled by `extra' read
\begin{align}
Z_\text{afund,extra}^{\vnl, \vnr}(\tilde M) \equiv &\; \prod_{A,B = 1}^N \prod_{r = 1}^{n_A^{\text{L}}} \prod_{s = 1}^{n_A^{\text{R}}} \frac{1}{f(i\varepsilon _2^{ - 1}({M_A} - {\tilde M_B}) + {\mathfrak{b}^2}(r - n_A^{\text{L}} - 1) + (s - n_A^{\text{R}} - 1))}  \label{def:Z-afund-extra} \\
Z_\text{vf,extra}^{\vec n^\text{L},\vec n^\text{R}} \equiv &\; \prod_{A,B = 1}^N \frac{\prod_{r = 0}^{n_A^{\text{L}} - n_B^{\text{L}} - 1} {\prod_{s = 0}^{n_A^{\text{R}} - n_B^{\text{R}} - 1} {f({\Delta _-}(r,s))\;f({\Delta _-}(r,s) - {\mathfrak{b}^2} - 1)} } }{\prod_{r = 1}^{n_B^{\text{L}} - n_A^{\text{L}}} {\prod_{s = 1}^{n_A^{\text{R}} - n_B^{\text{R}}} {f({\Delta _+}(r,s) - {\mathfrak{b}^2})\;f({\Delta _+}(r,s) + 1)} } } \label{def:Z-vf-extra}  \\
  & \times \prod_{A,B = 1}^N \prod_{r = 1}^{n_A^{\text{L}}} \prod_{s = 1}^{n_A^{\text{R}}} \frac{1}{f(i\varepsilon _2^{ - 1}({M_A} - { M_B}) + {\mathfrak{b}^2}(r - n_A^{\text{L}} - 1) + (s - n_A^{\text{R}} - 1))} \nn\\
Z_{{\text{bifund,extra}}}^{\vnl, \vnr} \equiv &\;\prod_{A,B = 1}^N {\prod_{r = 1}^{n_A^{\text{L}}} {\prod_{s = 1}^{n_A^{\text{R}}} \frac{1}{f(i\epsilon _2^{ - 1}({\Sigma '_B} - {M_A} + \hat M) + {\mathfrak{b}^2}r + s)} } } \label{def:Z-bifund-extra} \;,
\end{align}
where ${\Delta _\pm}(r,s) \equiv i\epsilon _2^{ - 1}(M_A-M_B) \pm {\mathfrak{b}^2}r - s$.

\subsection{Factorization for small \texorpdfstring{$N$}{N}-tuples of Young diagrams}\label{subapp:factorization-for-small-young-diagrams}
For $N$-tuples of Young diagrams that are not large, which we refer to as \emph{small}, a similar factorization of the summand of the instanton partition function occurs, but is more involved. A (tuple of) small Young diagram $\vec Y$, namely ${Y_{A\nl_A}} < \nr_A$ for some $A$, again defines two non-decreasing sequences of integers as in \eqref{smallms}. In particular, $\mathfrak{m}_{A\mu} ^\text{L} $ can be negative: for each $A$, we define $\bar \mu$ such that $\mathfrak{m}_{\bar \mu_A }^\text{L} \geqslant 0$, $\mathfrak{m}_{\bar \mu_A - 1}^\text{L} < 0$. For simplicity, we show the results for the SQCD instanton partition function. The summand of this instanton partition function evaluated at \eqref{def:Sigma-special-value}, \ie{}, $z_\text{vf}^{\vnl, \vnr}\ z_\text{afund}^{\vnl, \vnr}$, factorizes into, for small $N$-tuple of Young diagrams $\vec Y$,
\begin{multline}
z_{{\text{vf}}}^{\vnl, \vnr}(\vec Y^\text{small})\ z_\text{afund}^{\vnl, \vnr}(\vec Y^\text{small}, \tilde M^\varepsilon)  \\ =Z_{{\text{semi-vortex}}|\vec n^\text{L}}({\mathfrak{m}^\text{L}})\ Z_{{\text{vortex}}|\vec n^\text{R}}({\mathfrak{m}^\text{R}})\ Z_{{\text{vf,intersection}}}^{\vnl, \vnr}({\mathfrak{m}^\text{L}},{\mathfrak{m}^\text{R}})\ \big(Z_\text{afund,extra}^{\vnl,\vnr} (\tilde M)\ Z_\text{vf,extra.}^{\vnl, \vnr}\big)^{-1} 
  \label{IPF-factorization-small}
\end{multline}
where the `extra' are as before, and the intersection factor reads, again with $\Delta _\text{C} = i\epsilon _2^{ - 1}({M_A} - {M_B}) + (\nu  + \mathfrak{m}_{A\mu }^{\text{L}}) - {\mathfrak{b}^2}(\mu  + \mathfrak{m}_{B\nu }^{\text{R}})$,
\begin{multline}
Z_{{\text{vf,intersection}}}^{{{\vec n}^\text{L}},{{\vec n}^\text{R}}}({\mathfrak{m}^\text{L}},{\mathfrak{m}^\text{R}}) = \prod\limits_{A,B = 1}^N {\prod\limits_{\mu  = 0}^{n_A^{\text{L}} - 1} {\prod\limits_{\nu  = 0}^{n_B^{\text{R}} - 1} {\frac{1}{{f({\Delta _{\text{C}}} - {\mathfrak{b}^2})}}} } } \prod\limits_{\substack{A,B = 1 \\ A \ne B{\text{ or }}\bar \mu_A = 0 }}^N {\prod\limits_{\mu  = 0}^{n_A^{\text{L}} - 1} {\prod\limits_{\nu  = 0}^{n_B^{\text{R}} - 1} {\frac{1}{{f({\Delta _{\text{C}}} + 1)}}} } } \\
  \ \times \prod\limits_{A( = B) = 1|\bar \mu_A  > 0}^N {\left[ {\prod\limits_{\mu  = 0}^{\bar \mu _A - 1} {\prod\limits_{\nu  =  - \mathfrak{m}_{A\mu }^{\text{L}}}^{n_A^\text{R} - 1} {\frac{1}{{f({\Delta _{\text{C}}} + 1)}}} } \prod\limits_{\mu  = \bar \mu_A}^{n_A^{\text{L}} - 1} {\prod\limits_{\nu  = 0}^{n_B^\text{R} - 1} {\frac{1}{{f({\Delta _{\text{C}}} + 1)}}} } } \right]}\;, \label{smallintersectionfactor}
\end{multline}
and we defined
\begin{small}
\begin{align}
  & \;Z_{{\text{(semi-)vortex}|\vec n}}^\text{SQCDA}({\mathfrak{m}}) \nn \\
  = & \;\prod\limits_{A > B = 1}^N {\prod\limits_{\mu  = 0}^{{n_A} - 1} {\prod\limits_{\nu  = 0}^{{n_B} - 1} {f( - i\epsilon _2^{ - 1}{M_{AB}} + (\mu  - \nu ){\mathfrak{b}^2} - ({\mathfrak{m}_{A\mu }} - {\mathfrak{m}_{B\nu }}))} } } \prod\limits_{A = 1}^N {\prod\limits_{\mu  > \nu  = 0}^{{n_A} - 1} {f( + (\mu  - \nu ){\mathfrak{b}^2} - ({\mathfrak{m}_{A\mu }} - {\mathfrak{m}_{A\nu }}))} }  \nn \\
  & \times \prod\limits_{A,B = 1}^N {\prod\limits_{\mu  = 0}^{\bar \mu  - 1} {\prod\limits_{\nu  = 0}^{\bar \nu  - 1} {\frac{{( - i\epsilon _2^{ - 1}{M_{AB}} + (\mu  - \nu  - 1){\mathfrak{b}^2} + {\mathfrak{m}_{B\nu }})_{ - {\mathfrak{m}_{A\mu }}}^f}}{{(i\epsilon _2^{ - 1}{M_{AB}} + (\nu  - \mu  + 1){\mathfrak{b}^2})_{ - {\mathfrak{m}_{B\nu }}}^f}}} } }  \nn \\
  & \times \prod\limits_{A,B = 1}^N {\prod\limits_{\mu  = 0}^{\bar \mu  - 1} {\prod\limits_{\nu  = 0}^{\bar \nu  - 1} {\frac{{(i\epsilon _2^{ - 1}{M_{AB}} + (\nu  - \mu  - 1){\mathfrak{b}^2} + {\mathfrak{m}_{A\mu }})_{ - {\mathfrak{m}_{B\nu }}}^f}}{{(i\epsilon _2^{ - 1}{M_{AB}} + (\nu  - \mu  + 1){\mathfrak{b}^2} + 1)_{{\mathfrak{m}_{A\mu }}}^f}}\frac{{(i\epsilon _2^{ - 1}{M_{AB}} + (\nu  - \mu  - 1){\mathfrak{b}^2})_{{\mathfrak{m}_{A\mu }}}^f}}{{(i\epsilon _2^{ - 1}{M_{AB}} + (\nu  - \mu  + 1){\mathfrak{b}^2} + 1)_{ - {\mathfrak{m}_{B\nu }}}^f}}} } }  \nn \\
  & \times \prod\limits_{A,B = 1}^N {\prod\limits_{\mu  = \bar \mu }^{{n_A} - 1} {\prod\limits_{\nu  = \bar \nu }^{{n_B} - 1} {\frac{{( - i\epsilon _2^{ - 1}{M_{AB}} + (\mu  - \nu  - 1){\mathfrak{b}^2})_{{\mathfrak{m}_{B\nu }}}^f}}{{(i\epsilon _2^{ - 1}{M_{AB}} + (\nu  - \mu  + 1){\mathfrak{b}^2} - {\mathfrak{m}_{B\nu }} + 1)_{{\mathfrak{m}_{A\mu }}}^f}}} } } \nn \\
  & \times \prod\limits_{A = 1}^N {\prod\limits_{B( \ne A) = 1}^N {\prod\limits_{\mu  = 0}^{{n_A} - 1} {\frac{1}{{(i\epsilon _2^{ - 1}M_{AB} - \mu {\mathfrak{b}^2} + 1)_{\mathfrak{m}_{A\mu}}^f}}} } } \prod\limits_{A = 1}^N {\prod\limits_{\mu  = \bar \mu }^{{n_A} - 1} {\frac{1}{{( - \mu {\mathfrak{b}^2} + 1)_{{\mathfrak{m}_{A\mu} }}^f}}} } \nn\\
  & \times \frac{{\prod_{B = 1}^N {\prod_{A,\mu | \mathfrak{m}_{A\mu} \ge 0} {(1 - i\varepsilon _2^{ - 1}({M_A} - {{\tilde M}_B}) + (\mu  + 1)\mathfrak{b}^2 - {\mathfrak{m}_{A\mu }})_{{\mathfrak{m}_{A\mu }}}^f} } }}{{\prod_{B = 1}^N {\prod_{A,\mu | \mathfrak{m}_{A\mu} < 0} {(1 - i\varepsilon _2^{ - 1}({M_A} + {{\tilde M}_B}) + (\mu  + 1)\mathfrak{b}^2 )_{ - {\mathfrak{m}_{A\mu }}}^f} } }}\;.\label{smallfactorization}
\end{align}
\end{small}%

\begin{figure}[t]
  \centering
  \includegraphics[width=.2\textwidth]{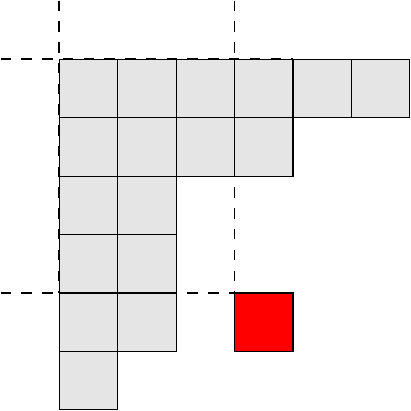}
  \caption{An example of a small Young diagram, with the ``forbidden box'' shown in red. The rectangular region, enclosed by the dashed-lines, is partially filled. In general, the intersection factor $Z_\text{vf,intersection}^{\vnl, \vnr}$ involves a product over those filled boxes.}
  \label{figure:small-young-diagram}
\end{figure}

We remark that in $Z_{{\text{vf,intersection}}}^{\vec n^\text{L},\vec n^\text{R}}$, the second line is in fact a product over the boxes filled inside the $n_A^\text{L} \times n_A^\text{R}$ rectangle, namely the gray boxes inside the region enclosed by the dashed lines in figure \ref{figure:small-young-diagram}. Also note that when all $\bar \mu_A = 0$, \eqref{smallintersectionfactor} turns into \eqref{intersection-factor-fund-large-diagram-HBL}, and the expression for $Z_{{\text{(semi-)vortex}}|\vnl}$ in \eqref{smallfactorization} reduces to the usual vortex partition function, since the small Young diagram has deformed into a large Young diagram.

\section{Poles and Young diagrams in 3d}\label{appendix:extra-pole-3d}
In this appendix we analyze the correspondence between poles in the three-dimensional Coulomb branch matrix model describing the worldvolume theory of intersecting codimension two defects, and (Young) diagrams. We will show that one can construct generic Young diagrams using a class of poles of the matrix model, which we call poles of type-$\hat{\nu}$, and the sum over the corresponding residues is precisely the instanton partition function evaluated at \eqref{def:Sigma-special-value}. All other classes of poles are spurious and their residues should cancel among themselves: we will indeed argue that this is the case by showing that they give rise to certain diagrams, consisting of boxes and anti-boxes, and that these diagrams pair up and the corresponding residues cancel each other. We will first consider generic intersecting defect theories on $S^3_{(1)} \cap S^3_{(2)}$ with gauge groups $U(n^{(1)})$ and $U(n^{(2)})$, sharing $n_\text{f} = n_\text{af} = N$.

\subsection{Poles of type-\texorpdfstring{$\hat{\nu}$}{nuhat}}

We recall from subsection \ref{subsubsection: matrixmodel5d/3d/1d} that the proposed matrix model that computes the partition function of the worldvolume theory of intersecting defects has an integrand of the form, see \eqref{matrixmodelS3US3},
\begin{equation}  \label{matrix-model-intersecting-3d}
Z^{(\mathcal T, S^3_{(1)}\cup S^3_{(2)}\subset S^5_{\vec \omega})}(\sigma^{(1)},\sigma^{(2)})=\frac{Z_{\text{1-loop}}^{(\mathcal T,S^5_{\vec\omega})} }{n^{(1)}!n^{(2)}!} \ Z^{S^3_{(1)}}(\sigma^{(1)}) \ Z_{\text{intersection}}(\sigma^{(1)},\sigma^{(2)})\   Z^{S^3_{(2)}}(\sigma^{(2)})\;,
\end{equation}
where $Z^{S^3_{(1)}}(\sigma^{(1)})$ denotes the integrand of $U(n^{(1)})$ SQCDA on $S^3_{(1)}$ with $\xi_\text{FI}^{(1)} > 0$, and similarly for $Z^{S^3_{(2)}}(\sigma^{(2)})$. Recall that the parameters entering the two three-sphere integrands satisfy various relations, see \eqref{parameterS3US3}. The intersection factor reads
\begin{equation}\label{intersection-factor-CBL}
  {Z_{{\text{intersection}}}}({\sigma ^{(1)}},{\sigma ^{(2)}}) = \prod\limits_{a = 1}^{n^{(1)}} {\prod\limits_{b = 1}^{n^{(2)}} {\prod\limits_ \pm  {{{\left[ {2i\sinh \pi (-{b_{(1)}}\sigma _a^{(1)} + b_{(2)}\sigma _b^{(2)} \pm \frac{i}{2}(b_{(1)}^2 + b_{(2)}^2))} \right]}^{ - 1}}} } }  \;.
\end{equation}

The Jeffrey-Kirwan-like prescription selects a large number of poles in the combined meromorphic integrand (\ref{matrix-model-intersecting-3d}). We now focus on the subclass of poles, defined in \eqref{polesnuhat_main}-\eqref{polesnuhat_main2}, and referred to as \emph{poles of type-$\hat \nu$}. It is useful to observe that $\mathfrak{n}^\text{L}$ and $\mathfrak{n}^\text{R}$ can be decoupled from the following discussion. Using the recursion relations of the double-sine function $s_b(iQ/2 + z)$ and the fact that $\sinh \pi i (x + \mathfrak{n}) = {( - 1)^\mathfrak{n}}\sinh \pi x$, they can be seen to give rise to $Z^{\mathbb{R}^2\times S^1_{(1\cap 3)}}_{\text{vortex}|\vec n_1}$ and $Z^{\mathbb{R}^2\times S^1_{(2\cap 3)}}_{\text{vortex}|\vec n_2}$, independent of the values of $\mathfrak{m}^\text{L}$ and $\mathfrak{m}^\text{R}.$ Therefore, without loss of generality, we will ignore the details of $\mathfrak{n}^\text{L}, \mathfrak{n}^\text{R}$.

It may be helpful to remark that the poles of type-$\hat \nu$, as defined in \eqref{polesnuhat_main}-\eqref{polesnuhat_main2}, can be obtained by solving the component equations
\begin{equation}
  \begin{aligned}
    {Z^{S_{(2)}^3}}:& \;{\sigma ^{(2)}} \to \sigma _{\mathfrak{m},\mathfrak{n}}^{(2)} \\
    Z_{{\text{intersection}}}:&\;{b_{(1)}}\sigma _{A0}^{(1)} = b_{(2)}\sigma _{A{{\hat \nu }_A}}^{(2)} + \frac{i}{2}b_{(1)}^2 + \frac{i}{2}b_{(2)}^2 \\
    {Z^{S_{(1)}^3}}: &\;\sigma _{A0}^{(1)} - m_A^{(1)} =  - i\mathfrak{m}_{A0}^\text{L}{b_{(1)}} - i \mathfrak{n}_{A0}^\text{L}b_\text{L}^{ - 1} \\
    & \; \sigma _{A(\mu  \geqslant 1)}^{(1)} = \sigma _{A(\mu  - 1)}^{(1)} + m_X^{(1)} - i\Delta \mathfrak{m}_{A\mu} ^\text{L}{b_{(1)}} - i\Delta \mathfrak{n}_{A\mu} ^\text{L}b_{(1)}^{ - 1} \;,
  \end{aligned}
  \label{type-I-pole-equations}
\end{equation}
with the requirement that $\mathfrak{m}_{A\hat \nu_A}^{\text{R}}=0$ (which automatically implies that for all $\mu\leq \hat \nu_A$ also $\mathfrak{m}_{A\mu}^\text{R}=0$ since the $\mathfrak m_{A\mu}^{\text{R}}$ are a non-decreasing sequence). One should also bear in mind the parameter relations ${b_{(1)}}m_A^{(1)} - b_{(2)}{m_A^{(2)}} = \frac{i}{2}(b_{(2)}^2 - b_{(1)}^2)$. Here we assigned to $\sigma ^{(2)}$ the poles locations defined in \eqref{old-pole-equations-3d}. As usual, for each $A$, $\sigma^{(1)}_{A0}$ should be solved either with the equation in the second or  third line.

The class with all $\hat{\nu}_A = -1$ is obtained from solving all $\sigma^{(1)}_{A0}$ via the equation in the third line, since $\sigma _{A({{\hat \nu }_A=-1})}^{(2)}$ does not exist. The resulting poles are simply (the union of) the poles $\sigma _{\mathfrak{m},\mathfrak{n}}^{(1)}$ and $\sigma _{\mathfrak{m},\mathfrak{n}}^{(2)}$ of ${Z^{S_{(1)}^3}}$ and ${Z^{S_{(2)}^3}}$ respectively, which  were discussed in detail in appendix \ref{subapp: S3 partition function}. Their residues are equal to the product of the summand of two SQCDA vortex partition functions times the intersection factor evaluated at the pole position. The remaining classes with at least one $\hat \nu _A \ge 0$ are then obtained by solving all four equations. Shortly, we will see that poles of type-$\{\hat \nu_A = -1\}$ are associated with large Young diagrams, while the remaining poles of type-$\hat{\nu}$ are associated with small Young diagrams. The poles of type-$\hat \nu$ are special cases of the more general poles that will be discussed in later appendices.

\subsection{Constructing Young diagrams\label{subapp:constructing-Young-diagrams}}
\begin{figure}[t]
\centering
\includegraphics[width=\textwidth]{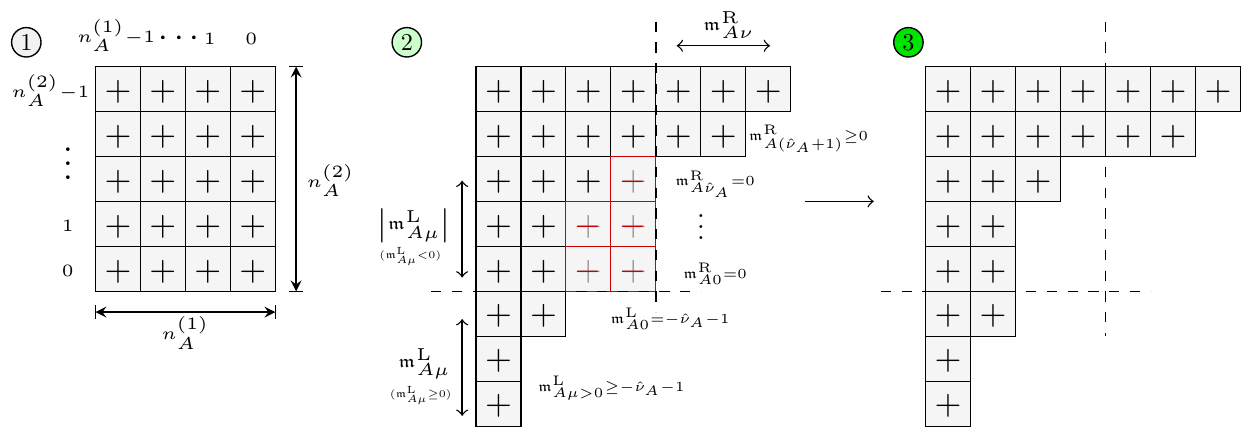}
\caption{ \label{figure:diagrams-and-poles}The three steps in constructing a Young diagram $Y$ using the combinatoric data from a pole of type-$\hat{\nu}$. Boxes with a black $+$ are normal boxes, while boxes with a red $-$ are anti-boxes. Coincident boxes and anti-boxes, \ie{}, the ones with red edges in the middle figure, annihilate to create vacant spots.} 
\end{figure}

We now construct Young diagrams associated with poles of type-$\hat{\nu},$ labeled by the integers $\mathfrak{m}^{\text{L/R}}$, through the following steps. We only present the construction of $Y_A$ for a given $A$, which can be repeated to construct the full $N$-tuple of Young diagrams $\{Y_A\}$. The procedure is also depicted in figure \ref{figure:diagrams-and-poles}.

\begin{itemize}
  \item[1.] Start with a rectangular Young diagram $Y_{\square}$ with $n_A^{(1)}$ columns and $n_A^{(2)}$ rows of boxes. The columns can be relabeled by $\mu$, starting as $\mu = n_A^{(1)}-1$ for the first column and decreasing towards the right in unit steps. Note that the $n_A^{(1)}$-th column has label 0, and columns to the right of it are negatively-labeled. Similarly, the rows can be labeled by $\nu$, starting as $\nu = n_A^{(2)}-1$ for the first row and decreasing in unit steps downwards. 
  \item[2a.] Consider each component $ \sigma _{A\nu} ^{(2)} = m_A^{(2)} + \nu m_X^{(2)} - i\mathfrak{m}_{A\nu} ^\text{R}b_{(2)} - i \mathfrak{n}_{A\nu}^\text{R} b_{(2)}^{-1}$. For each $\nu = 0, ..., n^{(2)} - 1$, attach an extra segment of $\mathfrak{m}^\text{R}_{A\nu}$ boxes to the $\nu$-th row at the right edge of $Y_{\square}$ extending towards the right.
  \item[2b.] Consider each component $ \sigma _{A\mu} ^{(1)} = \;{m^{(1)}} + \mu m_X^{(1)} - i\mathfrak{m}_{A\mu} ^\text{L}{b_{(1)}} - i\mathfrak{n}_{A\mu} ^\text{L}b_{(1)}^{ - 1}$. For each $\mu = 0, ..., n_A^{(1)} - 1$, if $\mathfrak{m}^\text{L}_{A\mu} > 0$, attach an extra segment of $\mathfrak{m}^\text{L}_{A\mu}$ boxes to the $\mu$-th column at the bottom edge of $Y_{\square}$ hanging downwards, or, if $\mathfrak{m}^\text{L}_{A\mu} < 0$, attach a segment of $\mathfrak{m}^\text{L}_{A\mu}$ \emph{anti-boxes} to the $\mu$-th column at the bottom edge standing upwards. 
  \item[3.] An anti-box annihilates a box at the same location, creating a vacant spot.
\end{itemize}

It is now obvious that, poles of type-$\{\text{all }\hat \nu _A = -1\}$ generate large Young diagrams, since all $\mathfrak{m}^{\text{L/R}}$ are non-negative. When there is at least one $\hat{\nu}_A \ge 0$, the poles of type-$\hat{\nu}$ generate small Young diagrams whose $n_A^{(1)}$-th column (labeled as $\mu = 0$) has length $ n_A^{(2)} - \hat{\nu}_A - 1 < n_A^{(2)}$.

\subsection{Residues and instanton partition function\label{subapp:residues-and-instanton-partition-function}}
\begin{figure}[t]
\centering
\includegraphics[width=1\textwidth]{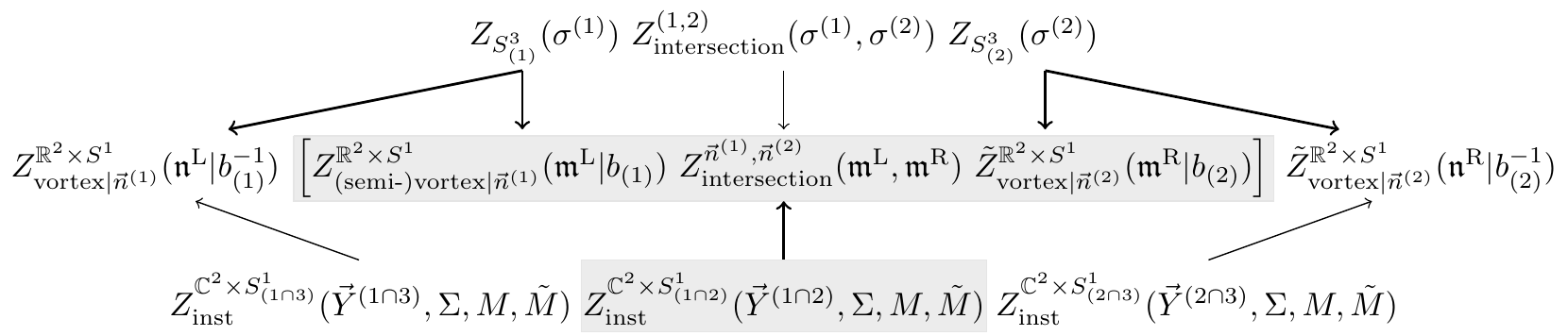}
\caption{ \label{figure:residues-to-residues} Schematic representation of how taking the sum over residues of \ref{sumRes}, drawn by the downward arrows, reproduces the result as obtained from the five-dimensional Higgsing analysis, depicted by the upward arrows. In the bold-face upward pointing arrow, we have omitted the `extra' factors, see appendix \ref{subapp:factorization-of-instanton-partition-function-large}. In the bold-face downward arrows, we have omitted the classical action and one-loop factors (at the Higgs branch vacuum $\vec n^{(1)}$ or $\vec n^{(2)}$ respectively).} 
\end{figure}

The correspondence between poles and Young diagrams in the previous subappendix does not stop at the combinatoric level: it also leads to an equality between residues of the matrix model and the summand of the instanton partition function evaluated at $\Sigma _A^{\vnl, \vnr}$ as in  \eqref{def:Sigma-special-value}. Namely, we will show that
\begin{equation}\label{sumRes}
  \sum_{{\sigma _{{\text{pole}}}} \in \{ {\text{poles of type-}}\hat{\nu} \}} \Res_{\sigma\rightarrow\sigma_{\text{pole}}}{Z^{(\mathcal T, S^3_{(1)}\cup S^3_{(2)}\subset S^5_{\vec \omega})}(\sigma^{(1)},\sigma^{(2)})}  = \text{right-hand side of equation \eqref{totalS5n1n2}}\;.
\end{equation}
Figure \ref{figure:residues-to-residues} indicates schematically how the various factors in the integrand (\ref{matrix-model-intersecting-3d}) reorganize themselves upon taking the residues of the poles of type $\hat \nu$ (ignoring the classical and overall one-loop factors, which are trivial to recover).

Let us present some more details. We start with the poles of type-$(-1)$. These poles are simply the familiar SQCDA poles (\ref{old-pole-equations-3d}), and the corresponding residues of ${Z^{S_{(1)}^3}}$ and ${Z^{S_{(2)}^3}}$ are just the summand of the relevant vortex partition functions multiplied by the classical action and one-loop determinant at the Higgs branch vacuum. The remaining factor $ Z_{\text{intersection}}(\sigma^{(1)},\sigma^{(2)} )$ can be trivially evaluated at the pole $\sigma_\text{pole}$, giving, with $\Delta _{\text{C}}(\mathfrak{m})$ as defined below \eqref{intersection-factor-bifund-large-diagram-HBL} and using the mass relation \eqref{3dparamsident},
\begin{equation}
{Z_{{\text{intersection}}}}({\sigma _{{\text{pole}}}}) = \prod\limits_{A,B = 1}^N {\prod\limits_{\mu  = 0}^{n_A^{(1)} - 1} {\prod\limits_{\nu  = 0}^{n_A^{(2)} - 1} {\frac{1}{{2i\sinh i\pi ({b^2_{(1)}\Delta _{\text{C}}}(\mathfrak{m}) + b_{(1)}^2)\;2i\sinh i\pi ({b^2_{(1)}\Delta _{\text{C}}}(\mathfrak{m}) - b_{(2)}^2)}}} } } \;.
\end{equation}
This is precisely the intersection factor appearing in \eqref{factorizationsvfbifund},\footnote{Since we are considering SQCD, we should set all Young diagrams $\vec Y'$ to be empty. In particular, $z_{\text{fund}}\rightarrow  1$ and $z_{\text{defect}}^{\text{L/R}}\rightarrow 1$. See footnote \ref{simplifytoSQCD}.} see \eqref{intersection-factor-fund-large-diagram-HBL}, with $f(x)=2i\sinh \pi i{b_{(1)}^2}x$. Summing the product of all factors just described over $\mathfrak{m}^{\text{L}}, \mathfrak{m}^{\text{R}}, \mathfrak{n}^{\text{L}}, \mathfrak{n}^{\text{R}}$ reproduces the sum over large diagrams in the right-hand side of \eqref{totalS5n1n2}. Note that we have inserted a trivial factor of one written as the ratio of the extra factors appearing in \eqref{factorizationsvfbifund}. One factor of this ratio completes the Higgsed instanton partition function (of the large $N$-tuples of diagrams), and the other one merges with the three-dimensional one-loop determinants at the Higgs branch vacuum to form the Higgsed five-dimensional one-loop determinant. Of course this should come as no surprise, since the matrix model integrand (\ref{matrix-model-intersecting-3d}) was designed to reproduce the instanton partition functions for large Young diagrams, when evaluated at these poles.

Next we consider the poles of type-$\hat{\nu}$ with some $\hat{\nu}_A > -1$, which we claim will reproduce the small Young diagram contributions to the instanton partition function. Define $\bar \mu_A$ as the smallest integer for which $\mathfrak{m}_{A \bar \mu_A }^{\text{L}} \geq 0,$ \ie{}, $\mathfrak{m}_{A(\bar \mu_A - 1)}^{\text{L}} < 0 \leq \mathfrak{m}_{\bar \mu_A }^{\text{L}}$. Notice that ${\hat \nu _A} \geqslant 0 \Leftrightarrow {\bar \mu _A} > 0$. At this point, we will suppress the details about $\mathfrak n^{\text{L}}$ and $\mathfrak n^{\text{R}}$, as their computational details are similar to the ones just presented for the large diagrams. We first compute the reside of the fundamental one-loop factor in $Z^{S^3_{(1)}}$. It reads
\small
\begin{align}
   & \Res_{\sigma \to \sigma_\text{pole}} \prod_{A=1}^N\prod_{a=1}^{n^{(1)}} s_{b_{(1)}}(\frac{iQ_{(1)}}{2}-\sigma^{(1)}_a +m^{(1)}_A) = {\Big[ {\Res_{z \to 0}{s_{b_{(1)}}}(\frac{iQ_{(1)}}{2} - z)} \Big]^{\sum\limits_{A = 1}^N {{\delta _{\bar \mu _A 0}}} }}\nn \\
  & \times \prod_{\substack{A,B = 1\\A \ne B}}^N {\prod_{\mu  = 0}^{n_A^{(1)} - 1} {{s_{b_{(1)}}}(\frac{iQ_{(1)}}{2} - m_{AB}^{(1)} - \mu m_X^{(1)})} }  \prod_{A = 1}^N {\prod_{\mu  = 1}^{n_A^{(1)} - 1} {{s_{b_{(1)}}}(\frac{iQ_{(1)}}{2} - \mu m_X^{(1)})} } \prod_{\substack{A = 1\\{{\bar \mu }_A} > 0}}^N {{s_{b_{(1)}}}(\frac{iQ_{(1)}}{2} - i{b_{(1)}})} \nn \\
  & \times \prod_{A = 1}^N {\prod_{\mu  = {{\bar \mu }_A}}^{n_A^{(1)} - 1} {\prod_{k = 1}^{\mathfrak{m}_{A\mu }^{\text{L}}} {\frac{1}{{2i\sinh \pi {b_{(1)}}( - \mu m_X^{(1)} + ik{b_{(1)}})}}} } }  \prod_{\substack{A,B = 1\\A \ne B}}^N {\prod_{\mu  = 0}^{n_A^{(1)} - 1} {\prod_{k = 1}^{\mathfrak{m}_{A\mu }^{\text{L}}} {\frac{1}{{2i\sinh \pi {b_{(1)}}( - m_{AB}^{(1)} - \mu m_X^{(1)} + ik{b_{(1)}})}}} } } \nn \\
  & \times \prod_{\substack{A = 1\\{{\bar \mu }_A} > 0}}^N {\left[ {\prod_{\mu  = 1}^{{{\bar \mu }_A} - 1} {\prod_{k = 0}^{ - \mathfrak{m}_{A\mu }^{\text{L}} - 1} {2i\sinh \pi {b_{(1)}}( - \mu m_X^{(1)} - ik{b_{(1)}})} } \prod_{k = 0}^{{{\hat \nu }_A} - 1} {2i\sinh \pi {b_{(1)}}( - i{b_{(1)}} - ik{b_{(1)}})} } \right]} \;. \label{fund-residue-poles-of-type-nuhat}
\end{align}
\normalsize We note that $s_{b_{(1)}}(iQ_{(1)}/2 - i b_{(1)}) = b_{(1)}$. Next we take the residue of one of the factors of \eqref{intersection-factor-CBL}
\begin{equation}
Z_\text{intersection,1} \equiv \prod_{a = 1}^{n^{(1)}} \prod_{b = 1}^{n^{(2)}} \Big(2i\sinh \pi \Big( b_{(2)} \sigma _b^{(2)} - {b_{(1)}}\sigma _a^{(1)} + \frac{i}{2}b_{(1)}^2 + \frac{i}{2}b_{(2)}^2 \Big) \Big)^{-1}\ .
\end{equation}
We also denote the other factor in \eqref{intersection-factor-CBL} as $Z_\text{intersection,2}$. We use again $b_{(1)}^2{\Delta _{\text{C}}}(\mathfrak{m}) = i b_{(1)} m^{(1)}_{AB} + b_{(1)}^2(\mathfrak{m}_{A\mu }^\text{L} + \nu ) - b_{(2)}^2(\mathfrak{m}_{B\nu }^\text{R} + \mu )$, and observe that
\begin{align}
  & \;\prod_{A( = B) = 1}^N {\left[ {\prod_{\nu  = 0}^{ - \mathfrak{m}_{A0}^{\text{L}} - 2} {2i\sinh i\pi ({\Delta _{\text{C}}}(\mathfrak{m}) + b_{(1)}^2)} \prod_{{\mu _A} = 1}^{{{\bar \mu }_A} - 1} {\prod_{\nu  = 0}^{ - \mathfrak{m}_{A\mu }^{\text{L}} - 1} {2i\sinh i\pi ({\Delta _{\text{C}}}(\mathfrak{m}) + b_{(1)}^2)} } } \right]} \nn \\
  = &\; \;\prod_{A( = B) = 1}^N {\left[ {\prod_{k = 0}^{{{\hat \nu }_A} - 1} {2i\sinh i\pi ( - (k + 1)b_{(1)}^2)} \;\prod_{\mu  = 1}^{{{\bar \mu }_A} - 1} {\prod_{k = 0}^{ - \mathfrak{m}_{A\mu }^{\text{L}} - 1} {2i\sinh \pi {b_{(1)}}( - \mu m_X^{(1)} - ik{b_{(1)}})} } } \right]} \nn\;.
\end{align}
The residue of $Z_\text{intersection,1}$ can then be written as 
\begin{align}
  & \Res_{\sigma \to \sigma_\text{pole}}Z_\text{intersection,1} = b_{(1)}^{ - \sum_A \delta_{\bar \mu_A 0}} \prod_{\substack{ A,B = 1 \\ A \ne B{\text{ or }}{{\bar \mu }_A} = 0 }} ^N {\prod_{\mu  = 0}^{n_A^{\text{L}} - 1} {\prod_{\nu  = 0}^{n_B^{\text{R}} - 1} {\frac{1}{{2i\sinh i\pi ({\Delta _{\text{C}}} + b_{(1)}^2)}}} } } \nn \\
  & \times \prod\limits_{\substack{A( = B) = 1 \\ {{\bar \mu }_A} > 0}}^N {\left[ {\prod\limits_{\mu  = 0}^{{{\bar \mu }_A} - 1} {\prod\limits_{\nu  =  - \mathfrak{m}_{A\mu }^{\text{L}}}^{n_A^{\text{R}} - 1} {\frac{1}{{2i\sinh i\pi ({\Delta _{\text{C}}} + b_{(1)}^2)}}} } \prod\limits_{\mu  = {{\bar \mu }_A}}^{n_A^{\text{L}} - 1} {\prod\limits_{\nu  = 0}^{n_B^{\text{R}} - 1} {\frac{1}{{2i\sinh i\pi ({\Delta _{\text{C}}} + b_{(1)}^2)}}} } } \right]}   \\
  & \times \prod_{A( = B) = 1}^N {{{\left[ {\prod_{k = 0}^{{{\hat \nu }_A} - 1} {2i\sinh i\pi ( - (k + 1)b_{(1)}^2)} \;\prod_{\mu  = 1}^{{{\bar \mu }_A} - 1} {\prod_{k = 0}^{ - \mathfrak{m}_{A\mu }^{\text{L}} - 1} {2i\sinh \pi {b_{(1)}}( - \mu m_X^{(1)} - ik{b_{(1)}})} } } \right]}^{ - 1}}} \;. \nn
\end{align}
Observe that the last line and $b_{(1)}^{\#}$ cancel against the last line in \eqref{fund-residue-poles-of-type-nuhat} and the products of $s_{b_{(1)}}(\frac{iQ_{(1)}}{2} - i b_{(1)})$. The factors in the second line are precisely a product over the filled boxes inside the $n_A^{(1)} \times n_A^{(2)}$ rectangular region, and, together with the leftover factor in the first line and $Z_\text{intersection,2} (\sigma_\text{pole})$, reproduce the intersection factor in the factorization result for small diagrams \eqref{IPF-factorization-small}. The leftover factors of \eqref{fund-residue-poles-of-type-nuhat} together with the residues of other one-loop factors combine into the ``(semi-) vortex'' partition function factor in \eqref{IPF-factorization-small}.

\subsection{Extra poles and diagrams\label{subapp:extra-poles-and-diagrams}}
The matrix model (\ref{matrix-model-intersecting-3d}) has more simple poles, which are selected by the JK prescription, than just those of type-$\hat{\nu}$. All of them assign to $\sigma^{(2)}$ poles of type $\sigma^{(2)}_{\mathfrak{m}, \mathfrak{n}},$ 
\begin{equation}\label{new-poles-general-3d1}
  \sigma _{j\nu }^{(2)} = \;  m_j^{(2)} + \nu m_X^{(2)} - i\mathfrak{m}_{j\nu }^{\text{R}}b_{(2)} - i\mathfrak{n}_{j\nu }^{\text{R}}b_{(2)}^{ - 1}\;, \qquad \qquad  \mathfrak{m}_{j\nu}^{\text{R}},\mathfrak{n}_{j\nu}^{\text{R}} \geqslant 0\;,
\end{equation}
while $\sigma^{(1)}$ are solutions to the component equations
\begin{equation}
  \begin{aligned}
    & & \sigma _a^{(1)} = & \; m_{{i_a}}^{(1)} - i\mathfrak{m}_a^{\text{L}}{b_{(1)}} - i\mathfrak{n}_a^{\text{L}}b_{(1)}^{ - 1}\;, & ~~~~ & \mathfrak{m}_a^{\text{L}},\mathfrak{n}_a^{\text{L}} \geqslant 0\\
    & \text{type adj. : } & \sigma _a^{(1)} = & \; \sigma _b^{(1)} + m_X^{(1)} - i\Delta \mathfrak{m}_{ab}\ b_{(1)} - i\Delta {\mathfrak{n}_{ab}}\ b_{(1)}^{ - 1}\;, & ~~~~ & \Delta {\mathfrak{m}_{ab}},\Delta {\mathfrak{n}_{ab}} \geqslant 0\\
    & \text{\color{black!30!green}type I: } & {b_{(1)}}\sigma _a^{(1)} = & \; b_{(2)}\sigma _b^{(2)} + \frac{i}{2}b_{(1)}^2 + \frac{i}{2}b_{(2)}^2 - i\mathfrak{n}_a^{\text{L}}\;, & ~~~~ &\mathfrak{n}_a^{\text{L}} \geqslant 0 \\
    & \text{\color{black!30!red}type II: } & {b_{(1)}}\sigma _a^{(1)} = & \; b_{(2)}\sigma _b^{(2)} - \frac{i}{2}b_{(1)}^2 - \frac{i}{2}b_{(2)}^2 - i\mathfrak{n}_a^{\text{L}}\;, & ~~~~ & \mathfrak{n}_a^{\text{L}} \geqslant 0 \;.
  \end{aligned}
  \label{new-poles-general-3d2}
\end{equation}
where at least one of the component $\sigma^{(1)}_a$ should be solved by a \typeI or \typeII equation (otherwise one just gets back the poles $\sigma^{(1)}_{\mathfrak{m}, \mathfrak{n}}$, $\sigma^{(2)}_{\mathfrak{m}, \mathfrak{n}}$, which are already discussed). Similar to those of the SQCDA partition functions on $S^3_b$, the poles specified by \eqref{new-poles-general-3d1}-\eqref{new-poles-general-3d2} can be characterized by forest-tree diagrams. However, there are now three possible types of links between two nodes, corresponding to the equations of type-adj., \typeI and {\color{black!30!red}type II}. Note that the poles of type-$\hat\nu$ discussed in the previous section, which gave rise to large and small Young diagrams, can be recovered as special cases of \eqref{new-poles-general-3d1}-\eqref{new-poles-general-3d2}.

For simplicity and clarity of the presentation, we consider the cases of $n_\text{f} = n_\text{af} = 1$, gauge groups $U(n^{(1)} = 2)$ on $S^3_{(1)}$ and $U(n^{(2)})$ on $S^3_{(2)}$. The flavor index $A$ is spurious in this case and will be omitted. For more general unitary gauge groups, the poles can be analyzed following exactly the same logic. We will show that the general simple poles not of type-$\hat\nu$ cancel among themselves. Again, we decouple the $\mathfrak{n}^\text{L}, \mathfrak{n}^\text{R}$ in the following discussions.

\begin{figure}[t]
\centering
\includegraphics[width=.9\textwidth]{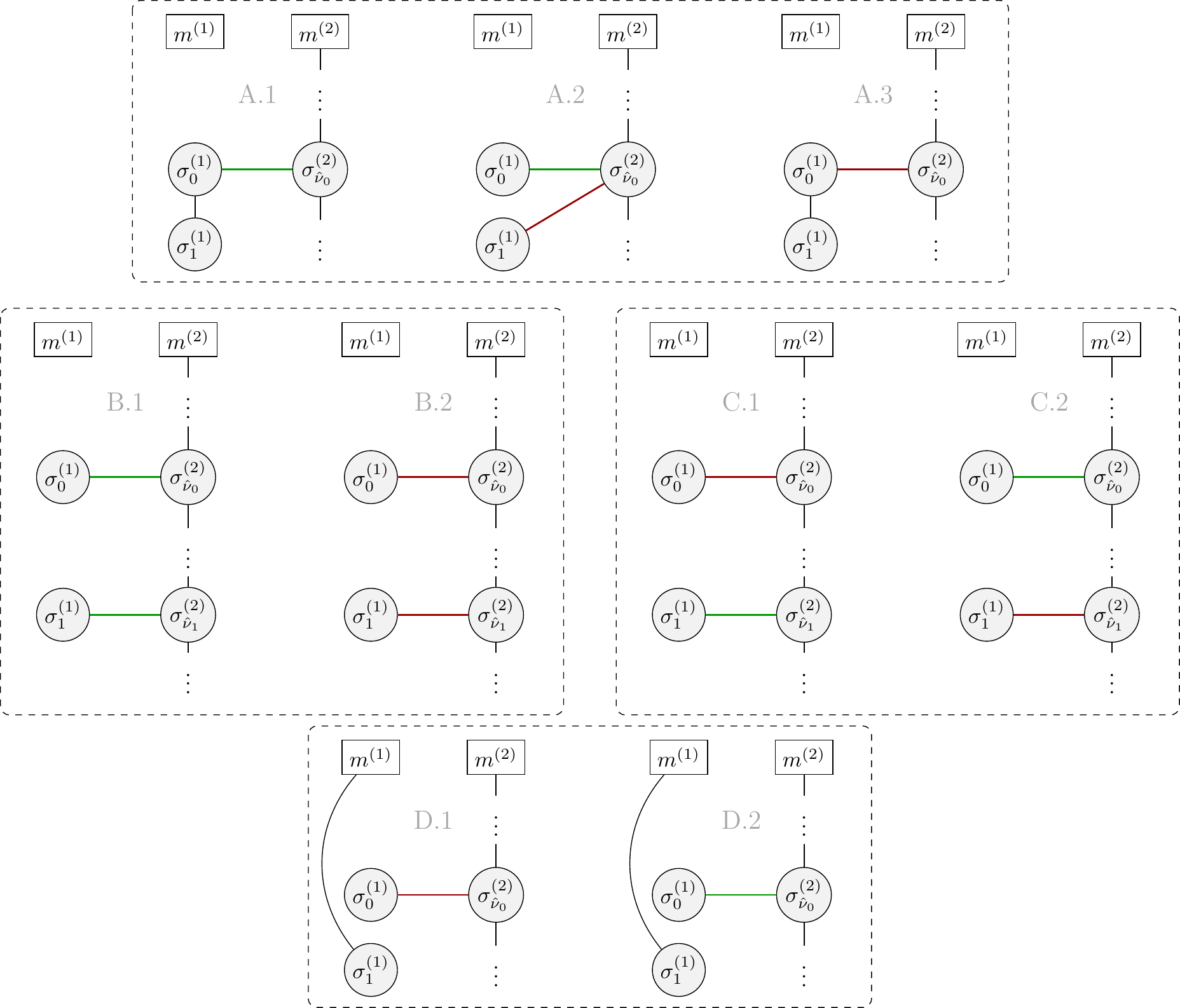}
 \caption{Classes of forests that describe the extra poles. Note that we have omitted some other classes that are obviously not contributing due to symmetry reason. {\color{black!30!green}Green} and {\color{black!30!red}red} lines correspond to \typeI and {\color{black!30!red}type II} equations, which are used to solve $\sigma^{(1)}_{0,1}$ in terms of component(s) of $\sigma^{(2)}$. Poles of type-$\hat \nu$ form a subclass of class A.1. The residues of poles corresponding to non-type-$\hat \nu$ diagrams enclosed within a dashed rectangle cancel each other.   \label{figure:forest-tree-extra-pole}}
\end{figure}

First of all, there are many families of poles. The components $\sigma^{(1)}_\mu, \sigma^{(2)}_\nu$, given by solving \eqref{new-poles-general-3d2}, can be written universally as
\begin{equation}
  \sigma _\mu ^{(1)} = {m^{(1)}} + {h_\mu }m_X^{(1)} - i\mathfrak{m}_\mu ^{\text{L}}{b_{(1)}}, \;\;\;\; \sigma _\nu ^{(2)} = { m^{(2)}} + \nu m_X^{(2)} - i\mathfrak{m}_\nu ^{\text{R}}b_{(2)}\;,
  \label{new-pole-general-U2Un}
\end{equation}
where $h_\mu$ (which later determines the horizontal position of the appended boxes) is closely related to the tree structure that describes the pole, and can be negative. In figure \ref{figure:forest-tree-extra-pole}, we list all classes of contributing forests that describe the above poles, and we tabulate the corresponding values of $h_\mu$ and $\mathfrak{m}_\mu^{\text{L}}$ in table \ref{Table}. Note that poles of type-$\hat \nu$ with $\hat \nu \ge 0$ all lie in class A.1.
\begin{table}
\centering
  \begin{tabular}{c | c | c | c | c}
    & $h_0$ & $\mathfrak{m}^\text{L}_0$ & $h_1$ & $\mathfrak{m}^\text{L}_1$ \\
    \hline
    \hline
    A.1 & $ {\color{black!30!green}- \mathfrak{m}^\text{R}_{\hat \nu_0}}$ & ${\color{black!30!green}- (\hat \nu_0 + 1)}$ & $1 - \mathfrak{m}^\text{R}_{\hat \nu_0}$ & $- (\hat \nu_0 + 1 - \Delta \mathfrak{m}^\text{L}_1),\;\; \forall \Delta \mathfrak{m}^\text{L}_{10}\in \mathbb{N}$ \\
    A.2 & ${\color{black!30!green}- \mathfrak{m}^\text{R}_{\hat \nu_0}}$ & ${\color{black!30!green}- (\hat \nu_0 + 1)}$ & $- \mathfrak{m}^\text{R}_{\hat \nu_0} - 1$ & $- \hat \nu_0 $ \\
    A.3 & ${\color{black!30!red}- \mathfrak{m}^\text{R}_{\hat \nu_0} - 1}$ & $ {\color{black!30!red}- \hat \nu_0}$ & $ - \mathfrak{m}^\text{R}_{\hat \nu_0}$ & $- (\hat \nu_0 - \Delta \mathfrak{m}^\text{L}_{10}),\;\; \forall \Delta \mathfrak{m}^\text{L}_{10}\in \mathbb{N}$ \\
    \hline
    B.1 ($\hat \nu_0 \ne \hat \nu_1$) & ${\color{black!30!green}- \mathfrak{m}^\text{R}_{\hat \nu_0}}$ & ${\color{black!30!green}- (\hat \nu_0 + 1)}$ & ${\color{black!30!green}- \mathfrak{m}^\text{R}_{\hat \nu_1}}$ & ${\color{black!30!green}- (\hat \nu_1 + 1)}$ \\
    B.2 ($\hat \nu_0 \ne \hat \nu_1$)& ${\color{black!30!red}- \mathfrak{m}^\text{R}_{\hat \nu_0} - 1}$ & ${\color{black!30!red}- \hat \nu_0}$ & ${\color{black!30!red}- \mathfrak{m}^\text{R}_{\hat \nu_1} - 1}$ & ${\color{black!30!red}- \hat \nu_1}$ \\
    \hline
    C.1 ($\hat \nu_0 \ne \hat \nu_1$)& ${\color{black!30!red}- \mathfrak{m}^\text{R}_{\hat \nu_0} - 1}$ & ${\color{black!30!red}- \hat \nu_0}$ & ${\color{black!30!green}- \mathfrak{m}^\text{R}_{\hat \nu_1} }$ & ${\color{black!30!green}- (\hat \nu_1 + 1)}$ \\
    C.2 (${\hat \nu_0 \ne \hat \nu_1}$)& ${\color{black!30!green}- \mathfrak{m}^\text{R}_{\hat \nu_0}}$ & ${\color{black!30!green}- (\hat \nu_0 + 1)}$ & ${\color{black!30!red}- \mathfrak{m}^\text{R}_{\hat \nu_1} - 1}$ & ${\color{black!30!red}- \hat \nu_1}$ \\
    \hline
    D.1 & ${\color{black!30!green}- \mathfrak{m}^\text{R}_{\hat \nu_0}}$ & ${\color{black!30!green}- (\hat \nu_0 + 1)}$ & $0$ & $\forall \mathfrak{m}^\text{L}_{10} \in \mathbb{N}$ \\
    D.2 & ${\color{black!30!red}- \mathfrak{m}^\text{R}_{\hat \nu_0} - 1}$ & ${\color{black!30!red}- \hat \nu_0}$ & $0$ & $\forall \mathfrak{m}^\text{L}_1 \in \mathbb{N}$
  \end{tabular}
\caption{\label{Table}Values of of $h_\mu$ and $\mathfrak{m}_\mu^{\text{L}}$ in the pole equation \eqref{new-pole-general-U2Un} corresponding to the tree-diagrams in figure \ref{figure:forest-tree-extra-pole}.}
\end{table}

It is easiest to look for potential cancellations by first inspecting the classical factor. The Coulomb branch classical factor on $S^3_{(1)} \cup S^3_{(2)}$ is ${Z_{{\text{cl.}}}} = \exp \left[ { - 2\pi i\xi _{\text{FI}}^{(1)}\sum_\mu  {\sigma _\mu ^{(1)}}  - 2\pi i\xi _{{\text{FI}}}^{(2)}\sum_\nu  {\sigma _\nu ^{(2)}} } \right]$. Substituting in \eqref{new-pole-general-U2Un}, and using the fact that ${\xi _{{\text{FI}}}^{(1)}{b_{(1)}}} = {\xi _{{\text{FI}}}^{(2)}b_{(2)}} \equiv {\xi _b}$, one has
\begin{equation}
  {Z_{{\text{cl.}}}} = N \exp \left[ { - 2\pi i\xi _{{\text{FI}}}^{(1)}m_X^{(1)}({h_0} + {h_1}) + 2\pi {\xi _b}\left( {\mathfrak{m}_0^{\text{L}} + \mathfrak{m}_1^{\text{L}} + \sum_{\nu  = 0}^{{n^{(2)}} - 1} {\mathfrak{m}_\nu ^{\text{R}}} } \right)} \right]\;,
\end{equation}
where $N$ denotes some common factors shared across all families of poles. Clearly, one necessary condition for two poles to potentially cancel is that they have equal classical contributions, and hence equal ${{h_0} + {h_1}}$ and ${\mathfrak{m}_0^{\text{L}} + \mathfrak{m}_1^{\text{L}} + \sum_{\nu  = 0}^{{n^{(2)}} - 1} {\mathfrak{m}_\nu ^{\text{R}}} }$.

\begin{figure}
 \centering
  \includegraphics[width=.6\textwidth]{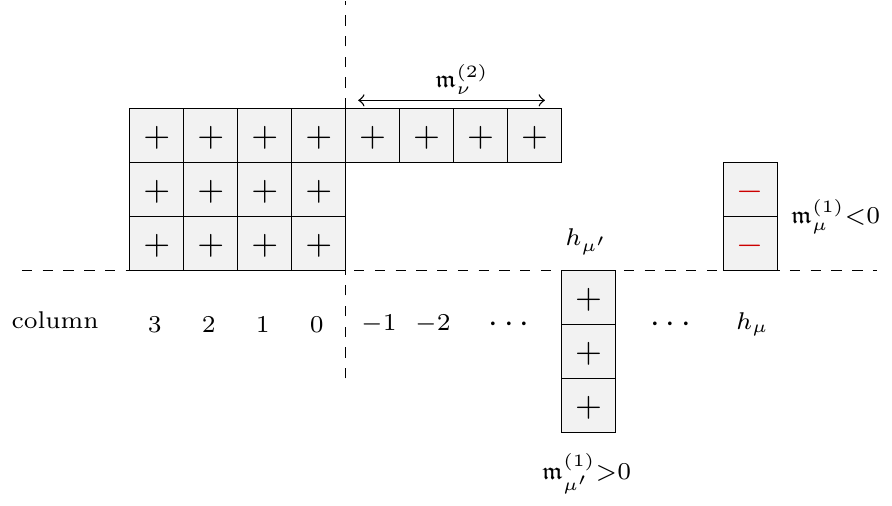}
  \caption{Construction of a general diagram associated with the poles \eqref{new-pole-general-U2Un}.}
  \label{figure:general-diagram-construction}
\end{figure}

\begin{figure}[t]
\centering
\includegraphics[width=\textwidth]{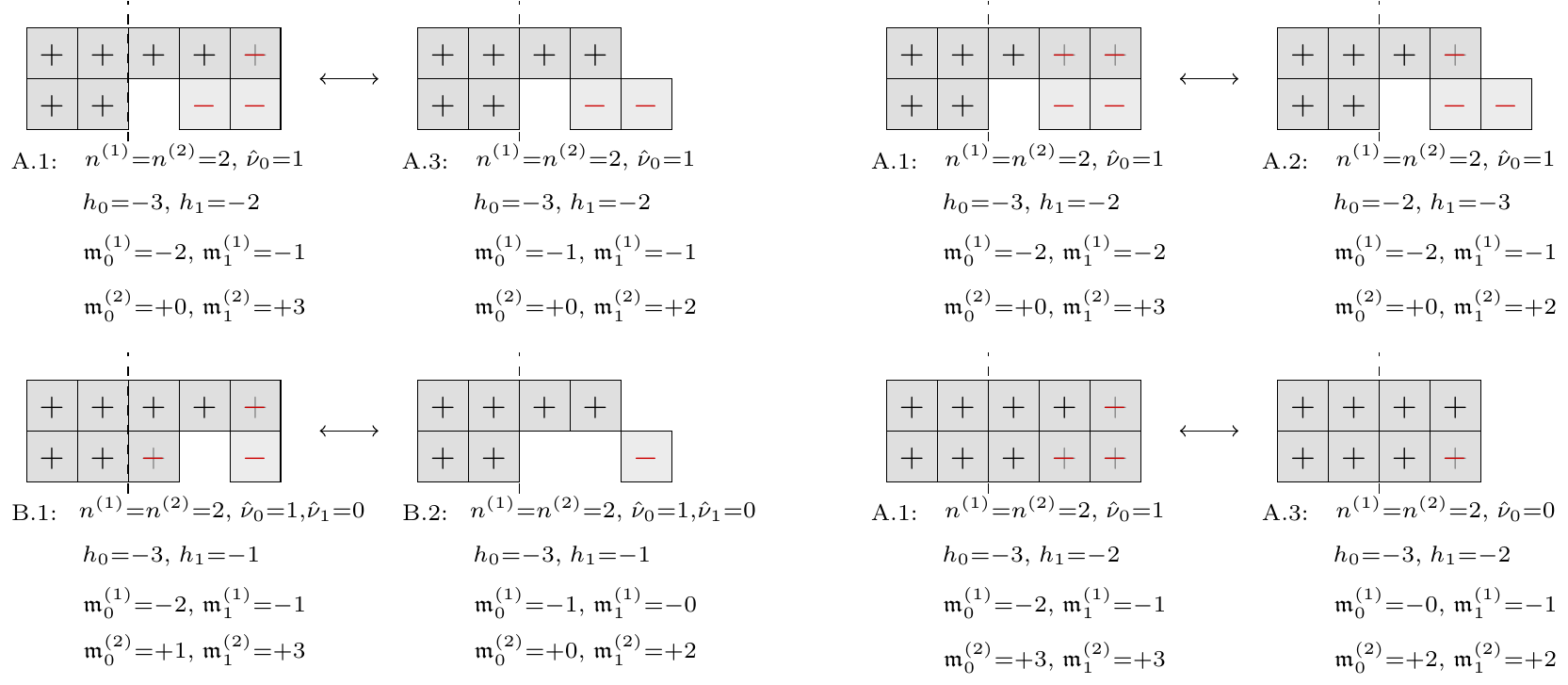}
  \caption{{Some examples of diagrams constructed from the indicated poles. The double arrows indicate that the residues from the related poles, which generate the same diagrams, are opposite.}}
  \label{figure:general-diagram-extra-poles}
\end{figure}

An excellent tool to pinpoint the canceling pairs of poles is again given by diagrams associated with the poles \eqref{new-pole-general-U2Un}. These diagrams consist of boxes and anti-boxes, and it is possible that anti-boxes survive after annihilation. The construction is a simple generalization of that in appendix \ref{subapp:constructing-Young-diagrams}, and is illustrated in figure \ref{figure:general-diagram-construction}: step 1. and 2a. are identical.
When it comes to appending vertical boxes or anti-boxes corresponding to $\sigma^{(1)}_\mu$, one should, generalizing 2b., append to the $h_\mu$-th column. Now that $h_\mu$ can be negative, these vertical segments of boxes can sit \textit{to the right} of $Y_{\square}$, and can have annihilation with the horizontal segments of boxes corresponding to $\sigma^{(2)}_\nu$. Figure \ref{figure:general-diagram-extra-poles} demonstrates a few examples of such diagrams, constructed from several poles.

It can be shown that if two poles contribute opposite residues, then their corresponding diagrams (after annihilation) must be the same. Moreover, given a pole not of type-$\hat \nu$ with associated diagram, one can always find another pole within the same class (A,B,C, or D) with the same diagram; hence they cancel\footnote{Note that poles of type-$\hat \nu$ with $\hat \nu \ge 0$ although being special case of A.1, do not have such canceling siblings, therefore they have non-zero contributions in the end.}. See figure \ref{figure:general-diagram-extra-poles} for some examples. In all these examples, the pairs of poles have indeed equal $h_0 + h_1$ and ${\mathfrak{m}_0^{\text{L}} + \mathfrak{m}_1^{\text{L}} + \sum_{\nu  = 0}^{{n^{(2)}} - 1} {\mathfrak{m}_\nu ^{\text{R}}} }$.

\section{Poles and Young diagrams in 2d}\label{appendix:extra-pole-2d}
In this appendix we study the poles and their residues of the matrix model computing the partition function of intersecting surface defects supported on $S^2_\text{L} \cup S^2_\text{R} \subset S^4_b$. Throughout the appendix we will use (sub-)superscripts L, R for quantities on $S^2_\text{L/R}$, and N, S for quantities associated to the north- or south-pole contributions. The main idea is very similar to the discussion in appendix \ref{appendix:extra-pole-3d}, but slightly more involved, due to the fact that the intersection between $S^2_\text{L}$ and $S^2_\text{R}$ have two connected components, namely the north and south poles. We will need to bring the contributions from both poles together to reproduce the square of the instanton partition function.

\subsection{Four types of poles}

Recall that for a theory $\mathcal{T}$ of $N^2$ free hypermultiplets in the presence of intersecting defects with $U(n^\text{L})$ SQCDA on $S^2_\text{L}$ and $U(n^\text{R})$ SQCDA on $S^2_\text{R}$ respectively, the partition function $Z^{(\mathcal{T}, S^2_\text{L}\cup S^2_\text{R} \subset S^4_b)}$ has integrand
\begin{equation}
  Z^{(\mathcal{T},S^2_\text{L}\cup S^2_\text{R} \subset S^4_b)}(\sigma^{\text{L}},\sigma^{\text{R}}) = \frac{Z_\text{1-loop}^{(\mathcal{T},S^4_b)}}{\nl ! \nr !} \sum\limits_{{B^{\text{L}}},{B^{\text{R}}}} {{Z^{S_\text{L}^2}}({\sigma ^{\text{L}}},{B^{\text{L}}}) \, \prod_\pm Z_{\text{intersection}}^\pm(\sigma^{\text{L}}, B^{\text{L}},\sigma^{\text{R}},B^{\text{R}})\;{Z^{S_\text{R}^2}}({\sigma ^{\text{R}}},{B^{\text{R}}})}\,, 
  \label{matrix-model-intersecting-2d}
\end{equation}
where the intersection factor is defined in \eqref{intersection-factor-CBL-S2}.

The combined meromorphic integrand \eqref{matrix-model-intersecting-2d} has many poles. Recall that $m_X^{\text{R}} = i{b^{ - 2}}$. It is easy to check that all poles take the form
\begin{equation}
  \begin{gathered}
    \left\{ \begin{gathered}
    i\sigma _{A\nu }^{\text{L}} + \frac{{B_{A\nu }^{\text{L}}}}{2} = im_A^{\text{L}} + h_\nu ^{\text{L}}im_X^{\text{L}} + \mathfrak{m}_{A\nu }^{\text{L}} \hfill \\
    i\sigma _{A\nu }^{\text{L}} - \frac{{B_{A\nu }^{\text{L}}}}{2} = im_A^{\text{L}} + h_\nu ^{\text{L}}im_X^{\text{L}} + \mathfrak{n}_{A\nu }^{\text{L}} \hfill \\ 
  \end{gathered}  \right. \hfill \\
    \left\{ \begin{gathered}
    i\sigma _{A\nu }^{\text{R}} + \frac{{B_{A\nu }^{\text{R}}}}{2} = im_A^{\text{R}} + h_\mu ^{\text{R}}im_X^{\text{R}} + \mathfrak{m}_{A\nu }^{\text{R}} \hfill \\
    i\sigma _{A\nu }^{\text{R}} - \frac{{B_{A\nu }^{\text{R}}}}{2} = im_A^{\text{R}} + h_\mu ^{\text{R}}im_X^{\text{R}} + \mathfrak{n}_{A\nu }^{\text{R}} \hfill \\ 
    \end{gathered}  \right.\;. \hfill
  \end{gathered}
  \label{def:abstract-pole-equations-2d}
\end{equation}

First of all, we define \emph{type-old} poles by simply taking the (union of) poles of $Z^{S^2_\text{L}}$ and $Z^{S^2_\text{R}}$ discussed in appendix \ref{subapp: S2 partition function}. Additionally, we introduce three special classes of poles, which we refer to as \emph{type-$\text{N}_{\hat \nu}^+$, $\text{S}_{\hat \nu}^+$} and \emph{$\text{NS}_{\hat \nu}^+$} poles. Their definition goes as follows. We start by selecting partitions $\vnl$, $\vnr$ of the ranks $\nl$, $\nr$: this corresponds to choosing a Higgs branch vacuum of the SQCDA theory living on $S^2_{\text{L}}$ and $S^2_{\text{R}}$ respectively. Next we select a set of integers $\{\hat \nu^\text{N/S}_A, A = 1, \ldots, N\}$, where each $\hat \nu^\text{N/S}_A \in \{-1, 0, \ldots, n^\text{R}_A - 1\}$ and $\sum_{A=1}^N \hat \nu^\text{N/S}_A > -N$. In the end we will sum over all such partitions $\vnl$, $\vnr$ and sets $\{\hat \nu^\text{N/S}_A\}$ to obtain all relevant poles. Then the three special types of poles are given by the abstract equations (\ref{def:abstract-pole-equations-2d}) with $h^{\text{R}}_\nu = \nu$, $h^{\text{L}}_\mu = \mu$, together with the following conditions:

\begin{itemize}
  \item Poles of type-$\text{N}_{\hat \nu^\text{N}}^+$:
  \begin{equation}
    \begin{aligned}
      & \mathfrak{m}_{A(n_A^{\text{R}} - 1)}^{\text{R}} \geqslant ... \geqslant \mathfrak{m}_{A(\hat \nu_A^\text{N}  + 1)}^{\text{R}} \geqslant \mathfrak{m}_{A\hat \nu_A^\text{N} }^{\text{R}} = \mathfrak{m}_{A(\hat \nu_A^\text{N}  - 1)}^{\text{R}} = ... = \mathfrak{m}_{A0}^{\text{R}} = 0, & & \mathfrak{n}_{A\nu} ^{\text{R}} \geqslant 0 & \\
      & \mathfrak{m}_{A(n_A^{\text{L}} - 1)}^{\text{L}} \geqslant ... \geqslant \mathfrak{m}_{A1}^{\text{L}} \geqslant \mathfrak{m}_{A0}^{\text{L}},& & \mathfrak{n}_{A\mu} ^{\text{L}} \geqslant 0 &\\
      & \mathfrak{m}^\text{L}_{A0} = - (\hat \nu^\text{N}_A + 1) \quad \text{if} \ \hat \nu_A^\text{N} \ge 0, \qquad \text{or} \qquad \mathfrak{m}^\text{L}_{A0} \ge 0 \quad \text{if}\ \hat \nu_A^\text{N} = -1\;.
    \end{aligned}
  \end{equation}
  \item Poles of type-$\text{S}_{\hat \nu^\text{S}}^+$:
  \begin{equation}
    \begin{aligned}
      & \mathfrak{n}_{A(\nra - 1)}^{\text{R}} \geqslant ... \geqslant \mathfrak{n}_{A(\hat \nu^\text{S}_A  + 1)}^{\text{R}} \geqslant \mathfrak{n}_{A\hat \nu^\text{S}_A }^{\text{R}} = \mathfrak{n}_{A(\hat \nu^\text{S}_A  - 1)}^{\text{R}} = ... = \mathfrak{n}_{A0}^{\text{R}} = 0, & & \mathfrak{m}_{A\nu} ^{\text{R}} \geqslant 0 & \\
      & \mathfrak{n}_{A(\nla - 1)}^{\text{L}} \geqslant ... \geqslant \mathfrak{n}_{A1}^{\text{L}} \geqslant \mathfrak{n}_{A0}^{\text{L}},& & \mathfrak{m}_{A\mu} ^{\text{L}} \geqslant 0 &\\
      & \mathfrak{n}^\text{L}_{A0} = - (\hat \nu^\text{S}_A + 1) \quad \text{if} \ \hat \nu_A^\text{S} \ge 0, \qquad \text{or} \qquad \mathfrak{n}^\text{L}_{A0} \ge 0 \quad \text{if}\ \hat \nu_A^\text{S} = -1\;.
    \end{aligned}
  \end{equation}
  \item Poles of type-$\text{NS}_{\hat \nu^\text{N}\hat \nu^\text{S}}^+$:
  \begin{equation}
    \begin{aligned}
      &\mathfrak{m}_{A(n_A^{\text{R}} - 1)}^{\text{R}} \geqslant ... \geqslant \mathfrak{m}_{A(\hat \nu_A^\text{N}  + 1)}^{\text{R}} \geqslant \mathfrak{m}_{\hat \nu_A^\text{N} }^{\text{R}} = \mathfrak{m}_{\hat \nu_A^\text{N}  - 1}^{\text{R}} = ... = \mathfrak{m}_0^{\text{R}} = 0 \;, \\
      &\mathfrak{n}_{A(n_A^{\text{R}} - 1)}^{\text{R}} \geqslant ... \geqslant \mathfrak{n}_{A(\hat \nu_A^\text{S}  + 1)}^{\text{R}} \geqslant \mathfrak{n}_{A\hat \nu_A^\text{S} }^{\text{R}} = \mathfrak{n}_{A(\hat \nu^\text{S}  - 1)}^{\text{R}} = ... = \mathfrak{n}_{A0}^{\text{R}} = 0 \;,\\
      &\mathfrak{m}_{A (n_A^{\text{L}} - 1)}^{\text{L}} \geqslant ... \geqslant \mathfrak{m}_{A1}^{\text{L}} \geqslant \mathfrak{m}_{A0}^{\text{L}}\;, \qquad \mathfrak{n}_{A(n_A^{\text{L}} - 1)}^{\text{L}} \geqslant ... \geqslant \mathfrak{n}_{A1}^{\text{L}} \geqslant \mathfrak{n}_{A0}^{\text{L}}, \\
      & \mathfrak{m}^\text{L}_{A0} = - (\hat \nu^\text{N}_A + 1) \quad \text{if} \ \hat \nu_A^\text{N} \ge 0, \qquad \text{or} \qquad \mathfrak{m}^\text{L}_{A0} \ge 0 \quad \text{if}\ \hat \nu_A^\text{N} = -1\\
      &\mathfrak{n}^\text{L}_{A0} = - (\hat \nu^\text{S}_A + 1) \quad \text{if} \ \hat \nu_A^\text{S} \ge 0, \qquad \text{or} \qquad \mathfrak{n}^\text{L}_{A0} \ge 0 \quad \text{if}\ \hat \nu_A^\text{S} = -1\;.
    \end{aligned}
  \end{equation}
\end{itemize}

A few remarks are in order. Poles of type-$\text{N}_{\hat \nu^\text{N} }^+$ come from solving the equations
\begin{equation}
  \left\{ \begin{gathered}
    i\sigma _{C\nu} ^{\text{R}} + \frac{1}{2}B_{C\nu} ^{\text{R}} = i{m_C^{\text{R}}} + \nu im_X^{\text{R}} + \mathfrak{m}_{C\nu} ^{\text{R}}\;, \qquad \text{with}\qquad \mathfrak{m}^{\text{R}}_{C\hat \nu_C^\text{N}} = 0, \qquad C = 1, \ldots, N \hfill \\
    i \sigma^\text{L}_{A0} + \frac{1}{2}B^\text{L}_{A0} - i m_A^\text{L}= + \mathfrak{m}_A{\color{gray}(\ge 0)} \hfill \\
    {b^{ - 1}}\left( {i\sigma _{A0}^{\text{L}} + \frac{1}{2}B_{A0}^{\text{L}}} \right) - b\left( {i\sigma _{A\hat \nu_A ^\text{N}}^{\text{R}} + \frac{1}{2}B_{{\hat \nu_A ^\text{N} }}^{\text{R}}} \right) + \frac{{b + {b^{ - 1}}}}{2} = 0 \hfill \\
    i\sigma _{A(\mu  + 1)}^{\text{L}} + \frac{1}{2}B_{A(\mu  + 1)}^{\text{L}} = i\sigma _{A\mu} ^{
    (\text{L})} + \frac{1}{2}B_{A\mu} ^{\text{L}} + im_X^{\text{L}} + \Delta {\mathfrak{m}_{A(\mu+1)\mu}{\color{gray} (\geqslant 0)}}\;, \qquad \mu = 0, \ldots n^\text{L}_A -1\;. \hfill 
  \end{gathered}  \right.
\end{equation}
If $\hat \nu_A^\text{N} = -1$ for a given $A$, one should use the equation in the second line to obtain $\sigma^\text{L}_{A0} + \frac{1}{2}B^\text{L}_{A0}$  ($\sigma^\text{R}_{A (\nu = -1)}$ does not exist anyway), otherwise the equation in the third line. If $\hat \nu_A^\text{N} = -1$ for all $A$, one simply recovers the poles of type-old which we define separately, and therefore we exclude such case when defining poles of type-$\text{N}_{\hat \nu^\text{N} }^+$. Among the solutions, most of those with $\mathfrak{n}^{\text{L}}_{A0} < 0$ are canceled by zeros of the fundamental one-loop determinant $Z^{S^2_\text{L}}$. Similarly for poles of type-$\text{S}_{\hat \nu^\text{N} }^-$. However, there are survivors from the cancellation, which involve simultaneous solutions to the set of equations
\begin{equation}
  \left\{ \begin{gathered}
    i\sigma _{C\nu} ^{\text{R}} + \frac{1}{2}B_{C\nu} ^{\text{R}} = i{m_C^{\text{R}}} + \nu im_X^{\text{R}} + \mathfrak{m}_{C\nu} ^{\text{R}}, \quad \mathfrak{m}^\text{R}_{B \hat \nu_C^\text{N}} = 0 \hfill \\
    i\sigma _{C\nu} ^{\text{R}} - \frac{1}{2}B_{C\nu} ^{\text{R}} = i{m_C^{\text{R}}} + \nu im_X^{\text{R}} + \mathfrak{n}_{C\nu} ^{\text{R}}, \quad \mathfrak{n}^\text{R}_{B \hat \nu_C^\text{S}} = 0 \hfill \\
    {b^{ - 1}}\left( {i\sigma _{A0}^{\text{L}} + \frac{1}{2}B_{A0}^{\text{L}}} \right) - b\left( {i\sigma _{\hat \nu_A^\text{N}}^{\text{R}} + \frac{1}{2}B_{\hat \nu_A^\text{N}}^{\text{R}}} \right) + \frac{{b + {b^{ - 1}}}}{2} = 0 \hfill \\
    {b^{ - 1}}\left( {i\sigma _{A0}^{\text{L}} - \frac{1}{2}B_{A0}^{\text{L}}} \right) - b\left( {i\sigma _{\hat \nu_A^\text{S}}^{\text{R}} - \frac{1}{2}B_{\hat \nu_A^\text{S}}^{\text{R}}} \right) + \frac{{b + {b^{ - 1}}}}{2} = 0\;. \hfill \\ 
  \end{gathered}  \right.
\end{equation}
Naively, simultaneous solutions to the last two equations seem to correspond to double poles of the integrand, since two separate intersection factors develop a pole. However, they are actually simple poles after canceling with the zeros of $Z^{S^2_\text{L}}$. These poles are called type-$\text{NS}_{\hat \nu^\text{N} \hat \nu^\text{S}}^+$ in the above classification: they have negative $\mathfrak{m}_0^{\text{L}/ \text{R}}, \mathfrak{n}_0^{\text{L}/ \text{R}}$ controlled by $\hat \nu^\text{N/S}$. The presence of these delicate poles forbids us to decouple $\mathfrak{n}$ from the discussion of $\mathfrak{m}$ as we did in the previous appendix.

It is clear that one can construct all pairs of $N$-tuples $(\vec Y^\text{N}, \vec Y^\text{S})$ from the four types of poles. The construction is essentially the same as outlined in appendix \ref{subapp:constructing-Young-diagrams}, where $\mathfrak{m}^\text{L/R}$ will now take care of $\vec Y^\text{N}$, and $\mathfrak{n}^\text{L/R}$ will take care of $ \vec Y^\text{S}$. More precisely, one has the correspondence
\begin{center}
  \begin{tabular}{c | c | c | c}
    type-old & type-$\text{N}^+_{\hat \nu^\text{N}}$ & type-$\text{S}^+_{\hat \nu^\text{S}}$ & type-$\text{NS}^+_{\hat \nu^\text{N} \hat \nu^\text{S}}$\\
    \hline
    (large, large) & (small, large) & (large, small) & (small, small)\\
    $Y^\text{N}_{An^{\text{L}}_A}, Y^\text{S}_{An^{\text{L}}_A} \ge 0$ & $Y^\text{N}_{A n^{\text{L}}_A} = n^{\text{R}}_A - \hat \nu_A^\text{N} - 1$ & $Y^\text{S}_{A n^{\text{L}}_A} = n^{\text{R}}_A - \hat \nu_A^\text{S} - 1$ & $Y^\text{N}_{A n^{\text{L}}_A} = n^{\text{R}}_A - \hat \nu_A^\text{N} - 1$\\
    & $Y^\text{S}_{An^{\text{L}}_A} \ge 0$ & $Y^\text{N}_{An^{\text{L}}_A}\ge 0$ & $Y^\text{S}_{An^{\text{L}}_A} = n^{\text{R}}_A - \hat \nu_A^\text{S} - 1$
  \end{tabular}
\end{center}
Exhausting all four types of poles, one recovers all possible pairs of $N$-tuples of Young diagrams. Again, the residues of the four types of poles sum up to the modulus squared $| Z_\text{inst} |^2$ of the instanton partition function, evaluated at the specific value of its gauge equivariant parameter, which appears in the full $S^4_b$ partition function.

\subsection{Extra poles and diagrams}

There are many extra poles in the integrand \ref{matrix-model-intersecting-2d} selected by the JK prescription, besides the four types of poles discussed above. For simplicity, here we only present the cancellation in the simplest case of $n^{\text{L}} = n_\text{f} =n_\text{af}= 1$. The main idea is very similar to the discussion in appendix \ref{subapp:extra-poles-and-diagrams} and techniques to analyze more general cases can be found there as well.

There are four types of extra poles selected by the JK prescription (we recycle the naming appearing in the previous subsection):
\begin{center}
  \begin{tabular}{c|c|c|c|c|c|c}
    & $h_0^{\text{L}}$ & $\mathfrak{m}^{\text{L}}_0$ & $\mathfrak{n}^{\text{L}}_0$ & $h_\nu^{\text{R}}$ & $\mathfrak{m}^{\text{R}}_\nu $& $\mathfrak{n}^{\text{R}}_\nu$ \\
    \hline
    type-N$^+_{\hat \nu^\text{N}}$ & $- \mathfrak{m}^{\text{R}}_{\hat \nu^\text{N}}$ & $- (\hat \nu^\text{N} + 1)$ & $\in \mathbb{Z}$ & $\nu$ & $\ge 0$ & $\ge 0$ \\
    type-N$^-_{\hat \nu^\text{N}}$ & $- (\mathfrak{m}^{\text{R}}_{\hat \nu^\text{N}} + 1)$ & $- \hat \nu^\text{N} $ & $\in \mathbb{Z}$ & $ \nu $& $\ge 0$ & $\ge 0$ \\
    type-S$^+_{\hat \nu^\text{S}}$ & $- \mathfrak{n}^{\text{R}}_{\hat \nu^\text{S}}$ & $\in \mathbb{Z}$ & $- (\hat \nu^\text{S} + 1)$ & $\nu$ & $\ge 0$ & $\ge 0$ \\
    type-S$^-_{\hat \nu^\text{S}}$ & $- (\mathfrak{n}^{\text{R}}_{\hat \nu^\text{S}} + 1)$ & $\in \mathbb{Z}$  & $- \hat \nu^\text{S} $ & $ \nu $& $\ge 0$ & $\ge 0$ \\
  \end{tabular}
\end{center}
It is straightforward to verify that the residues of poles of type-N$^+_{\hat \nu^\text{N}}$ cancel those of type-N$^-_{\hat \nu^\text{N}}$, and similarly between type-S$^+_{\hat \nu^\text{S}}$ and -S$^-_{\hat \nu^\text{S}}$. Again, from the four types of poles one can construct pairs of general diagrams consisting of boxes and anti-boxes. Two poles contribute opposite residues when their corresponding pairs of diagrams coincide (taking into account of annihilation between coincident boxes/anti-boxes).

\vfill

{
\bibliographystyle{utphys}
\bibliography{ref}
}

\end{document}